\documentclass[preprint]{aastex}
\usepackage{color}
\usepackage{natbib} 
\usepackage{graphicx}
\usepackage{multirow}

\usepackage{url} 

\begin{document}

\title{The Origin of Major Solar Activity - Collisional Shearing Between Nonconjugated Polarities of Multiple Bipoles Emerging Within Active Regions}


\author{Georgios Chintzoglou\altaffilmark{1,2}}

\email{gchintzo@lmsal.com}

\and

\author{Jie Zhang\altaffilmark{3}}

\and

\author{Mark C. M. Cheung\altaffilmark{1,4}}

\and

\author{Maria Kazachenko\altaffilmark{5}}

\altaffiltext{1}{Lockheed Martin Solar and Astrophysics Laboratory,
                        3176 Porter Dr, Palo Alto, CA 94304, USA}
\altaffiltext{2}{University Corporation for Atmospheric Research,
       Boulder, CO 80307-3000, USA}
\altaffiltext{3}{Department of Physics and Astronomy, George Mason University, Fairfax, VA, 22030, USA}
\altaffiltext{4}{Stanford University, Stanford, CA, USA}

\altaffiltext{5}{Space Sciences Laboratory, University of California, Berkeley, CA, USA}

\begin{abstract}
	Active Regions (ARs) that exhibit compact Polarity Inversion Lines (PILs) are known to be very flare-productive. However, the physical mechanisms behind this statistical inference have not been demonstrated conclusively. We show that such PILs can occur due to the collision between two emerging flux tubes nested within the same AR. In such multipolar ARs, the flux tubes may emerge simultaneously or sequentially, each initially producing a bipolar magnetic region (BMR) at the surface. During each flux tube's emergence phase, the magnetic polarities can migrate such that opposite polarities belonging to different BMRs collide, resulting in shearing and cancellation of magnetic flux. We name this process ``collisional shearing'' to emphasize that the shearing and flux cancellation develops due to the collision. Collisional shearing is a process different from the known concept of flux cancellation occurring between polarities of a single bipole, a process that has been commonly used in many numerical models.
High spatial and temporal resolution observations from the Solar Dynamics Observatory for two emerging ARs, AR11158 and AR12017, show the continuous cancellation of up to 40\% of the unsigned magnetic flux of the smallest BMR, which occurs at the collisional PIL for as long as the collision persists. The flux cancellation is accompanied by a succession of solar flares and CMEs, products of magnetic reconnection along the collisional PIL. Our results suggest that the quantification of magnetic cancellation driven by collisional shearing needs to be taken into consideration in order to improve the prediction of solar energetic events and space weather.
\end{abstract}

\section{Introduction}\label{INTRODUCTION}
The extraordinary discovery of the magnetic nature of our star and its sunspots by~\citet{Hale_1908} has forever transformed our understanding on the universal role magnetic fields play in the solar system and beyond~\citep{Harvey_1999}. Following this discovery, the theory of magnetic reconnection was proposed in the 1940s~\citep{Giovanelli_1946} to explain the catastrophic release of magnetic energy associated with solar flares (then observed from the ground, in lines like H$\alpha$). Since then, a lot of progress has been made towards uncovering the physics involved, thanks to the improvement of ground-based and space-borne instrumentation, as well as the advent of computational power allowing for physics-based 3D numerical simulations. However, our understanding on what triggers solar activity is still limited in part due to our inability to predict the photospheric evolution of magnetic fields and also due to our inability to observe the full magnetic field vector in the magnetized atmosphere of our star -- the solar corona~\citep{Cargill_2009}. Thus, accurately predicting solar activity is as of now not possible.

\subsection{Observational Background}

\begin{figure*}
        \includegraphics[width=\linewidth]{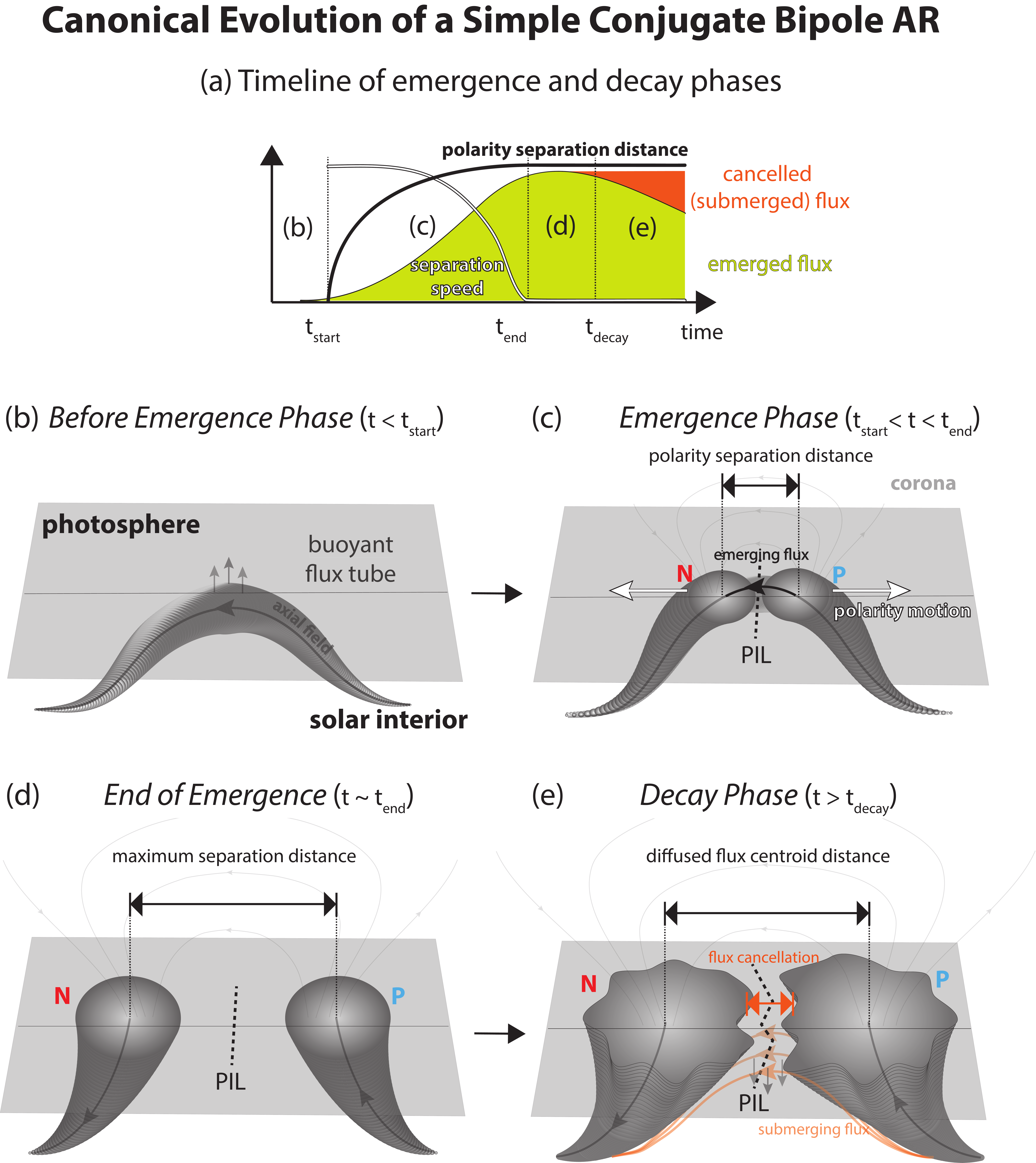}
\caption{
	Illustration of the formation and evolution of a simple, bipolar AR from the emergence of an isolated magnetic flux tube. The local photospheric surface is shown as a grey transparent plane. Top panel (a) summarizes graphically the evolution of magnetic flux (green color) over the lifetime of the bipolar AR. Initially (after the onset of emergence, t$_\mathrm{start}$, (c)), the two opposite polarities (N and P) form a PIL and self-separate rapidly (separation distance and speed are overplotted as black and white curves repsectively). The self-separation of opposite flux progressively slows down until the end of the flux emergence phase, where the bipole has assumed its final (typically maximum) self-separation at the surface (t$_\mathrm{end}$, (d)). The decay phase begins (t$_\mathrm{decay}$, (e)) when surface effects spread and transport the flux from the initially compact polarity concentrations and lead them to cancel out and submerge at the PIL, leading to a reduction of flux observed at the surface. 
}\label{FIG_MODEL}
\end{figure*}

Sunspots often come in pairs; one of positive polarity where the magnetic field is pointing outwards from the surface and one of negative polarity with magnetic field pointing back to the solar interior. In addition, 90\% of these sunspot pairs have the same east-west polarity orientation on the solar surface, a trend that reverses between the north and south hemispheres (Hale's hemispheric polarity rule) and also with every solar cycle (22-year Hale cycle). This observation has led to the understanding that the simple case of bipolar sunspot pairs results from the emergence of 3D ``$\Omega$-shaped'' magnetic flux tube structures from the solar interior -- the Solar Convection Zone (SCZ; the outer 28\% of the solar radius spanning $\approx$ 198\,Mm below the surface; \citealt{Spruit_1974, Christensen-Dalsgaard_etal_1991, Cheung_Isobe_2014}). These magnetic flux tube structures are believed to be part of a toroidal system of magnetic fields created by a dynamo process that acts at a depth in the SCZ \citep{Charbonneau_2005}; small perturbations lead these magnetic flux tubes to develop an $\Omega$-shape and buoyantly traverse the entire SCZ (by means of the magnetic buoyancy instability; \citealt{Parker_1955}). Once such an $\Omega$-shaped flux-tube emerges, its cross-section with the photosphere causes it to manifest itself in the form of a conjugated magnetic polarity pair, i.e. a positive and a negative magnetic polarity originating from the axial field of the same magnetic flux tube (Figure~\ref{FIG_MODEL} b,c). During the emergence phase, these two conjugated polarities are seen to separate from each other (we call this hereafter ``self-separation'') until the end of the emergence, when they assume their maximum separation and final positions on the rotating surface of the Sun (Figure~\ref{FIG_MODEL} d). At all times during the evolution of the system, the two opposite polarities are naturally divided by a line where the magnetic polarity sign flips, the so-called polarity inversion line (in short, PIL). A PIL is a curve whose shape changes with time, depending on the relative position, shape and proximity (bipole compactness) of the self-separating polarities.

During the emergence of buoyant magnetic flux tube structures and the subsequent formation of polarities (manifested as sunspots), there is an increased likelihood for solar activity to occur in the form of plasma heating and high-energy radiation from solar flares and Coronal Mass Ejections (CMEs; \citealt{Schrijver_2007, Schrijver_2009}). Areas where magnetic field concentrations are seen on the solar surface are therefore called ``Active Regions" (ARs). Often, these ARs turn out to be the source of the most energetic events in the solar system and have the potential in making the near-Earth space a hazardous place for advanced technological systems and human space travel.

Depending on the complexity of the emerging magnetic fields, the resultant ARs may be composed of one or more sunspots giving the impression of sunspot groups. The statistical classification of sunspot groups in terms of their complexity~\citep{Hale_Nicholson_1938} has shown that the likelihood of flare productivity scales with increasing complexity of their photospheric magnetic distribution~(e.g. \citealt{Gallagher_Moon_Wang_2002, Georgoulis_Rust_2007}). The most flare-productive of these magnetic distributions is the class of so-called ``$\delta$-spot" ARs~(introduced by \citealt{Kunzel_1960}). These are clusters of sunspot groups in such close proximity to each other that two or more sunspot umbrae of opposite polarity are seen to share a single penumbra at the PIL. Due to this proximity, these sunspot groups exhibit a compact or strong (spatial) gradient PIL~\citep{Schrijver_2007}. Since the determination of the high flare productivity of $\delta$-spots, continuing investigations attempt to explain the origin behind this activity, both through observations~\citep{Zirin_Liggett_1987, Schrijver_2007} and modeling~\citep{Linton_etal_1998, Linton_etal_1999, Toriumi_etal_2014, Takasao_etal_2015, Fang_Fan_2015}. However, there are inherent limitations in both approaches since the exact subsurface 3D structure of ARs is not known. In addition, for the case of \emph{eruptive} flares (i.e. flares associated with CMEs), several possibilities have been proposed regarding magnetic configurations which may lead to eruptions. The theories suggest that the core structure of the CME, the Magnetic Flux Rope (MFR), is formed either on-the-fly (e.g.,~\citealt{Antiochos_etal_1999, Lynch_etal_2008}) or exists \emph{before} the eruption (e.g., ~\citealt{Kliem_Torok_2006}) with increasing evidence for the latter case (e.g. \citealt{Zhang_etal_2012, Patsourakos_etal_2013, Chintzoglou_etal_2015} and references therein). The eruptive MFR may be formed at different layers of the solar atmosphere, ranging from the photosphere/lower chromosphere to the corona, or even ``bodily'' emerge from the SCZ~(as in \citealt{Okamoto_etal_2008, MacTaggart_Hood_2010}; however, bodily emergence cannot happen due to the pressure stratification at shallow depths below the photosphere;~\citealt{Cheung_Isobe_2014}). 


Apart from the appearance of magnetic fields on the Sun after their emergence, observations often reveal the disappearance/annihilation of small magnetic elements, a process known as \emph{photospheric cancellation}. Cancellation was first observed in the 1980s as the mutual disappearance of small, closely-spaced magnetic field patches of opposite polarity in the quiet Sun (QS) and decaying ARs~\citep{Livi_Wang_Martin_1985, Martin_Livi_Wang_1985, Zwaan_1987}. A cancellation event is regarded as a manifestation of either the emergence of a small $\bigcup$-shaped flux tube \citep{Parker_1984, Spruit_etal_1987} or the submergence of a small $\Omega$ flux tube~\citep{Zwaan_1987}. In order for a flux tube to submerge under the photosphere, it must overcome its upward magnetic buoyancy force (in addition, convective downflows can pump small loops down; \citealt{Abbett_etal_2004}). This can be obtained if the downward magnetic tension force of a loop-like flux tube becomes significant (small radius of curvature). \citet{Parker_1979} showed that this occurs if the separation of the bipole's polarities at the photosphere is less than $\approx$ 0.9 Mm, which is consistent with the observations (cancelling polarities in close contact to each other at the resolution limit of modern magnetograms; for HMI this is 1$\arcsec$ equivalent to $\approx$ 0.76\,Mm at the center of the solar disk). 

Decaying ARs are an apparent source of continuous cancellation of opposite magnetic elements as the latter diffuse away from their decaying source polarities. ARs typically enter their decay phase just after the emergence phase stops (see cartoon model in Figure~\ref{FIG_MODEL} panel a; also compare panels d and e). Naturally, magnetic flux is dispersed away from its initial concentrations, leading decaying ARs to lose their large-scale magnetic complexity and appear largely bipolar (\citealt{vanDriel-Gesztelyi_Culhane_2009}). These diffuse bipoles can often be seen over multiple solar rotations. They further decay by additional spreading (due to turbulent diffusion, differential rotation, and meridional flow) and magnetic cancellation along their extended PIL, until they become completely indistinguishable from the background magnetic field of the QS. If the decaying bipole AR is isolated, the PIL separating the diffuse polarities is named internal and if there are several decaying ARs in close proximity to each other, they also form external PILs (\citealt{MacKay_etal_2008, Karna_etal_2017}). Often, a filament is seen above the resultant PILs of those decaying bipoles, and, in many cases, these filaments erupt causing a CME. Typically, prior to the CME, Soft X-ray imaging shows sigmoidal magnetic field structures, which are regarded as the manifestation of a coronal flux rope~\citep{Rust_Kumar_1996}. However, while decaying ARs can indeed produce eruptive events, the most energetic events originate in still evolving, emerging ARs~\citep{Schrijver_2009}.

\subsection{Theoretical Background and Previous Modeling Efforts}\label{THEORY}
Apart from the \emph{CSHKP} ``standard'' eruptive model (acronym acknowledging collectively the work of~\citealt{Carmichael_1964, Sturrock_1966, Hirayama_1974, Kopp_Pneuman_1976}), a plethora of numerical models has been produced towards improving the theoretical understanding of major solar activity. These works can be organized in three modeling classes: (1) data-inspired, (2) data-constrained, and, (3) data-driven models.

\subsubsection{Data-Inspired Models}

Thanks to increases in computational power over the last two decades, we saw a multitude of 3D MHD models that simulated an eruption from first principles. These models can be described as \emph{data-inspired}, since their numerical setups do not directly implement the observational data but rather are inspired by the latter \citep{Cheung_Isobe_2014}. These models can be further split into two sub-categories: (a) \emph{emergence-based} models, and (b) \emph{cancellation-based} (or \emph{diffusion-based}) models.

The majority of the data-inspired models consider an emerging, simple bipolar AR. These emergence-based 3D MHD models start with a single twisted flux tube under the photosphere that is made buoyant to emerge in a stratified atmosphere containing the top parts of the SCZ all the way into the corona; the tube emerges and a conjugated polarity pair forms on the photosphere, followed by a successful eruption (e.g. \citealt{Fan_2001, Manchester_etal_2004, Gibson_Fan_2006, Archontis_Torok_2008, MacTaggart_Hood_2009, Archontis_Hood_2010, Roussev_etal_2012, Leake_etal_2013, Magara_2015, Syntelis_etal_2017}; also see review by \citealt{Cheung_Isobe_2014}). Typically, in these models, a compact PIL is formed between the conjugated polarities, accompanied by shearing motions along its length. This typically lasts for as long as the conjugated pair is self-separating during the emergence process.

The sub-category of cancellation-based 3D numerical models is also capable of producing eruptions by allowing flux cancellation between the model-prescribed, conjugated polarities of a single bipole (e.g. \citealt{vanBallegooijen_Martens_1989, Martens_Zwaan_2001, Amari_etal_2003b, Mackay_vanBallegooijen_2006a, Aulanier_etal_2010, Amari_etal_2011, Zuccarello_etal_2016}). This is obtained either (1) by shearing the field at the PIL and introducing convergence and diffusion at the PIL, or, (2) by only imposing shear and diffusion at the PIL (but no convergence of the polarity centroids) in order to simulate the cancellation process. The first, and representative theoretical model of this sub-category, known as the \emph{photospheric cancellation scenario}~\citep{vanBallegooijen_Martens_1989}, invokes (i) shearing flows, and (ii) convergence of magnetic elements of opposite polarity to produce cancellation by means of magnetic reconnection occurring at the photosphere (or up to a few photospheric pressure scale heights). This model has been used to explain the creation of filament channels and MFRs in the corona. In observational studies, evidence for the validity of this mechanism is commonly found in discrete locations at PILs of decaying ARs where small opposite polarity patches cancel out in sync with the appearance (above the PIL) of sigmoidal coronal structures~\citep{Rust_Kumar_1996, Green_etal_2011, Savcheva_etal_2012b}.

\subsubsection{Data-Constrained Models}

In this category, models utilize observations of the photospheric magnetic field to model the 3D coronal magnetic field and thus we name them as \emph{data-constrained} models. These models are commonly employed to estimate the 3D coronal magnetic field above an AR as the solution of a boundary value problem (with the photospheric magnetic field as the bottom boundary) either by neglecting the presence of currents in the corona, i.e. potential field extrapolation~\citep{Schmidt_1964}, or by assuming that currents are aligned with the magnetic field and therefore are in force balance, i.e. the special case of linear force-free field (in short, LFFF; e.g. computed via fast Fourier transforms;~\citealt{Alissandrakis_1981}) and the more general case of the non-linear force-free field (NLFFF, e.g. computed via optimization techniques~\citealt{Wheatland_etal_2000, Wiegelmann_etal_2006}). Given a magnetogram image series of an AR, it is possible to construct models for each time frame and study the evolution of the coronal magnetic field of an AR, despite that these model outputs represent static equilibria. In the force-free models, it is assumed that the photospheric surface is force-free and torque-free, which is in general not true, but there are techniques that impose this condition (e.g.~\citealt{Wiegelmann_etal_2006}). However, the NLFFF is known to reproduce magnetic flux ropes in modeled eruptive ARs, in agreement with observational evidence in the multi-million degree corona (e.g.~\citealt{Chintzoglou_etal_2015}). 

An alternative approach to achieving an NLFFF field is by starting from the more realistic non-force-free and non-torque-free initial condition (already contained in photospheric vector magnetic field observations) and by iteratively reducing the magnetic stress in the model via the Magnetofrictional method (in short, MF;~\citealt{Yang_etal_1986}) using a simplified set of MHD equations in vacuum. The MF method modifies the initial model field by assuming an ad-hoc inductive velocity, $\mathbf{v}$, parallel to the Lorentz force, eventually relaxing the field into a NLFFF state. This method has been also used to insert a flux rope structure into a potential field model of an AR (e.g. to simulate a filament channel as seen in observations, which naturally cannot be modeled as a potential field) and obtain an NLFFF solution by relaxing it to a static NLFFF equilibrium state (the so-called ``flux-rope insertion method'', e.g.~\citealt{vanBallegooijen_2004, Savcheva_etal_2012}). Other methods take a more computationally expensive approach and evolve the field by solving the full set of the MHD equations and by introducing necessary assumptions about the thermodynamics of the solar atmosphere (e.g.~\citealt{Jiang_Feng_2013, Jiang_etal_2013}, or zero-$\beta$, e.g.~\citealt{Inoue_etal_2015, Inoue_etal_2018}). However, all these approaches generate static equilibria which are not applicable to eruptions.

\subsubsection{Data-Driven Models}

All data-constrained models (such as the Potential, LFFF, NLFFF, Flux-rope insertion, MHD relaxation methods) produce static equilibria. The category of \emph{data-driven} models is distinctly different than that of data-constrained models, since the latter use data only from one time instant. \citet{Cheung_DeRosa_2012} presented the possibility of moving a step forward from solutions of static equilibria and produce evolutionary models, by driving the MF with vector magnetic field observations at the bottom boundary. In this article we implement the data-driven approach of \citet{Cheung_DeRosa_2012} to properly capture the dynamics in the evolution of magnetic fields (similarly to \citealt{Cheung_etal_2015, Fisher_etal_2015}). This approach is necessary to overcome the limitations in studies based on static solutions, which are
unable to capture the dynamic processes that stress and
energize magnetic fields in the solar atmosphere.

\subsection{Remaining Issues between Observations and Modeling}


In this article, we present the analysis of observations of two highly flare- and CME-productive ARs of an overall quadrupolar nature. An evolving quadrupolar configuration (bipole-bipole interaction) essentially represents the next step in terms of ``simplified realism'' which is needed in order to capture the complex processes leading to energetic flares and eruptions. Both of the flare- and CME-productive quadrupolar ARs we present in this article had a long period of time in their emergence phase during which there was virtually no activity at all. Flaring activity occurred only when their individual bipoles, with their conjugated polarity pair ``self-separating'' naturally due to the emergence process, collided with their \emph{nonconjugated} opposite polarities. The bipoles may emerge either (a) simultaneously or (b) sequentially, driving the collision with their self-separation. The flare (and CME) activity did not occur above the self-PIL of the bipoles (as it is typically produced in the aforementioned simulations) but rather above the PIL segment which formed due to the collision of (opposite) nonconjugated polarities. Here, we call this PIL between the colliding nonconjugated polarities the ``\emph{collisional PIL}''. The multi-day flare activity clusters lasted for as long as the collision was taking place and culminated with eruptive flares. In this study, we examine the amount of flux cancelled in collisional PILs and the timing of flux cancellation with respect to flare activity.

In the \citet{Zirin_Liggett_1987} paper, the authors proposed three ways through which a flare-productive delta spot develops.
(1) Emergence of a single complex AR formed below the surface. This is in direct contrast with our findings. We find that the complexity is caused by the interaction of multiple relatively simple flux tubes that emerge through the surface simultaneously or sequentially.
(2) Emergence of satellite spots near a large older spot. This is possibly similar to our scenario of sequential emergence.
(3) Collision of spots of opposite polarity from different bipoles. This is likely similar to our scenario of simultaneous emergence.
Interestingly, \citet{Zirin_Liggett_1987} found that cases (1) and (2) are the cause of large flares, but not case (3).
Also, the authors agreed with the findings of \citet{Kunzel_1960} that the complexity is correlated with flare activity, which we also generally agree. However, the \citet{Zirin_Liggett_1987} paper does not explicitly recognize that the flux cancellation is the key ingredient of flaring activity. Furthermore, that article lacks quantitative analyses of the physical parameters during the observed bipole-bipole interaction.

Contrary to the approach of data-driven and data-constrained modeling (where actual data are utilized as a basis), the aforementioned data-inspired types of numerical experiments (emergence-based and cancellation-based simulations) aim to reproduce the conditions leading to solar eruptions from first principles, typically considering only a single bipole (the latter imposed mainly by computational power constraints). However, despite the soundness of their simplistic approach (the latter mainly due to computational power constraints), our observations of two well-studied ARs are in contrast with these scenarios, (a) at least during the phases of simple bipolar AR emergence and (b) as they evolve to maturity (but before they enter the decay phase and the cancellation of diffused polarities takes place). It is also important to note that even the standard eruptive flare model, the CSHKP model, invokes a simple bipole, also not consistent with our findings.

This article is organized as follows: in \S~\ref{OBS} we present the observations and describe the general evolution of the ARs as well as the various quantities used to characterize their evolution. In \S~\ref{RESULTS} we introduce the method of conjugate flux deficit and report the results of our analysis. In \S~\ref{MODEL} we outline the physics of collisional shearing together with an evolutionary data-driven model of the corona above a collisional PIL, followed by a detailed discussion of the wider implications of collisional shearing in \S~\ref{DISCUSSION}. We conclude in \S~\ref{CONCLUSION}.

\section{Observations}\label{OBS}
The Sun is rotating about its axis in $\sim$25 days at the equator. However, because of our changing vantage point from the Earth as it orbits the Sun, its apparent (synodic) rotation period is $\sim$28 days. Therefore, we are only able to study the magnetic evolution of ARs for $<$14 days. In addition to this limitation, ARs appear to emerge at random longitudes and as a result, the chance of observing their complete evolution becomes rather low (e.g. ARs may begin to emerge on the west visible hemisphere of the sun, but then the solar rotation progressively takes them out of sight). In fact, while in the current solar cycle there have been several ARs that were extremely flare-productive, only a small fraction of these flare-productive ARs has been observed since the beginning of their emergence phase; many of them appeared on the east limb of the solar disk rather complex and already developed, entering their decay phase. Therefore, in order to properly answer the important question of ``\emph{what is the way ARs evolve to produce extreme solar activity?}'', it is \emph{imperative} to observe such ARs from the very moment of their birth all the way to their decay phase. Here, we investigate the evolution of two such well-observed, well-studied, flare- and CME-productive solar ARs catalogued by the US National and Atmospheric Administration (NOAA) with AR numbers 11158~\citep{Schrijver_etal_2011, Jiang_etal_2012, Sun_et_al_2012a, Liu_Schuck_2012, Cheung_DeRosa_2012, Chintzoglou_Zhang_2013, Tziotziou_etal_2013, Toriumi_etal_2014, Aschwanden_etal_2014, Kazachenko_etal_2015, Inoue_etal_2018} and 12017~\citep{Judge_etal_2014, Kleint_etal_2015, Liu_etal_2015, Rubio_da_Costa_etal_2016}. AR11158 began emerging in the east hemisphere of the solar disk on 10 February 2011 (heliographic coordinates E53$\degr$S20$\degr$) and AR12017 on 22 March 2014 (E65$\degr$N10$\degr$).
 
The complete emergence period of these ARs has been uninterruptedly recorded in high spatial (0$\farcs$5 pix$^{-1}$) resolution observations of the photospheric magnetic field and Doppler velocity field with the Helioseismic and Magnetic Imager (HMI;~\citealt{Schou_et_al_2012}) onboard the \emph{Solar Dynamics Observatory} (\emph{SDO};~\citealt{Pesnell_etal_2012}). The HMI instrument captures a 4096 $\times$ 4096 pixel full disk map of the photospheric
magnetic field projected along the line of sight (LOS) every
45\,s, and every 720\,s, the photospheric vector magnetic field by
composing the full Stokes vector, $\mathbf{S}$=(\emph{I,Q,U,V}). The Stokes data are then fed into the HMI data pipeline~\citep{Hoeksema_etal_2014} and are inverted (assuming a Milne-Eddington atmosphere) producing full-disk Doppler velocities, line continuum intensities and the photospheric vector magnetic field observations as maps of the magnitude of the magnetic field, the inclination and its azimuth, $\textbf{B}$=($|\mathbf{B}|$,$\theta$,$\phi$). In addition, the 180$\degr$ ambiguity for the azimuth maps, $\phi$, is resolved as described in~\citet{Hoeksema_etal_2014}, which allows the decomposition of $\textbf{B}$ into a radial, north-south and east-west component on a sphere, ($\mathrm{B_r,B_t,B_p}$). The final observing period we considered spans  $\approx$ 10 days for AR 11158, from 10-Feb-2011 21:58 UT until 19-Feb-2011 05:58 UT (1000 full-disk vector magnetogram frames and Dopplergrams at 12 min cadence) and $\approx$ 11 days for AR12017, from 22-Mar-2014 06:58 UT until 01-Apr-2014 21:58 UT (1263 frames respectively), fully covering the entire evolution of these ARs until they entered their decay phase.  

For calculating the amount of magnetic flux, $\Phi$, that perforates the solar surface, the natural choice is to integrate the radial component of the photospheric magnetic field vector, $B_r$ over the area, $A$, occupied by the magnetic flux concentration as: 

\begin{equation}
\Phi=\int B_r dA
\end{equation}

\noindent For the magnetic field measurements, the noise level for the $B_r$ (from HMI vector maps) is estimated to be that of $|\mathbf{B}|\sim$100\,G (1-$\sigma$ level; \citealt{Hoeksema_etal_2014}) and for the $\mathrm{B_{LOS}}$ at around 10\,G \citep{Couvidat_etal_2016}. It is reported \citep{Hoeksema_etal_2014} that computing the flux from the B$_r$ can lead to spurious increases of this integral toward the limb. This is due to the increased noise levels in the measurements (primarily due to the measurement of the transverse component of the field), especially for locations $\Theta\gtrsim$ 30$\degr$ from the disk center (also producing artificial peaks for the total unsigned flux at $\Theta \approx$ 60$\degr$; \citealt{Hoeksema_etal_2014}, Figure 5 and discussion therein). In other words, while the AR is not emerging, the measured flux would seem to increase rapidly as the AR transits the solar disk past $\Theta=$ 30$\degr$, giving the erroneous impression that it is still emerging. In order to mitigate this effect, one should either (1) compute the flux using only pixels with sufficient magnetic signal ($|\mathbf{B}|\gtrsim$ 100\,G), or (2) use line-of-sight magnetograms, $\mathrm{B_{LOS}}$, for the flux calculation. For our purpose, we chose approach (2) (more details below). 

The major activity of AR11158 (that is, flux emergence, flares and CME eruptions) was clustered around the central meridian during its transit, i.e. between Stonyhurst longitudes $30\degr$E to $30\degr$W (roughly within the acceptable limit of $\Theta\lesssim$ 30$\degr$). Therefore, according to our previous discussion on noise considerations, it is safe to use the radial component of the field, $B_r$, derived from the vector HMI observations, which is nevertheless the natural choice for measuring the magnetic flux. On the other hand, AR12017 began emerging as an inactive bipole at the east limb and transited the solar disk uneventfully, until a week later, when a second bipole emerged within the AR when it was at a Stonyhurst longitude $30\degr$W. This heliographic longitude is at the limit of calculating accurately the flux from the $B_r$ product from the vector HMI observations. Therefore, for the case of AR12017 we chose approach (2), i.e. we used the LOS series. Our choice for implementing approach (2) is justified further since imposing a threshold would impact our ability to characterize PILs. In addition, the observed LOS field is not truly radial but rather projected to the LOS. To partially account for this modulation, in our analysis the observed LOS field, $\mathrm{B}^\mathrm{raw}_\mathrm{LOS}$, was deprojected from the LOS at each pixel assuming the field is radially oriented, as $\mathrm{B_{LOS}}=\mathrm{B}^\mathrm{raw}_\mathrm{LOS}/\mu$, where $\mu$ is the cosine of the angle between the LOS and the local normal at the solar surface.

Before the flux calculations, we first created cutout maps from the full-disk $\mathrm{B_r}$ and $\mathrm{B_{LOS}}$ products with a FOV of 400$\arcsec\times$300$\arcsec$ (large enough to contain an AR) commoving with the AR guiding center at the solar differential rotation rate (adopting the differential rotation coefficient values from \citealt{Snodgrass_Ulrich_1990}). We transformed the cutout maps from their native helioprojective system of coordinates into a local Cartesian system of coordinates by remapping the data into a Cylindrical Equal Area (CEA) projection, which preserves the pixel scale in the entire cutout FOV (approach detailed in \citealt{Hoeksema_etal_2014}). The final size of the cutout series in physical units is 250\,Mm$\times$130\,Mm ($\times$1000 frames) at a time cadence of 12\,min for the duration of the observations. 

We then calculate the flux of each polarity in the CEA projection as

\begin{equation}
	\Phi=\int \mathrm{B} dx dy
\end{equation}

\noindent where $\mathrm{B=B_{LOS}}$ for the case of AR12017, $\mathrm{B=B_r}$ for AR11158 respectively, and $dx=dy=$ 0.36$\times$10$^8$\,cm, the physical pixel scale of the CEA magnetograms. Given the evolution the ARs undergo during their transit period over the solar disk, the relative position of their conjugate polarities changes significantly over time. This evolution is morphing the AR's magnetic spatial distribution as well as the shape of its PIL(s). We track the proper motions of the self-separating polarities of the bipoles in the CEA FOV via their flux-weighted centroids, calculated inside a radius of 5\,Mm around the peak intensity of each of the polarities at each frame in the magnetogram series. This tracking method produced the best results in terms of tracking accurately the motions of the polarities in a smooth and uniform way, by ensuring centroids are not biased/influenced by the presence of like-signed flux in the near vicinity of the major individual polarities (especially when like-signed strong polarities reach close proximity). We use the centroid information to advance the positions of the masks used in the calculation of the magnetic flux for each polarity. It is relatively straightforward to track and isolate the individual polarities with minimal intervention (results in \S~\ref{RESULTS}). The polarity separation distance (or ``self-separation'') of each bipole is measured via the same centroid information. 

We also produced a second CEA dataset per AR with the magnitude of the full photospheric magnetic vector, $|\mathbf{B}|$, same in FOV and cadence with the previously described CEA map series. We used the $|\mathbf{B}|$ CEA maps to infer the 3D subphotospheric configuration of the bipoles during their emergence stage with the Time-Stacking Method, described in \citet{Chintzoglou_Zhang_2013}.

When the polarities collide, they compress against each other and develop a deformation. To characterize this deformation we fit an ellipse around the polarities. There is no expectation for the polarities to be strictly elliptical, however, they can be minimally represented by an elliptical ``blob'' as a measure of their deviation from an overall undisturbed, circular shape.  We perform the fitting on an image series of magnetogram bitmaps (binary maps above a fixed magnetic field threshold value) so as to show the core field of the polarity, corresponding to the sunspot umbra (threshold choice is -1100 G for AR11158 and -1400 G for AR12017). We preferred magnetogram bitmaps instead of using continuum images for the sunspots due to the ease of separating opposite polarities by their different polarity signs. Nevertheless, our choices for the thresholds correspond well to the shapes of the sunspot umbras seen in images of the continuum. The \emph{oblateness} of an ellipse is defined as

\begin{equation}
	        f=\frac{(a-b)}{a}\times100\%
\end{equation}

\noindent where $a$ and $b$  the lengths of the semimajor and semiminor
axes of the ellipse. It reaches a lower limit, $f=0\%$, for the case of a perfect circle, and an upper limit $f=100\%$ for a degenerate ellipse, i.e. a linear segment of length $a$. The magnitude of the oblateness quantifies the magnitude of the deformation of the polarities and thus the severity of the collision. The uncertainty associated with measuring the polarity oblateness was estimated by considering the impact of the SDO's orbital velocity variation in the magnetic field measurements of the HMI data (e.g. \citealt{Hoeksema_etal_2014}; we found it to be up to a maximum $\sim$10\% peak-to-peak variation in flux), together with the increasing noise levels in the measurements approaching $\Theta\gtrsim$ 30$\degr$ from disk center. Such variations could impact the appearance and/or the oblateness of the polarity as measured in the magnetogram bitmaps above a fixed threshold. The maximum uncertainty in the oblateness was found to be $\sim$10\%.

The use of Doppler velocities at the photospheric surface can help differentiate between the different cases of magnetic cancellation at a PIL. If during a cancellation event the disappearing magnetic flux is submerging at the PIL as a small $\Omega$-loop structure, a dominant redshift is expected to be found at that PIL. Inversely, if in that PIL a blueshift is found instead, this may suggest the emergence of a $\bigcup$-loop structure through the photosphere. Thus, proper zero-point calibration of the raw Doppler maps, $\mathrm{v_{raw}}$, obtained by the HMI instrument is important for our investigation. We correct the full-disk Doppler maps for the \emph{SDO}'s velocity vector, $\mathrm{v_{SDO}}$, (which is known very well, to an accuracy of a few cm), and for the differential rotation, $\mathrm{v_{rotation}}$, the North-South ``rotation'' (instrumental effect due to heating of the front door of HMI; \citealt{Scherrer_pc_2018}; private communication), $\mathrm{v_{N/S\ rot}}$, and the red limb shift effect, $\mathrm{v_{red\ limb\ shift}}$, we performed a least-squares fit to obtain time-dependent global fit coefficients for the aforementioned effects. The calibrated Dopplergram series, $\mathrm{v_{Doppler}}$, is obtained as:


\begin{equation}
	\mathrm{v_{Doppler}=v_{raw}-v_{SDO}-v_{rotation}-v_{N/S\ rot}-v_{red\ limb\ shift}}
\end{equation}

\noindent with the data remapped to heliographic as a CEA projection. The FOV in each frame of the CEA Doppler series is identical to the $\mathrm{B_{LOS}}$ and vector $\mathrm{B_r}$ magnetogram CEA series. With regards to the convective blueshift, our experimentation (following the method by \citealt{Welsch_etal_2013}) showed that it alters the zero-point of our mean Doppler velocities by only $\sim$+0.1\,km s$^{-1}$. See \S~\ref{RESULTS} for discussion.

The activity these ARs caused in the corona was recorded by the \emph{SDO}'s Atmospheric Imaging Assembly (AIA;~\citealt{Lemen_etal_2011}) at $\sim0\farcs6$ pix$^{-1}$ every 12\,s, in six passbands of extreme ultraviolet emission (EUV): 171\,\AA\ containing emission from \ion{Fe}{9} (corresponding to plasma temperature T$\approx$0.6\,MK), 193\,\AA\ with \ion{Fe}{12} and \ion{Fe}{24} (1.6 and 20\,MK), 211\,\AA\ \ion{Fe}{14} (2.0\,MK), 335\,\AA\ \ion{Fe}{16} (2.5\,MK), 94\,\AA\ \ion{Fe}{18} (6.0\,MK) and 131\,\AA\ with \ion{Fe}{8} and \ion{Fe}{21} (0.4 and 10\,MK). The formation of filament channels in the core of the ARs was monitored in 304\,\AA\ imaging by \emph{SDO}/AIA (\ion{He}{2}; T$\approx$50,000\,K), in same time-cadence and resolution with the aforementioned passbands. To inspect the spatial distribution and evolution of flare ribbons we used imaging by the 1600\,\AA\ \emph{SDO}/AIA passband in the ultraviolet (UV) part of the spectrum that contains emission from \ion{C}{4} 1550\,\AA\ ($\approx$100,000\,K; chromosphere-corona transition region) and the UV continuum (photosphere). Individual flaring events seen in those wavelengths were detected both by automated detection modules and human observers, and were registered in the Heliophysics Events Knowledgebase (HEK; \citealt{Hurlburt_etal_2012}). Solar activity in higher energies, was monitored in soft X-rays flux (between $\lambda$ 1-8\,\AA) by the National Oceanic and Atmospheric Administration's (NOAA) \textit{Geostationary Operational Environmental Satellites (GOES)} 13 and 15 (the latter after 2012) and in 6-12\,keV photon energies ($\lambda \approx$ 1-2\,\AA) with the \textit{Reuven Ramaty High Energy Solar Spectroscopic Imager} (\textit{RHESSI};~\citealt{Lin_etal_2002}). The GOES and RHESSI X-ray flare event lists also provide the time and location for most of the flares. During the Spring and Fall season, the GOES satellite undergoes a daily eclipse period, when the spacecraft's solar view is obstructed by the Earth. The GOES eclipse period can last from 1-3 hours over the Spring and Fall season. On the other hand, RHESSI, which is in a 38$\degr$ inclined Low Earth Orbit, suffers from an hourly eclipse period as well as passing through the Earth's South Atlantic Anomaly (SAA). This produces hourly data gaps of irregular length, typically 30-40\,min long depending on the overlap of the eclipse and SAA transit period. Using RHESSI data alone, the determination of the flare centering on the solar disk becomes a challenging task. A combination of the GOES time-series with RHESSI mitigates this observational coverage problem. 

The determination of the centering for each flare (magnitudes $>$ C1.0) is vital to assess the importance of these ARs in contributing to the solar activity (as measured by the GOES and RHESSI flux). To alleviate the issues due to the aforementioned data gaps as well as potential erroneous entries, we use the HEK database as a third source of centering for flares, in which the flare centering is determined by the coronal EUV image series from SDO/AIA~\citep{Hurlburt_etal_2012}. That way we obtained three sources to intercorrelate flare events with time and per AR. For the intercorrelation of the event lists we allowed a maximum tolerance of 30 minutes between GOES, RHESSI and HEK events and centerings within 20$\degr\times$15$\degr$ from the guiding center of the AR. Once the relevant flares per AR were selected, we adopted the centering derived from the mean centering for each common intercorrelated flare event in the GOES, RHESSI and HEK event lists (or a subset of them, depending on whether there was a matching entry in all three event lists). Finally, we categorized the importance of all flare events above C1.0 magnitude into two groups: (a) sub-flaring events (flare magnitudes between C1.0 to C9.9) and (b) major flares (for any flare magnitude $\geq$M1.0). This way we obtained a comprehensive list of timing and centering of flare events, which allowed us to assess which location (or PIL) within the AR was associated with the energy release. The results for AR11158 and AR12017 are presented in Tables~\ref{TABLE_1} and~\ref{TABLE_2}, respectively. We also complemented the flare centering information by also noting which flaring events were eruptive (column with the plane-of-the-sky CME speeds obtained from the CDAW online catalog at \url{https://cdaw.gsfc.nasa.gov/CME_list/};~\citealt{Yashiro_etal_2004}). Composing a comprehensive dataset with flare times, flare centering and CME occurrences is necessary in assessing whether activity originates in the self-PIL of the bipoles or somewhere else in the AR. The answer to this question has significant implications on what drives solar activity (see \S~\ref{DISCUSSION}).  



\subsection{The Formation of Collisional Polarity Inversion Lines}\label{CPIL}

\begin{figure*}
	\includegraphics[width=\linewidth]{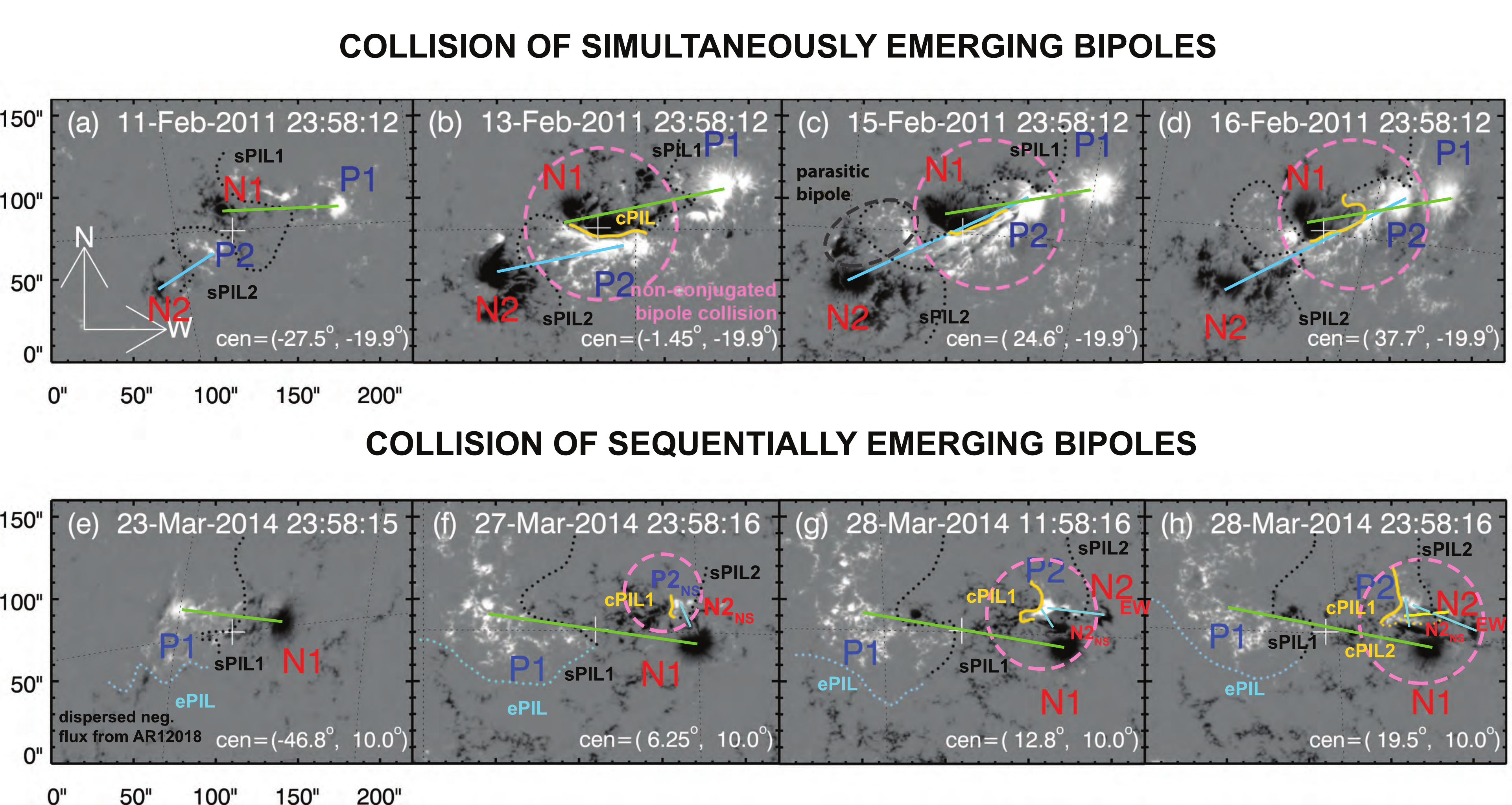}
\caption{
	The emergence stage and subsequent evolution of two flare- and CME-productive ARs. Top panels (a)-(d): Representative snapshot LOS magnetograms showing the evolution of AR 11158 as observed by SDO/HMI. Both bipoles emerge simultaneously with their conjugated polarities (annotated; N1 P1 and N2 P2) naturally separating from each other at their respective self-PILs (sPIL1 and sPIL2). As a result of the natural separation and orientation of the bipoles, a collision occured between the negative polarity of the westward flux tube (N1) and the leading polarity of the eastward flux tube (P2). Due to their opposite polarity signs, a collisional PIL (cPIL; yellow curve) is formed. Thus the overall PIL of AR11158 is composed of the integral system of sPIL2/cPIL/sPIL1. The apparent self-separation distance for each bipole is shown with a green and blue line respectively. Note that a smaller parasitic bipole emerged later on (panel (c); dashed circle). Bottom panels (e)-(h): Selected snapshot LOS magnetograms for AR 12017. Here the emergence of the two flux tubes is sequential (N1 P1 and N2 P2 emergence onsets separated by four days). A cPIL develops between P2 and N1. The greyscale magnetograms in all panels are saturated at $\pm$1,000 G.
	(An animation of this figure is available in the online version of the journal.)
}\label{FIG_1}
\end{figure*}

Qualitatively, both of the ARs we studied exhibit a general quadrupolar configuration either from the very beginning or after some time as they evolve (Figure~\ref{FIG_1} panels a,g). AR11158 was initially (end of February 10) composed of two emerging bipoles with each of their conjugated polarities growing in size and magnetic flux content over time (conjugate polarities denoted with the same number; P1 N1 for bipole 1 and P2 N2 for bipole 2). In addition, as they were emerging, both bipoles were increasing their conjugated polarity separation (i.e. their ``self''-separation) rather uneventfully but still in relatively close proximity to each other (Figure~\ref{FIG_1}a). At that time, NOAA classified AR11158 with a $\beta$ magnetic configuration since only the N1 P1 bipole was strong enough to produce sunspots, thus it appeared in white light observations as a singular bipole AR all the way until the end of February 11 (therefore, a simple, ``monotonic'' PIL could be drawn between the sunspots in this AR). We show the flux evolution of each polarity of the two bipoles in Figure~\ref{FIG_FLUX}a, also noting the daily NOAA Mt Wilson classification of the AR. Suddenly, after the beginning of February 12, polarity P1 was seen to displace towards the north direction by 10\,Mm for the following $\sim$60\,h (similar displacement for N1; see polarity tracking for P1 and N1 in Figure~\ref{FIG_2}a). During the end of the second day of emergence (February 12) a second, stronger episode of emerging flux occurred within each pre-existing bipole almost simultaneously (Figure~\ref{FIG_FLUX}a). We use the term ``second episode'' to describe the subsequent emergence of a secondary bipole \emph{within} the self-PIL of the pre-existing (or first episode) bipole. On the other hand, if subsequent emergence occurs right next to a pre-existing polarity but in the exterior part of the BMR, we call this emergence ``parasitic''. 

During the aforementioned second episodes, each bipole accelerated its self-separation motions (Figure~\ref{FIG_1} and Figure~\ref{FIG_FLUX}a). As a result of this additional self-separation, the two \emph{nonconjugated} polarities (i.e. not belonging to the same bipole; N1 and P2) collided and formed a high spatial gradient (or compact) PIL, evidently as a result of their extremely close proximity. Since this compact PIL formed as a result of the collision of these nonconjugated polarities, we introduce the term \emph{``collisional PIL''} (in short, \emph{``cPIL''}). We make this distinction to differentiate from the PIL portion that still separates opposite flux from the original conjugated polarities (i.e. the self-PILs separating N1 from P1, N2 from P2; in short, \emph{``sPIL1'' and ``sPIL2''}) of the individual bipoles). The full PIL of the AR is then the integral system of sPIL2/cPIL/sPIL1 (as shown in Figure~\ref{FIG_1}b,c). Despite all this dynamic evolution, it is striking that the Mt Wilson classification for AR11158 has been a mere $\beta$ up to February 13. As it can be seen in Figure~\ref{FIG_FLUX}a, initially, both bipoles have very comparable fluxes. What is different is the cohesion of the individual magnetic concentrations -- while N1P1 had formed a strong sunspot, N2P2 appears more of as a collection of pores instead of a well-formed pair of sunspots. This justifies the classification of AR11158 only as $\beta$ until the beginning of February 13. In the meantime, on February 13, a small parasitic bipole emerged right next to polarity N2 with its parasitic negative polarity grazing next to N2 and its positive polarity moving in between N2 and N1. During the course of this parasitic bipole emergence (lasted for about one day from February 13 to February 14; Figure~\ref{FIG_FLUX}a) a narrow-width (or jet-like) non-radial eruption occurred (studied in detail by \citealt{Sun_etal_2012b}). Here, the collisional PIL between N2 and P1 is dominating in terms of activity, and thus for simplicity we do not consider the activity associated with the small parasitic emergence (more in \S 3). The flaring activity centered at the location of the parasitic emergence is shown in Figure~\ref{FIG_FLUX}a with grey vertical lines.  NOAA promoted AR11158 to a $\beta\gamma$ configuration on February 14 until February 16 when the AR was classified as $\beta\gamma\delta$, with the additional emergence of two bipoles within sPIL2 (to simplify the analysis due to these additional bipoles, here we only consider the period up to 12:00~UT of February 15, i.e. before they emerged).  

We extract the location and shape of the cPIL by performing a gradient operation on the CEA magnetogram series described in \citet{Schrijver_2007}. The cPIL is calculated when opposite polarity elements in the region of interest (i.e. the collision site) are at a distance of $\leq$ 5 pixels or 1.8\,Mm. The use of 100\,G threshold for the magnetograms in calculating the gradient with the method of \citet{Schrijver_2007} yielded consistent results than higher thresholds that tend to weaken the sensitivity of the method in extracting the cPIL. To measure the length of collisional PIL, we perform a ``thinning operation'' (using a standard IDL image processing routine named \emph{morph\_thin}) on the extracted PIL until it becomes 1-pixel thick; then the cPIL length is found by summing the pixels and multiplying by the pixel scale (0.36\,Mm pix$^{-1}$ in the CEA magnetograms). The uncertainty associated with that method was estimated by considering the impact of the SDO's orbital velocity variation in HMI data (e.g. \citealt{Hoeksema_etal_2014}; we found it to be up to $\sim$10\% peak-to-peak variation in flux) given the choice of a single threshold. The uncertainty in the measured cPIL length is of order $\sim$10-15\%. We also use the cPIL information as way to define the onset of collision, $\mathrm{t_{collision}}$. We routinely define that moment from the time of the frame in our magnetogram series when the cPIL's continuous length definitely exceeds 40\,Mm; for AR11158 this occurred on 04:00 UT, February 13, 2011. This definition for $\mathrm{t_{collision}}$ appears to provide a consistent onset time with that determined visually from the CEA magnetogram observations.

The self-separation of the main bipoles in AR11158 continues well into the following days (Figure~\ref{FIG_FLUX}c), with P2 rapidly overtaking N1 and reaching P1 (Figure~\ref{FIG_1} and accompanying animation). Note that N1 and P2 have antiparallel speed orientations. The self-separation speeds begin with a maximum value of $\sim$ 400\,m s$^{-1}$ on February 13 and gradually reduce as the collision progresses. Also, the compact cPIL gradually weakens as P2 overtakes N1 and the cPIL is only due to N1 and some dispersed flux trailing from P2 (Figure~\ref{FIG_1} panels c to d). Note that as the collision takes place, the polarities involved (N1 and P2) progressively deform into oblate shapes with their semi-major axes aligned with the collisional PIL (see the animation of Figure~\ref{FIG_1} and the following section). 

The onset, duration and location of collision of these two simultaneously emerging bipoles within AR11158 correlate well, both in terms of space and time, with a cluster of flare activity and several CMEs (see \S~\ref{DISCUSSION}). Similar behavior is seen in AR12017. In this case, the first bipole N1 P1 emerged within about two days (between March 22 to end of March 23) was classified by NOAA as a $\beta$ magnetic configuration for each of these two days respectively. Since the end of its first emergence episode, the conjugated polarities have almost assumed 70\% of their maximum self-separation (Figure \ref{FIG_1}e,f and Figure~\ref{FIG_FLUX}b,d). On March 24 03:00 UT a second episode, shorter in duration ($\sim$ 20\,h) and lower flux content than the first, occurred within the same sPIL of N1P1 (Figure~\ref{FIG_1}f right by sPIL1; also see Figure~\ref{FIG_FLUX}b,d). While AR12017 was classified as a $\beta\gamma$ AR due to this new episode, emergence occurred rather uneventfully (no collision of opposite polarities and flare activity) by adding $\sim$ 2$\times$10$^{21}$\,Mx in the AR's total unsigned flux. With the end of this second episode (end of March 25 and beginning of 26) the bipole reached its maximum self-separation of 87\,Mm and the AR has reverted back to a $\beta$ configuration, since there was only a simple sPIL separating the polarities. Meanwhile, the positive polarity P1 is seen to lose its cohesion and decay, dispersing its flux rapidly after the second emergence episode. The dispersed flux was seen to cancel slowly but steadily with diffuse negative flux in the area south of P1. Part of that negative flux originated from the neighboring AR12018 (trailing just 5$\degr$ south of AR12017) which quickly entered its decay phase. The resulting PIL extends from the south end of sPIL1 and separates the opposite fluxes between these two different ARs; thus it is an \emph{external PIL} (in short ePIL; Figure~\ref{FIG_1}e,f,g,h). A total of $\sim$ 2$\times$10$^{21}$\,Mx in P1 cancelled over a period of five days (between March 25-30; blue dashed line in Figure~\ref{FIG_FLUX}b). However, except for a few subflaring (low C-class) events on March 23 during the first emergence episode, centered in the sPIL of AR12017, no flaring occurred until $\approx$5 days later (Figure~\ref{FIG_FLUX}b). 

Interestingly, after March 26 the leading polarity, N1, suddenly appears to become oblate (develops a semi-major or principal axis along the east-west direction) and also migrates towards the south by about 8\,Mm (compare N1 polarity's location with local line of latitude in Figure~\ref{FIG_1} f,g and the accompanying animation). The southward migration speed was slow and constant at $\approx$ 35\,m s$^{-1}$ for a duration of about 60\,h (see displacement in N1 polarity tracking Figure~\ref{FIG_3}a; similar to the one we saw before the second episode in N1 P1 of AR11158). Remarkably, on March 28 (Figure~\ref{FIG_1} g) magnetic flux begins to emerge (Figure~\ref{FIG_FLUX}b,d) at the location N1 occupied before its southward migration. This parasitic flux emergence appears to be composed of two bipoles; first, due to a bipole oriented along the north-south direction and a second bipole oriented along the east-west direction (see the animation of Figure 2). It also appears that the positive flux of both bipoles emerges at almost the same location mixing/piling up together; this renders the use of any straightforward method for tracking the individual positive flux an impractical task. We thus treat these positive polarities, say P2$^\mathrm{NS}$ and P2$^\mathrm{EW}$, as a single positive polarity P2. That polarity forms a collisional PIL (``cPIL1'') with the dispersed negative flux next to pre-existing sunspot N1. The $\mathrm{t_{collision}}$ for the parasitic positive flux with pre-existing negative flux in AR12017 was at 03:00 UT March 28, 2014. On the other hand, it is rather easy to distinguish the negative polarities associated with P2 since they do not mix together or mix with the dispersed negative flux of N1. Thus, we can measure their fluxes separately for the north-south and the east-west bipole's negative polarities (we will be referring to as N2$^\mathrm{NS}$ and N2$^\mathrm{EW}$ respectively). The negative east-west polarity seems to develop a circular shape (resembling a small sunspot to the west of sunspot N1), while the negative north-south polarity lies just north of N1 and develops an irregular elongated shape, resembling an east-west oriented negative ``patch'' in the HMI magnetogram observations (e.g. as shown in Figure~\ref{FIG_1}g). However, due to the inseparable boundaries between their conjugate positive polarities (i.e. P2) we group these bipoles as one, thus the total conjugate negative flux, N2, is

\begin{equation}
\Phi_\mathrm{N2}=\Phi_{\mathrm{N2}^\mathrm{NS}}+\Phi_{\mathrm{N2}^\mathrm{EW}}
\end{equation}

\noindent If there was no cancellation of P2 with N1, the above should also equal P2, i.e.

\begin{equation}
|\Phi_\mathrm{N2}|=\Phi_\mathrm{P2}
\end{equation}

The parasitic emergence (first episode) is followed by a second episode, i.e. a stronger east-west oriented bipole emerging within the sPIL of N2 P2 (Figure~\ref{FIG_FLUX}b). At that time (March 29) NOAA classified the AR as $\beta\delta$ ($\beta$ since the AR retained an overall $\beta$ appearance but also $\delta$ due to polarity P2 right next to N1). Our natural choice is to group this second episode with the east-west oriented N2 P2 bipole of the first emergence episode. The justification is that the new episode's positive flux merged with the first episode's P2 flux and its negative flux merged with that of N2$^\mathrm{EW}$. From the moment of the second emergence episode of N2 P2, the new positive polarity separated from its conjugate quite rapidly (Figure~\ref{FIG_FLUX}d), and as a result of its close proximity, P2 was forced to collide and grind against the nearby dispersed negative flux N1 and with the negative flux of the north-south bipole, N2$^\mathrm{NS}$. This essentially produced two collisional PILs, i.e. one oriented along the north-south direction,``cPIL1'', which is the one already formed after the emergence of the first parasitic episode next to the N1 dispersed flux, and ``cPIL2'', which formed after the collision between (a) the second parasitic episode's new positive flux and (b) the first episode's N2$^\mathrm{NS}$ flux. The latter occupies the space between P2 and the sunspot N1 and it is oriented along the east-west direction. Irrespective of whether the emergence has occurred within a previous self-PIL of the first episode, this PIL segment, cPIL2, formed as a result of collision between newly emerging flux and pre-existing flux (dashed orange line, Figure~\ref{FIG_1}g,h). The sequential emergence in two episodes of the N2 P2 bipole in AR12017 coincided with strong clusters of significant flare activity and CMEs (Figure~\ref{FIG_FLUX}b,d; see \S~\ref{RESULTS}).

This dynamic evolution can be summarized best in a 3D representation (2D in space and 1D in time; Figures~\ref{FIG_2}b and~\ref{FIG_3}b) by means of the Time-Stacking Method~\citep{Chintzoglou_Zhang_2013}. In this space-time representation, the proper motions of magnetic polarities during the flux emergence phase (i.e. the mutual or ``self''-separation of the conjugated polarities as the two $\Omega$-shaped flux tubes emerge) are naturally visualized into a self-consistent picture of individual bipoles, in which an emerging subsurface configuration is driving the evolution of the system during its emergence. Due to the strong stratification of pressure in the subphotospheric layers, an emerging magnetic flux tube is not expected to break through the photosphere as a solid structure but rather emerge laterally deformed (``pancaked''; \citealt{Cheung_Isobe_2014}). Nevertheless, the \emph{overall large-scale morphology} and connectivity is maintained as also typically seen in 3D MHD simulations of AR emergence (e.g.~\citealt{Fan_2009, Cheung_etal_2010, Toriumi_etal_2014}). The collisional PIL forms exactly at the locations where the magnetic flux tubes collide with each other as they emerge.



\section{Collisional Shearing - Spatio-temporal Correlation with Major Activity}\label{RESULTS}

Activity emanating from solar ARs is frequently associated with magnetically unstable structures, i.e. MFRs and/or electric current sheets (allowing magnetic reconnection to release free magnetic energy). An intrinsic limitation in predicting solar eruptive activity is associated with the inability to know the 3D pre-flare (or pre-eruption) magnetic configuration in ARs. What is certain is that accumulations of magnetic flux, in the form of magnetic flux tubes, must reach the photospheric surface before even creating the conditions for extreme activity in the corona. However, not all ARs are flare productive, and, in actuality, only a subset of them are CME-productive. For instance, about 10\% of the ARs in a solar cycle will ever produce major flares (i.e. flare classes $>$M1.0; \citealt{Georgoulis_Rust_2007}). In our case, both AR11158 and AR12017 were extremely flare-productive as well as CME-productive.

The space-time representation (Figure~\ref{FIG_2}b and Figure~\ref{FIG_3}b) captures the 3D nature of collision between emerging magnetic flux tubes as they form flare-productive ARs. In fact, two types of collision (photospheric ``bipole-bipole'' interaction) are possible based on the timing of the emergence of the two flux tubes: (i) \emph{simultaneous}, i.e. when two flux tubes are emerging simultaneously and in very close proximity to each other (AR11158; Figure~\ref{FIG_2}a,b), and (ii) \emph{sequential}, when the first flux tube has already emerged with its sunspots fully developed before a second flux tube collides with its opposite-signed polarity as it emerges (AR12017; Figure~\ref{FIG_3}a,b). In the simultaneous case, the collision occurs as a result of the relative proximity and orientation/tilt of the two bipoles on the heliographic plane. In the sequential collision, the conjugated polarities of the first (pre-existing) bipole N1 P1 are already formed and developed (i.e. the first flux tube's emergence has ceased for days), and have entered their decay phase. In addition, in the latter case, the collision is observed to occur in a different way, since the second bipole N2 P2 emerges at the location originally occupied by the mature leading sunspot N1. Thus, we surmise that the collision first started in 3D under the surface (manifested by the progressively increasing oblateness of N1 and its $\approx$ 8\,Mm displacement towards the south). 

Both cases of photospheric bipole-bipole interaction (simultaneous and sequential emergence) provide the conditions for convergence and shearing along the cPIL. The convergence of opposite-signed magnetic elements is expected to cause cancellation of flux along the cPIL. This is due to the fact that cPILs are high spatial gradient PILs, which means that magnetic elements meeting on each side of the cPIL can reach a critical distance for submergence ($\approx$ 0.9\,Mm). Likewise, the colliding polarities, as they overtake each other while self-separating, naturally produce strong shearing motions along the cPIL. The characteristic way of how convergence and shearing occurs in two colliding bipoles is fundamentally different than the one often simulated by the emergence of a simple conjugate bipole. In simulations of a simple conjugate bipole, the major activity occurs in its self-PIL (in this case there is no collisional PIL, since there is no collision happening in a single conjugate bipole). In either ``emergence-based'' or ``cancellation-based'' numerical experiments, all simulated major activity is typically emanating from the sPIL in a solitary conjugate bipole. Here, both of our ARs under study did not exhibit conjugate bipoles with compact sPILs (except when the two bipoles collided and formed a cPIL), and virtually no major activity (i.e. flare clusters, CMEs) occurred along the sPIL of any bipole of these two ARs. To summarize the characteristic way proper motions of emerging bipoles drive convergence and shearing, we coin the term \emph{collisional shearing}, to emphasize that cancellation occurs at the cPIL due to the collision between nonconjugated polarities of multiple bipoles. In the following subsections we provide a detailed analysis for each AR elucidating the local physics and the process of collisional shearing forming cPILs.     

\subsection{The Conjugate Flux Deficit Method}\label{METHOD}

ARs that are composed of more than one bipole may eventually become sites of collisional shearing due to the self-separation and relative orientation of the bipoles. If collision occurs between like-signed polarities then no cPIL will form, not to mention that the local 3D orientation of the magnetic fields is not favorable for reconnection. However, if the collision occurs between opposite-signed polarities, a cPIL will form. If the proximity is too close, magnetic flux cancellation via submergence is also expected to occur. Characterizing cPILs is very much related to the cancellation process. It is therefore necessary to be able to quantify cancellation occurring in cPILs in order to help understand why these are the kernels of extreme activity in ARs.  

In decaying ARs, it is rather easy to quantify cancellation over a period $\Delta t$; it is as simple as measuring the time evolution of the total (unsigned) magnetic flux, $\Phi_\mathrm{tot}$, and finding the amount it dropped over that period after it reaches its peak value ($t=t_\mathrm{peak}$ presumably at the end of emergence phase of the AR), i.e.

\begin{equation}
	\Phi_\mathrm{canceled}=\Phi_\mathrm{tot}(t=t_\mathrm{peak})-\Phi_\mathrm{tot}(t_\mathrm{peak}<t<t_\mathrm{peak}+\Delta t)
\end{equation}

In addition, in decaying ARs cancellation can be directly visible in magnetogram image series as canceling discrete patches along the PIL. However, in emerging ARs the situation is rather complex. It is very hard to observe the ongoing cancellation (unless it is rather visible as small discrete cancelling polarity elements; such cases are typically associated with Ellerman bombs in sPILs)  and what fraction of flux is canceled during the emergence stage in complex ARs still remains as a big question. In compact PILs (and thus cPILs) the proximity of strong polarities can be so close that any such individual cancellation events would be spatially underresolved in observations by current magnetograph instruments. 

In Figure~\ref{FIG_DEFICIT} we outline a method that allows to measure cancellation in complex ARs during their emergence phase. In the case of an emerging solitary bipole (Figure~\ref{FIG_DEFICIT}a), the magnetic flux evolves as shown in the plots for each polarity -- it is increasing during the emergence process and the positive and negative flux is found to be \emph{balanced}. Therefore, if we subtract the (unsigned) negative flux from its conjugate positive, 

\begin{equation}\label{DEFICIT_EQ}
\mathrm{\Delta_{N1P1}=|\Phi_\mathrm{N1}|-\Phi_\mathrm{P1}}, 
\end{equation}

\noindent it should yield zero for the duration of the emergence of this solitary bipole. However, if two bipoles are emerging in close proximity to each other in such ways that opposite polarities are bound to collide, flux cancellation occurs (between polarities P2 and N1; Figure~\ref{FIG_DEFICIT}b) and a \emph{reduction} of flux occurs in the colliding polarities, evident if compared against the flux of the conjugate polarities (in this case comparing N1 with P1 and N2 with P2). Thus we use the term \emph{conjugate flux deficit} (Figure~\ref{FIG_DEFICIT}c), symbolized with $\Delta$, to emphasize that due to cancellation there would always be a reduction of flux of the colliding polarities, instead of an increase. The higher the deficit $\Delta$ for a colliding conjugate bipole, the lower the flux becomes in the nonconjugated colliding polarities (e.g. N1 and P2 in Figure~\ref{FIG_DEFICIT}b). Moreover, while the two bipoles may have different total unsigned flux (i.e. one bipole may be stronger in terms of flux than the other), the $\Delta$ measured between the two polarities would also have the following property:

\begin{equation}\label{DEFICIT_EQUIVALENCE_EQ}
\mathrm{\Delta_{N1P1}=-\Delta_{N2P2}=|\Delta|}.
\end{equation} 

\noindent since $\Delta$ is defined by subtracting positive flux from the (unsigned) conjugate negative for either of the bipoles. Due to the ``differential'' nature of our proposed detection method, the cumulative cancellation of flux can be measured via eq.~(\ref{DEFICIT_EQ}) irrespectively of whether the colliding bipoles are emerging or have stopped to emerge. The flux cancellation rate, $d\Phi/dt$ is derived from the slope of the deficit (Figure~\ref{FIG_DEFICIT}c) as

\begin{equation}\label{FLUX_RATE_EQ}
\frac{d\Phi}{dt} = - \frac{d|\Delta|}{dt}.
\end{equation}

The basis of the proposed approach in tracking the flux content in evolving magnetic polarities is much simpler than the approach taken by \citet{Tarr_Longcope_2012} or \citet{Tarr_Longcope_2013} in which the authors track individual polarities (with necessary assumptions on how they are connected). The condition here is that the non-colliding conjugate polarities are adequately isolated from other bipoles so that they can serve as a ``calibration'' or reference flux level. This allows the robust inference of the deficit. However, instrumental effects may also inhibit clean measurements of $\Delta$ or parasitic flux may also have an impact in the measurement of the imbalance. In the following paragraphs of this subsection, we discuss the possible sources of uncertainty in the measurements of the deficit.

We chose not to apply a noise threshold in the magnetogram series since this impacts our ability to track weak flux that cancels and also characterize the cPILs (such as in AR12017 where P2 cancels with the dispersed N1 flux). A noise threshold may also impact other locations of intermediate magnetic intensity where magnetic flux may cancel (such as in external PILs). It is critical to be able to properly resolve all locations of cancellation in an AR's neutral line. Nevertheless, we have applied noise thresholds at different levels and determined that for the $\mathrm{B_r}$ series (vector data) progressive noise thresholds of up to $|\mathbf{B}|$= 100\,G (1-$\sigma$ noise level; \citealt{Hoeksema_etal_2014}) reduce the maximum deficit by 20\% while a higher threshold of $|\mathbf{B}|$= 200\,G (corresponding to a 2-$\sigma$ noise level), reduces the deficit by 30\%. For the $\mathrm{B_{LOS}}$, a choice of a very conservative noise threshold of 50\,G has minimal impact to the deficit of about 3\%. One has to apply even higher threshold values for the $\mathrm{B_{LOS}}$, e.g. 200\,G, to see the deficit reducing to about 24\% of the non-thresholded value, although such noise threshold is unreasonably high given the sensitivity of the $\mathrm{B_{LOS}}$ which is $\sim$ 10\,G \citep{Couvidat_etal_2016}. Nevertheless, for AR 12017, such reduction for the deficit at that threshold in the $\mathrm{B_{LOS}}$ series is happening after March 30, where the N2 and P2 flux concentrations begin to disperse (i.e. they are entering their decay phase; see animation of Figure 2). This means that once the flux begins to disperse, it progressively fills wider areas with lower intensities or even becomes under-resolved and thus is measured as weaker in the magnetograms. This may cause a gross underestimation of the deficit. In AR12017, the N2 polarity loses its cohesion and the deficit decreases. Similarly, in AR11158, the N2 polarity has already began to decay, especially the south part of N2 that belonged to the first emergence episode of N2 P2 (see animation of Figure 2). Imposing the noise threshold for $|\mathbf{B}|$=100\,G masks out significant amount of negative flux, effectively reducing the deficit. Thus, attempting to reduce the instrumental noise in the vector data, has an impact in the deficit, but has much less of an impact in the deficit from $\mathrm{B_{LOS}}$ data. Nevertheless, this constitutes a source of uncertainty for the flux deficit calculation, $u_\mathrm{noise}$. 

A second source of uncertainty in measuring the deficit is due to deviations from flux balance in the entire FOV of the magnetogram data, $u_\mathrm{balance}$. This may be due to instrumental effects or due to either including flux in the FOV that doesn't belong to the target AR or failing to include flux that belongs to the AR. In both the $\mathrm{B_r}$ and the $\mathrm{B_{LOS}}$ data series in this study, the magnetograms are balanced to about 2.5\% to 3\%.

Last, a third source of uncertainty comes from difficulties in grouping the polarities in individual flux emergence episodes, $u_\mathrm{grouping}$. We estimate this uncertainty by attempting two different groupings in the negative fluxes of the sequential emergence in AR12017. For instance, if the negative patch above sunspot N1, i.e. polarity N2$^\mathrm{NS}$, is treated as part of N1, the maximum measured deficit is reduced by $\approx$50\% as compared to the deficit when that patch is considered part of N2. Thus, for AR12017 we estimate an uncertainty for $\Delta$ of 

\begin{equation}
	u_{\Delta}^\mathrm{AR12017}=\sqrt{u_\mathrm{noise}^2+u_\mathrm{balance}^2+u_\mathrm{grouping}^2}=\sqrt{(3\%)^2+(3\%)^2+(50\%)^2}=50.2\%
\end{equation}

As for AR11158, it is certain that the colliding polarities (N1 and P2) are part of two separate colliding bipoles (due to their proper motions and origin; hence they are nonconjugated), thus we remove the uncertainty for grouping. However, since we are using flux calculated from $\mathrm{B_r}$ from vector magnetogram data, the total uncertainty for $\Delta$ is 

\begin{equation}
	u_{\Delta}^\mathrm{AR11158}=\sqrt{u_\mathrm{noise}^2+u_\mathrm{balance}^2+u_\mathrm{grouping}^2}=\sqrt{(20\%)^2+(3\%)^2+(0\%)^2}=20.2\%
\end{equation}

Special care must be taken in fully isolating the polarities of the colliding bipoles. Visual inspection and validation is also necessary during the interpretation of the results. Typically, the measurements are easy to make in the early stages of emergence, until tertiary (i.e. subsequent) bipoles are too hard to exclude when measuring the flux. On the other hand, the latter does not pose a problem if the fluxes of such tertiary parasitic emerging bipoles can be absorbed separately in the positive and negative flux of one of the two main bipoles (see Section 3.2 and following sections).

Last we should also mention the following possibility. If the submerged field (that cancelled at the collisional PIL) accumulates under the surface and the colliding polarities drift past each other and away from the initial cancellation region (i.e. the collisional PIL), then the cancelled flux (or a fraction of it) may be free to emerge again (i.e. to ``re-emerge''). If such re-emerging flux at/near the cPIL is included in the like-signed flux of each of the colliding polarities, then the cumulative cancelled flux represented by the deficit will appear to reduce (denoting emergence).

\subsection{AR11158: Collisional Shearing Forced by Simultaneous Emergence}

In Figure~\ref{FIG_4} we show the time evolution of the magnetic flux measured at the photospheric surface for one of the conjugated bipoles (N2 P2) that is nested within AR11158 (Figure~\ref{FIG_4}a). As we already showed in Figure~\ref{FIG_FLUX}a, flux emergence in AR11158 occurred in two-episodes simultaneously for each of the two bipoles. Even at the end of the first emergence episode, the polarities N1, P1, N2, and P2 still remain well-isolated from each other (Figure~\ref{FIG_1} a). During this time, we see little difference between the flux of the conjugated polarities N2 and P2 (Figure~\ref{FIG_4}; also between N1 and P1 as shown in Figure~\ref{FIG_FLUX}a); each bipole's flux is \emph{balanced} to about 2.5\%, considering that AR11158 is at an angular distance of $\Theta\gtrsim$30\,$\degr$ from disk center and the noise is significantly high in the radial component of the vector magnetic field, $\mathrm{B_r}$ (see discussion in \S~\ref{OBS}). Conveniently, at around the end of the first episode, the AR progresses towards disk center distances below the 30\,$\degr$ limit. Then a stronger, second flux emergence episode begins in both bipoles. Following that moment, the rapid self-separation of N2 P2 (and similarly N1 P1) drives the P2 polarity to collide with N1. The leading polarities move faster (P1, P2) than the trailing conjugate polarities (N1, N2). Thus the velocity shear at the PIL between P2 and N1 can be determined by the proper motions of the polarities (via the self-separation speed measurements in Figure~\ref{FIG_FLUX}) and it is quite high; in Figure~\ref{FIG_FLUX}c (solid red curve) $v_\mathrm{N2P2}$ is at $\sim$400 m s$^{-1}$ for the day of February 13 and after February 14 it gradually drops to $\sim$200 m s$^{-1}$ (February 15) and $\sim$100 m s$^{-1}$ (February 16th). In tandem with the rapid shear, upon impact (February 13 03:00 UT), an imbalance develops almost immediately between the flux of each bipole's conjugated polarities (see Figure~\ref{FIG_4} a). This imbalance is captured in the $\mathrm{\Delta_{N2P2}}$ (solid curve in Figure~\ref{FIG_4} c) and it occurs because P2 flux is gradually reduced compared to its conjugate N2 flux (evident in Figure~\ref{FIG_4} a). 

During the first 7 days of evolution considered here for AR11158, the shape of its PIL system changed significantly due to the proper motions of the colliding polarities. The collision between P2 and N1 lasted for three days after the collision onset in the beginning of February 13. Remarkably, out of the total of 38 flares that occurred in AR11158 (both confined and eruptive; Table~\ref{TABLE_1}), an impressive number of 30 flares were produced/centered in the collisional segments of the PIL in the area between N1 and P2. The remaining 8 flares (each marked with an asterisk in Table~\ref{TABLE_1}) are spatio-temporally correlated with the sequential emergence of a small parasitic bipole near polarity N2 (marked with a dashed ellipse in Figure~\ref{FIG_1} c). Indeed the emergence of this bipole can be considered as a sequential-type emergence (e.g. the kind of emergence we consider in the following subsection \S~\ref{sequential}) producing a second collisional PIL between N2 and the positive parasitic polarity. However, the negative parasitic polarity merged with N2 making the detailed analysis challenging. Thus, to simplify the analysis, we also absorb the parasitic bipole's positive and negative flux in the fluxes of bipole N2 P2 by adding the positive flux in the respective flux of P2. In other words, we treat that new bipole as part of bipole N2 P2 (having sPIL2 in common) so that mutual cancellation in the sPIL2 could not affect the measurement of the deficit $\Delta$. Further justification for doing so is the dominance of activity at the collisional PIL (30 out of 38 flares occurred there) resulting from the collision of N1 with P2 (see Figure~\ref{FIG_2} and Table~\ref{TABLE_1}). 

For each flare event, we cross-checked the centering we provide in Table~\ref{TABLE_1} with centering inferred by the temporal evolution and spatial distribution of the flare ribbons seen best in imaging taken in the  \emph{SDO}/AIA 1600\,\AA\ passband. Our inspection for the period after February 15 showed that 10 out of these 30 flares (marked in Table~\ref{TABLE_1}) were centered in the collisional PIL that formed due to another tertiary sequential emergence in the wake of P2's tail. During this two-day period, negative polarity elements rapidly moved and merged with polarity N2, while colliding with the dispersed flux that is trailing behind P2. We speculate that the origin of this tertiary emergence is due to the cancelled flux that re-emerged in the wake of P2 polarity's collision with N1. No matter what the origin of this tertiary emergence is, the polarity proper motions create the conditions for a collisional PIL at the surface enriching the flare activity in AR11158.

In the case where there are no subsequent parasitic emergence events, the deficit $\Delta$ effectively measures the cumulative cancelled flux between the colliding nonconjugated polarities N1 and P2. This measurement (calculated for each conjugate bipole) strongly indicates that cancellation of magnetic flux is taking place along the cPIL. The cancellation of flux is independent of the flux increase (due to the emergence process) and it kicks in only at the onset of collision (onset shown with an arrow in Figure~\ref{FIG_4}a). The deficit $\Delta_\mathrm{N2P2}$ reaches its peak value within a day ($\Delta\approx$ 0.9$\times$10$^{21}$\,Mx) near the time when the collision distance (measured from the flux centroid of the colliding polarities/sunspots) is minimum ($\sim$ 10\,Mm; Figure \ref{FIG_4}d). Over this period of one day the rate of flux cancellation calculated with equation~(\ref{FLUX_RATE_EQ}) is $\sim$ -8.3$\times$10$^{15}$\,Mx s$^{-1}$. Due to the complexity of the AR (that would require us to track multiple polarities to measure the deficit accurately), we restrict our analysis for the time period up to February 15 during which only four polarities (and those of the tertiary bipole emerging above N2 absorbed in the N2 P2 bipole) are present. 

The increase in flux deficit is accompanied by a dominance of average redshifts over blueshifts at the collisional PIL (values at $\approx$ 0.35\,km s$^{-1}$, higher than the average trend of the nearby QS $\sim$0.2\,km s$^{-1}$; dashed curves Figure~\ref{FIG_4} b). Indeed, the average blueshifts seem more comparable in magnitude to that of the QS. We interpret this dominant average redshift signal at the cPIL to be consistent with the submergence into the photosphere of post-reconnected loops. In this case, the reconnection associated with cancellation would occur (on average) above the formation height of the Fe I 6173\,\AA\ line used by HMI, which is at a height of $\approx$0.3\,Mm \citep{Norton_etal_2006}. In Figure~\ref{FIG_DOPPLER} we show the evolution of the Doppler signal at the location of the collisional PIL. In Figure~\ref{FIG_DOPPLER}a we show how the photospheric Doppler velocity field appears in a large patch between P2 and N1, before the onset of collision ($\mathrm{t_{collision}}$). The granulation pattern is dominant and the velocities are the same with the nearby QS, suggesting that nothing special occurs at that location before $\mathrm{t_{collision}}$. However, after $\mathrm{t_{collision}}$, (Figure~\ref{FIG_DOPPLER}b,c,d) the photospheric velocities along the cPIL are predominantly positive, signifying downflows as measured at the formation height of the Fe I 6173\,\AA\ line. Some small patches appear along the length of the cPIL suggesting that the cancellation occurs across different segments along the cPIL (or at a lower height compared to that of the Fe I line). This pattern continues for as long as the collision takes place.

The collision is evidently so intense that forces the colliding sunspots to become severely oblate; The south fragment of N1 (produced during the second episode within N1 P1) exceeds 60\% oblateness right after the minimum collision distance (suggesting maximum compression) and drops back to 40\% after the end of the close encounter, which is still a very high number. Furthermore, the collisional PIL maintains a more or less consistent length of $\approx$ 60\,Mm and it quickly reaches this value in only 12 hours after the onset of collision ($\mathrm{t_{collision}}$).

Just 12 hours after the onset of deficit increase and the dominance of redshifts at the cPIL (Figure~\ref{FIG_4} b,c) a flare activity cluster is triggered, also originating from the cPIL (Figure~\ref{FIG_2} a,b). The emergence of N2 P2 continues well into February 16. The deficit plot for N2 P2 reveals a total cumulative amount for the cancelled flux of $\Delta \approx$ 1 $\times$10$^{21}$\,Mx for the period up to February 15. This number amounts to $\approx$ 6\% of the total unsigned flux of a perfectly balanced N2 P2 and $\approx$ 8\% for a balanced N1 P1. After February 15, the cPIL weakens as P2 flux rushes away from N1 and conglomerates right by the location of P1 polarity (Figure~\ref{FIG_1}c,d,~\ref{FIG_2}b). Moreover, the activity in AR11158 continues well into February 17, following additional rapid proper motions from parasitic emergence on February 15 and 16 (righ by or within the previous cPIL and in the tail of P2), and on February 17 (emergence and collision in the location of previous parasitic bipole emergence by N1 and N2 polarities).

Remarkably, a sigmoidal structure appears at the cPIL $\approx$ 5 hours after $\mathrm{t_{collision}}$, becoming progressively more coherent (Figure~\ref{FIG_EUV1}a,b). The structure is discernible as a dark feature in 304\,\AA\ passband (dark feature due to free-free absorption in EUV by cooler plasma; also in other coronal passbands such as 171\,\AA\ ) and is aligned with the cPIL. This signifies a location where cool chromospheric filament material can be supported in locations where the magnetic field has a ``convex-up'' or ``dipped'' local geometry, consistent with the presence or progressive formation of a weakly-twisted MFR structure. However, this growing structure is accompanied by transient brightenings along its length, similar to the observations reported by~\citet{Chintzoglou_etal_2015}. We surmise that this MFR structure above the cPIL inevitably interacts/reconnects with the immediately neighboring ambient sheared-arcade field, and/or that the brightenings signify reconnection between individual stressed magnetic structures comprising the sigmoidal structure. This period of transient brightenings of weakly coherent structures seen in the corona across compact PILs has been previously associated with transient reconnection events eventually producing more monolithic structures (e.g.~\citealt{Chintzoglou_etal_2015}). Indeed, this transient coronal activity accompanies the photospheric evolution in our observations to the point where coherent monolithic structures are seen best in the 10\,MK passband of AIA/\emph{SDO} (Figure~\ref{FIG_EUV1}b,d and the accompanying animation).

\citet{Toriumi_etal_2014} considered possible scenarios for the subsurface magnetic structure of AR11158. In both cases modeled using numerical MHD simulations, a pair of BMRs emerged simultaneously. In one of the cases (Case 1 therein), the BMRs emanate from a single flux tube underneath the surface, whereas in the other case (Case 2 therein) the two are disjoint. Neither model reproduces the fact that a deficit develops between the nonconjugated polarities. For Case 1, instead of a deficit there is an excess of flux during the collision, and for Case 2 the flux was balanced. While for the former (Case 1) it is not thoroughly investigated why there was an increase, the latter (Case 2) remained in flux balance because collision between the nonconjugated polarities simply did not occur; the lateral distance between the two flux tubes was large enough that the nonconjugated polarities never approached each other significantly for cancellation to happen. see Section 5.3 for a discussion).


\begin{deluxetable}{rccccc}
\tablecolumns{6}
\tablewidth{0pc}
\tablecaption{Flare events and CMEs produced during the evolution of NOAA AR11158}

\tablehead{
\colhead{\#} &\colhead{Flare Start (UT)} & \colhead{Carrington Longitude\tablenotemark{\dagger}} & \colhead{Heliographic Latitude\tablenotemark{\dagger}}& \colhead{Flare Class} &\colhead{v$_\mathrm{CME}\tablenotemark{\ddagger}$ (km\,s$^{-1}$)}}
\startdata
       1 & 2011-02-13 13:44 & 33$\fdg$3 & -20$\fdg$5 & C4.7 & - \\
       2 & 2011-02-13 17:28 & 34$\fdg$2 & -20$\fdg$6 & M6.6 & 373 \\
       3 & 2011-02-13 21:17 & 36$\fdg$9 & -19$\fdg$6 & C1.1 & - \\
       4 & 2011-02-14 02:35 & 29$\fdg$9 & -20$\fdg$9 & C1.6 & -* \\
       5 & 2011-02-14 04:29 & 34$\fdg$2 & -20$\fdg$5 & C8.3 & - \\
       6 & 2011-02-14 06:51 & 32$\fdg$2 & -20$\fdg$6 & C6.6 & -*\\
       7 & 2011-02-14 12:41 & 29$\fdg$8 & -20$\fdg$6 & C9.4 & -* \\
       8 & 2011-02-14 13:47 & 33$\fdg$4 & -20$\fdg$7 & C7.0 & - \\
       9 & 2011-02-14 17:20 & 30$\fdg$6 & -20$\fdg$3 & M2.2 & 229* \\
      10 & 2011-02-14 19:23 & 29$\fdg$2 & -20$\fdg$5 & C6.6 & -* \\
      11 & 2011-02-14 23:14 & 32$\fdg$1 & -21$\fdg$2 & C1.2 & - \\
      12 & 2011-02-14 23:40 & 37$\fdg$1 & -18$\fdg$6 & C2.7 & - \\
      13 & 2011-02-15 00:31 & 29$\fdg$5 & -20$\fdg$8 & C2.7 & -* \\
      14 & 2011-02-15 01:44 & 32$\fdg$9 & -20$\fdg$3 & X2.2 & 669 \\
      15 & 2011-02-15 04:27 & 31$\fdg$0 & -20$\fdg$9 & C4.8 & -* \\
      16 & 2011-02-15 10:02 & 32$\fdg$8 & -21$\fdg$0 & C1.0 & - \\
      17 & 2011-02-15 14:32 & 30$\fdg$3 & -20$\fdg$8 & C4.8 & -* \\
      18 & 2011-02-15 19:30 & 33$\fdg$2 & -20$\fdg$4 & C6.6 & -**\\
      19 & 2011-02-15 22:49 & 36$\fdg$0 & -19$\fdg$4 & C1.3 & - \\
      20 & 2011-02-16 01:32 & 32$\fdg$4 & -20$\fdg$5 & M1.0 & - \\
      21 & 2011-02-16 01:56 & 24$\fdg$6 & -46$\fdg$1 & C2.2 & -**\\
      22 & 2011-02-16 05:40 & 33$\fdg$0 & -20$\fdg$9 & C5.9 & - \\
      23 & 2011-02-16 06:18 & 33$\fdg$5 & -20$\fdg$5 & C2.2 & -**\\
      24 & 2011-02-16 09:02 & 33$\fdg$8 & -21$\fdg$1 & C9.9 & -** \\
      25 & 2011-02-16 10:25 & 33$\fdg$7 & -20$\fdg$6 & C3.2 & -** \\
      26 & 2011-02-16 11:58 & 32$\fdg$7 & -21$\fdg$2 & C1.0 & - \\
      27 & 2011-02-16 14:19 & 32$\fdg$7 & -20$\fdg$9 & M1.6 & -** \\
      28 & 2011-02-16 15:27 & 34$\fdg$0 & -20$\fdg$5 & C7.7 & -** \\
      29 & 2011-02-16 19:29 & 33$\fdg$9 & -20$\fdg$9 & C1.3 & - \\
      30 & 2011-02-16 20:11 & 35$\fdg$2 & -21$\fdg$0 & C1.1 & - \\
      31 & 2011-02-16 21:06 & 33$\fdg$5 & -20$\fdg$5 & C4.2 & -** \\
      32 & 2011-02-16 23:02 & 38$\fdg$4 & -18$\fdg$3 & C2.8 & -** \\
      33 & 2011-02-17 01:41 & 33$\fdg$7 & -21$\fdg$1 & C6.1 & -** \\
      34 & 2011-02-17 04:09 & 33$\fdg$9 & -20$\fdg$7 & C1.5 & - \\
      35 & 2011-02-17 05:33 & 34$\fdg$7 & -20$\fdg$5 & C1.0 & - \\
      36 & 2011-02-17 06:44 & 33$\fdg$8 & -21$\fdg$1 & C1.2 & - \\
      37 & 2011-02-17 08:09 & 34$\fdg$5 & -20$\fdg$0 & C1.2 & - \\
      38 & 2011-02-17 09:24 & 33$\fdg$0 & -20$\fdg$9 & C1.9 & - \\
\enddata
\tablenotetext{\dagger}{Flare centering obtained by intercorrelating the GOES, RHESSI and HEK event lists.}
\tablenotetext{\ddagger}{CME velocity (linear fit) obtained from CDAW CME catalog.}
\tablenotetext{*}{Activity occurred in separate collisional PIL of parasitic (sequential) emergence near polarity N2.}
\tablenotetext{**}{After the middle of February 15, activity is seen to occur due to collisional shearing of tertiary emergence within the wake of P2.}
\label{TABLE_1}
\end{deluxetable}

\clearpage

\subsection{AR12017: Collisional Shearing Forced by Sequential Emergence}\label{sequential}

In Figure \ref{FIG_5} we show the results of our analysis for AR12017. Five days after the initial emergence of the first emergence episode forming N1 P1 (March 22, 2014) the AR is entering its decay phase. This is revealed by its trailing magnetic polarity P1, which appears very dispersed (see the animation or compare panels f,g in Figure \ref{FIG_1}). On March 24, a second flux emergence episode occurred within the sPIL1 and no collision between opposite polarities occurred. No flaring activity accompanied the self-separation of the second episode in N1 P1. The negative flux originating from to the leading polarity of that second episode dispersed east of the negative sunspot in N1.

As we discussed in detail in \S~\ref{CPIL}, in the beginning of March 28, a N2 P2 north-south oriented bipole and an east-west oriented bipole begin to emerge at the north of N1 (Figure~\ref{FIG_DOPPLER}e) in close proximity to the dispersed negative flux (in part originating from the second episode of N1 P1 and also due to the decay of N1; dispersed flux is also labeled as N1). Their positive polarities appear to be concentrating at the same location, rendering them impossible to separate. As for their conjugate negative polarities, they moved to locations spaced far apart, which allowed the measurement of their fluxes and the tracking of their motions with ease. We label them as N2$^\mathrm{NS}$ for the negative polarity of the north-south bipole (Figure~\ref{FIG_DOPPLER}e), and N2$^\mathrm{EW}$ for the negative polarity of the east-west bipole (Figure~\ref{FIG_DOPPLER}f). Adding them together amounts the total N2 flux (solid red line in Figure~\ref{FIG_5}a).

However, if one measures the newly emerging N2 P2 bipole flux, it would show a spurious imbalance (excess negative flux bias of about 5$\times10^{20}$\,Mx) due to a pre-existing flux distribution belonging to N1 next to the area of the emergence of N2 P2 (imbalance shown in Figure~\ref{FIG_FLUX}b). In order to remove this bias from our measurements of N2 flux, for consistency we correct both N2 and P2 fluxes by subtracting their corresponding pre-existing flux bias from the first measurement seen in Figure~\ref{FIG_5}a, which in this case was the initial flux level at about two days before the emergence of N2 P2. This produces a minimal imbalance until the beginning of N2 P2 emergence (e.g. Figure~\ref{FIG_5}c shows a flat line for the deficit $\Delta$). Most importantly, after the onset of emergence of the parasitic bipole N2 P2, the flux was balanced. Only upon collision of P2 with N1 the flux deficit began to increase, starting at around 03:00 UT of March 28, 2014 (Figure~\ref{FIG_5}c). A cluster of subflaring activity with two major eruptive flares accompanied the continuous increase in the flux deficit $\Delta$. The shear motions in the collisional PIL between P2 and the dispersed N1 persisted as additional emerging flux piled up at P2 pushing it eastwards. Despite that the self-separation was primarily due to the westward motion of N2$^\mathrm{EW}$ and not due to P2 (Figure~\ref{FIG_FLUX}b; self-separation speed $\sim$400-500 m s$^{-1}$), the P2 polarity was still under collision with the nonconjugated negative flux N1 at the cPIL1 (e.g. Figure~\ref{FIG_DOPPLER}f). Rapid cancellation was seen to develop as shown in the deficit plot (Figure~\ref{FIG_5}c). Imaging in EUV wavelengths from AIA shows that a low-lying filamentary helical structure (of similar shape and length with the cPIL), activated abruptly in the collisional PIL, immediately after 10:00 UT and persisted up until its eruption (Figure~\ref{FIG_EUV2}a,b,c,d and associated animation). This structure above the cPIL undergoes multiple brightenings along its length suggesting that reconnection occured within or beneath the structure. This evolution is fully consistent with that of a similar structure observed in the collisional PIL of AR11158 (see previous subsection).

At the end of March 28, the rapid cancellation and the associated flare activity cluster (and CMEs; see Table~\ref{TABLE_2} and Figure~\ref{FIG_3}) is succeeded by a much stronger emergence event occurring within the self-PIL of N2 P2. This new bipole is also oriented along the east-west direction and emerges at the sPIL2 part that was shared by the previous east-west bipole. Thus we consider this to be a second emergence episode of the first parasitic east-west bipole. For this reason, and also because it is impossible to isolate the new positive and negative flux from the pre-existing, we group the respective fluxes with the P2 and N2$^\mathrm{EW}$ flux (blue solid curve and red dotted curve in Figure~\ref{FIG_5}a). In this second episode, the newly emerging positive flux appeared to move rapidly to the east and at the same time collided with the pre-existing positive P2 flux (north of it) and also with the opposite-signed N2$^\mathrm{NS}$ polarity at the south. Moreover, it pushed the pre-existing positive flux to collide and shear with the N1 dispersed flux at the east. By tracking the magnitude of the velocity of the newly emerging and rapidly moving P2 we find that the velocity shear at the collisional PIL is in the range of $\sim$150-200 m s$^{-1}$ and persists for 48 hours after the emergence onset. Note that (as in AR11158) the second emergence event \emph{did not immediately lead to flaring activity}. In fact, this emergence onset precedes the second cluster of flare activity on March 29 by at least 8 hours.

These multiple parasitic emergence events in AR12017 present a great opportunity to demonstrate our proposed flux-deficit method as we apply it to more complex and evolving configurations. As we said above, during the emergence of the second episode, the positive polarity of the secondary east-west oriented bipole collides with the pre-existing north-south bipole's negative polarity N2$^\mathrm{NS}$ forming a cPIL2. Due to limitations in measuring the flux in individual polarities that eventually merge we ended up grouping these emergence episodes into a single bipole (i.e. N2 P2). However, if collision occurs between polarities of two bipoles that now have been grouped into one bipole, then the deficit in the newly grouped bipole would not change and thus would not yield the canceled flux as collision progresses. This is equivalent to the case where a collisional PIL between two colliding bipoles (either emerging simultaneously or sequentially; for instance, that in Figure~\ref{FIG_DEFICIT}b) effectively becomes a ``self-PIL'' of one ``bipolar AR'' (i.e. equivalent with going from the configuration shown in Figure~\ref{FIG_DEFICIT}b to that in Figure~\ref{FIG_DEFICIT}a due to the grouping of the two bipoles; therefore, we would not be able to measure cancellation between the two grouped bipoles). Note, however, that in the meantime the AR12017 P2 flux has been cancelling continuously with N1 flux at cPIL1, which is captured by the deficit (shown by the black solid line in Figure~\ref{FIG_5}c), irrespective of the cancellation due to collision within the groupped N2$^\mathrm{EW}$ bipole. This discussion shows that by grouping together bipoles that at some point happen to collide, there is a danger of eventually underestimating the cancelled flux in an emerging and developing complex AR. On the other hand, by ``absorbing'' a tertiary bipole via grouping is a good way to ignore its influence if this is desired and would simplify the analysis (for example, because such collision caused minor activity, or because it is away from main cPIL/activity core of the AR). That was the case in the previous subsection where we ignored the sequential parasitic emergence in AR11158. Here, however, we cannot ignore the collision between those bipoles because it happens in the very core of AR12017. 

We tackle this limitation by directly measuring the drop in flux in either one of the two polarities that collide. The polarity that is the easiest to isolate and measure its flux is the N2$^\mathrm{NS}$ (flux shown with a red dashed line in Figure~\ref{FIG_5}a). In Figure~\ref{FIG_5}a we see that indeed the emergence of the second episode coincides with the moment when N2$^\mathrm{NS}$ reaches its peak flux at 18:00 UT of March 28 2014 ($\mathrm{t=t_{peak}}$). The drop in N2$^\mathrm{NS}$ flux indeed occurs right after the emergence of the second east-west bipole (Figure~\ref{FIG_5}a), specifically when the newly emerging positive flux (``P2$^\mathrm{EW}$'') collides with N2$^\mathrm{NS}$. This N2$^\mathrm{NS}$ flux that has canceled is the correction to the deficit, $\Delta_\mathrm{corr}$, and is measured as

\begin{eqnarray}
\Delta_\mathrm{corr}=\left\{
  \begin{array}{cc}
  0, & t<t_\mathrm{peak}\\
 \Phi_\mathrm{N2^\mathrm{NS}}(t=t_\mathrm{peak})-\Phi_\mathrm{N2^\mathrm{NS}}(t>t_\mathrm{peak}), & t\geq t_\mathrm{peak} \\
  \end{array}\right.   
\end{eqnarray}

\noindent The correction term that contains the missing cancelled flux is then applied to the old deficit as

\begin{equation}
	\Delta_\mathrm{new}=\Delta_\mathrm{old}+|\Delta_\mathrm{corr}|
\end{equation}

This step is reflected in the flux deficit plot for N2 P2 (the deficit is in solid black $\Delta_\mathrm{N2P2}$ and the correction $\Delta_\mathrm{corr}$ is shown as a dotted-dashed red line; Figure~\ref{FIG_5}c). Indeed, after applying the correction for the cancellation forced by the emergence of the second episode of the east-west bipole, the deficit is progressively corrected for up to 4$\times10^{20}$\,Mx (about 25\%), yielding a total of 2$\times10^{21}$\,Mx of flux cancelled during this extended emergence phase in AR12017.

The first cluster occurs during a period of cancellation with a rate of $\sim$ -1.85$\times$10$^{16}$\,Mx s$^{-1}$, leading to a peak deficit at $\Delta\approx$ 1.1$\times$10$^{20}$\,Mx over 12\,h. Before the second activity cluster, the (corrected) deficit reaches a maximum of $\Delta\approx$ 1.7$\times$10$^{20}$\,Mx with a cancellation rate of $\sim$ -2.4$\times$10$^{16}$\,Mx s$^{-1}$ over a period of 8\,h. The total flux cancelled due to the P2 N1 collision (deficit at the end of activity on March 29) is $\Delta\approx$ 2$\times$10$^{21}$\,Mx. We find that the total cancelled flux amounts to $\sim$ 14\% of the total unsigned flux for AR12017 but also $\sim$ 40\% of the unsigned flux of a perfectly balanced secondary bipole N2 P2. The latter is quite a significant number and is more meaningful than the percentage related to the total unsigned flux of the whole AR since not all of the flux from the pre-existing bipole is participating in the collision. 

The average Doppler velocities at the collisional PIL (Figure~\ref{FIG_5}b) show clear dominance for redshifts (mean $\approx$ 0.6\,km s$^{-1}$ at around the time of the activity), which are also dominant with respect to the average Doppler velocities in the nearby QS. The QS Doppler shifts result from the photospheric granulation velocity field. In Figure~\ref{FIG_DOPPLER}e,f we demonstrate visually the Doppler velocity measurements at cPIL1. After the second emergence episode, cPIL2 is formed right by cPIL1 producing an extended region of strong downflows is revealed (Figure~\ref{FIG_DOPPLER}g). A brief increase of the average blueshift seems to occur between March 29 and 30, following the second flux emergence, which also slightly strengthened the P2 polarity. During the self-separation of this newly emerging bipole, the
new P2 polarity moved farther away from sunspot N1 to the
east. As it moved to the east, it began to develop a penumbra in
the south. The Evershed flow in the weak penumbra is directed towards the LOS (the AR is in the North hemisphere), resulting in strong blueshifts at that segment of the cPIL. However, the penumbra dissolves within a day (see the animation of Figure 2) possibly due to the natural decay of the P2 polarity, which on March 30 is seen to weaken and disperse completely. This is also seen in Figure~\ref{FIG_DOPPLER}h (inset figure showing blue patch in Doppler velocities within the south-east side of the cPIL2). Nevertheless, the redshifts seem consistent and dominate over the period covering the collision of the polarities. Doppler redshifts reach maximum speeds as high as 2.5 km s$^{-1}$ at the time of collision, signifying strong downflows. It is possible that these flows may be even stronger and the HMI instrument could not properly measure them. At any rate, this consistency between reduction of flux and persistently strong average redshifts strongly suggests that we observe the submergence of magnetic flux as it is cancelling at the collisional PIL.

In Figure~\ref{FIG_5}d we can see that the shape of the polarity N2$^\mathrm{NS}$ transforms from moderately elliptical into a strongly elongated shape of 80\% oblateness within 6-10 hours before the onset of the eruptive flares. This fact, together with the 8\,Mm southward displacement of the N1 sunspot suggests that collision had started from below the photospheric surface. The collision distance (defined as the distance between the flux centroids of the colliding nonconjugated polarities of the second emergence episode's positive polarity and N2$^\mathrm{NS}$) reaches a minimum value of $\sim$ 12\,Mm at the time of high oblateness. This rapid change in the collision distance is due to the rapid anti-parallel shearing motions between the colliding polarities. The tracking of the N2$^\mathrm{NS}$ centroid becomes impossible in the middle of March 29 since it has weakened and has almost completely dissolved. The second activity cluster ends with an eruptive X1.0 flare on March 29, 17:35 UT.

The length of cPIL1 grows rapidly to about 50\,Mm within the first half of March 28 (similar to the growth of AR11158 cPIL length). Then cPIL1 becomes conjoined with the cPIL2 that forms due to the second episode of the east-west bipole; as a result, a single continuous long cPIL is produced (reaching a length of about 80\,Mm). Note that while cPIL2 lengthens, cPIL1 becomes harder to track its length, primarily due to the increased noise levels of the HMI instrument for $\Theta\gtrsim$30\,$\degr$. This led HMI to underresolve both the weak polarity elements of N1 and the positive flux that disperses from P2 and becoming weaker. Thus, it is possible that the length of cPIL1 is underestimated after the first half of March 28. At the same time, cPIL2 is more robustly extracted since it separates areas of stronger and more closely spaced opposite magnetic flux (i.e. the newly emerging positive flux and the elongated strip of N2$^\mathrm{NS}$). Nevertheless, the overall shape of the conjoined system of cPILs visually corresponds to the filament channel system seen in Figure~\ref{FIG_EUV2}c,d,e,f panels which corresponds to the location of the pre-eruptive structure before the eruptive flares at the end of March 28 and March 29. On March 30 The cPIL length is briefly seen to reach again a large value ($\sim$ 90\,Mm) as P2 moves away from N1 and becomes surrounded by dispersed negative flux (belonging to N1 first and second episodes) in the east and northeast of P2. This is created a cPIL system that almost encircled P2, effectively doubling the cPIL length. However, at that time P2 has weakened significantly and has lost its cohesion -- a manifestation of a rapid transition to the N2 P2 bipole's decay phase. An eruptive flare follows in the decay phase, 12 hours into March 30, marking the end of eruptive activity for AR12017.

\begin{deluxetable}{rccccc}
\tablecolumns{6}
\tablewidth{0pc}
\tablecaption{Flare events and CMEs produced during the evolution of NOAA AR12017}

\tablehead{
\colhead{\#} & \colhead{Flare Start (UT)} & \colhead{Carrington Longitude\tablenotemark{\dagger}} & \colhead{Heliographic Latitude\tablenotemark{\dagger}}& \colhead{Flare Class} &\colhead{v$_\mathrm{CME}$\tablenotemark{\ddagger} (km\,s$^{-1}$)}}
\startdata
1&2014/03/23 01:20 &  140$\fdg$8 &  9$\fdg$3 & C1.7 & -*\\
2&2014/03/23 05:35 &  140$\fdg$5 &  11$\fdg$3 & C2.6 & -*\\
3&2014/03/23 14:09 &  140$\fdg$0 &  10$\fdg$2 & C1.3 & -*\\
4&2014/03/23 18:37 &  140$\fdg$6 &  11$\fdg$2 & C1.3 & -*\\
5&2014/03/28 12:55 &  147$\fdg$7 &  11$\fdg$1 & C1.1 & -\\
6&2014/03/28 16:03 &  147$\fdg$7 &  11$\fdg$4 & C1.0 & -\\
7&2014/03/28 17:30 &  147$\fdg$4 &  11$\fdg$5 & C1.2 & -\\
8&2014/03/28 17:56 &  147$\fdg$9 &  11$\fdg$1 & C2.3 & -\\
9&2014/03/28 19:04 &  147$\fdg$4 &  10$\fdg$8 & M2.0 & 420\\
10&2014/03/28 23:14 &  147$\fdg$4 &  11$\fdg$2 & C1.0 & -\\
11&2014/03/28 23:44 &  146$\fdg$2 &  10$\fdg$6 & M2.6 & 514\\
12&2014/03/29 07:50 &  146$\fdg$4 &  10$\fdg$2 & C2.1 & -\\
13&2014/03/29 09:53 &  144$\fdg$3 &  9$\fdg$1 & C1.4 & -\\
14&2014/03/29 10:48 &  146$\fdg$6 &  11$\fdg$0 & C1.7 & -\\
15&2014/03/29 12:41 &  147$\fdg$6 &  11$\fdg$9 & C1.5 & -\\
16&2014/03/29 13:18 &  147$\fdg$6 &  10$\fdg$6 & C1.0 & -\\
17&2014/03/29 14:26 &  142$\fdg$9 &  10$\fdg$1 & C3.3 & -**\\
18&2014/03/29 16:20 &  148$\fdg$5 &  11$\fdg$4 & C1.1 & -\\
19&2014/03/29 17:35 &  146$\fdg$3 &  10$\fdg$5 & X1.0 & 528\\
20&2014/03/30 11:48 &  146$\fdg$8 &  10$\fdg$2 & M2.1 & 487***\\
21&2014/03/30 21:10 &  147$\fdg$0 &  10$\fdg$2 & C7.6 & -***\\
\enddata
\tablenotetext{\dagger}{Flare centering obtained by intercorrelating the GOES, RHESSI and HEK event lists.}
\tablenotetext{\ddagger}{CME velocity (linear fit) obtained from CDAW CME catalog.}
\tablenotetext{*}{Minor activity occurred in early phase of N1 P1 emergence.}
	\tablenotetext{**}{Event occurred during decay phase in sPIL1 as suggested from flare ribbons seen in 1600\,\AA\ from \textit{SDO}/AIA.}
\tablenotetext{***}{Event occurred during decay phase.}
\label{TABLE_2}
\end{deluxetable}

\section{The Physics of Collisional Shearing and its Relation to Major Solar Activity}\label{MODEL}

The observations we report in this article have significant implications regarding our theoretical understanding of the mechanisms causing major solar activity. First, for each AR we studied there was a phase during which rapid emergence of new flux was ongoing and there was virtually no major activity produced in the corona. This is true for the first (double) flux emergence episode of AR11158 (Figure 7), but also for AR12017, which initially begins its life as a strong bipole (only producing four minor subflaring events in its self-PIL and then remained inactive for a period of 5 days; Figure 10). Evidently, the emergence phase doesn't lead immediately to flaring activity or eruptions, despite that the most intense activity is expected during that stage in the evolution of complex ARs~\citep{Schrijver_2009}. What stands out clearly is the contrast of activity before and after the onset of collision. The collision occurs between nonconjugated polarities of the opposite sign. Magnetic polarities are guided into a collision course due to their relative tilt and self-separation. The colliding polarities slip past each other with several hundreds of meters per second (average shear velocities between 200-400 m s$^{-1}$ measured in both fundamentally different ARs).

Just before the onset of collision, the relative proximity due to the convergence of the opposite nonconjugated polarities permits the development of a direct connectivity between them. As a result of the ``frozen-in effect''~\citep{Alfven_1942}, the connectivity in the atmosphere behaves as if it is ``anchored'' to footpoints in the respective polarities at the high-density photosphere. Therefore, after the onset of the collision phase, the continuous proper motions of the nonconjugated polarities are driving the shear of the ``line-tied'' field about (and above) the collisional PIL. We name this process \emph{collisional shearing} to emphasize the fact that the shear and convergence develop simultaneously due to the proper motions of the polarities during the collision. Cancellation naturally occurs in the collisional PIL, resulting in connectivity changes of the sheared arcade into a helical (poloidal) structure (as suggested by \citealt{vanBallegooijen_Martens_1989} at least for the formation of filament channels in decaying ARs). We present this idea in a cartoon model in Figure~\ref{FIG_6}, where photospheric reconnection occurs above the collisional PIL (shown with yellow color). Initially, the arcades converge and also begin to shear due to the collision. This brings them into such proximity and relative field line orientation (near the sheared loop footpoints along the cPIL) that reconnection begins at a height above the cPIL. The reconnection produces two post-reconnection connectivities, i.e. (1) predominantly axial structures aligned along the cPIL (magenta dashed line in Figure~\ref{FIG_6}a), and (2) loops of small radius of curvature (black line in Figure~\ref{FIG_6}a) which submerge below the photosphere (flux cancellation at the cPIL is currently underresolved by contemporary magnetograph instrumentation). With ongoing collisional shearing driving the photospheric reconnection, additional flux is wrapped around the axial structure (i.e. is added as poloidal flux) and an equal amount of flux is submerged under the surface. The location of the changes in connectivity due to the cancellation scenario at the cPIL coincides with confined sub-flaring and major flare activity at the cPIL (and \emph{not} at the sPILs). This is confirmed by visual inspection of EUV imaging (e.g. Figures~\ref{FIG_EUV1}, \ref{FIG_EUV2}) and by the three different sources for the centering of flare events (i.e. the GOES, RHESSI and HEK intercorrelated events in Table~\ref{TABLE_1} and Table~\ref{TABLE_2}, and Figures~\ref{FIG_2}, \ref{FIG_3}). 

The ambient/large-scale magnetic field configurations are expected to interact with the presence of a stressed magnetic flux system in the corona, the MFR, that forms due to photospheric cancellation driven by collisional shearing (cartoon Figure~\ref{FIG_6} c). The MFR is manifested by a sigmoidal structure in coronal EUV imaging in both ARs, which eventually becomes unstable and lifts and drives a CME. In addition, the Doppler measurements provide us with additional insight about the process. The fact that blueshifts and redshifts co-exist in the same collisional PIL has several implications regarding the physical processes involved. Indeed the cPIL average redshift dominates over the cPIL blueshift and also the average Doppler signatures of the QS, suggesting submergence at the collisional PIL. However, average blueshifts are also present in the cPIL (see Figure~\ref{FIG_DOPPLER}). This suggests the possibility that cancellation occurs in a fragmented fashion along the PIL (i.e. small ``pockets'' where sheared flux is oriented appropriately for photospheric reconnection to occur), which may then produce several stressed and twisted segments of MFR structures above the collisional PIL. These in turn may interact with each other via reconnection in the chromosphere and corona that is manifested as transient brightenings, forcing them to merge into a more monolithic system (e.g. \citealt{Chintzoglou_etal_2015}). Another possibility is that blueshifts represent locations where reconnection occurs below the formation height of the Fe I line. However, photospheric reconnection alone cannot explain the transient activity in the corona by itself as can be explained in the former case of fragmented formation of MFRs which would interact with each other or with sheared field near them, producing flare reconnection in the corona. In addition, an MFR can also lose poloidal flux by reconnecting with nearby flux systems if it is confined to stay above a PIL \citep{Chintzoglou_etal_2017}.  

In Figure~\ref{FIG_MF} we show results from an evolutionary Cartesian MF model (\citealt{Cheung_DeRosa_2012, Fisher_etal_2015}) for the coronal field of AR11158. The magnetic field is evolved (data-driven) by electric fields, $\mathbf{E}$, calculated with Faraday's law of induction 

\begin{equation}
	\frac{\partial\mathbf{B}}{\partial t}=-\nabla\times\mathbf{E}
\end{equation}

\noindent at the bottom boundary (photosphere) directly from the full photospheric vector magnetic field observations, $\mathbf{B}$, as $\mathbf{E}=-\mathbf{v}\times\mathbf{B}+\eta\mathbf{J}$ (Ohm's law for a conducting fluid), where $\eta$ the magnetic diffusivity, $\mathbf{v}$ the inductive velocity, and $\mathbf{J}$ the electric current density. The details of the electric field and energy flux analysis in AR11158 can be found in~\citet{Kazachenko_etal_2015}. The vector fields were projected onto a Cartesian system of reference. The simulation domain is 512$\times$512$\times$192 pixels$^3$ with $dx=dy=$0.36\,Mm and $dz=1.5dx$. The Cartesian MF code solves the following form of the induction equation 

\begin{equation}
\frac{\partial\mathbf{A}}{\partial t}=\mathbf{v}\times\mathbf{B} - \eta\mathbf{J}
\end{equation}

\noindent where

\begin{equation}
\mathbf{B} = \nabla\times\mathbf{A}
\end{equation}

\begin{equation}
\mathbf{J} = \nabla\times\mathbf{B}
\end{equation}

\begin{equation}
\mathbf{v} = \frac{\mathbf{J}\times\mathbf{B}}{\nu}
\end{equation}

\noindent where $\mathbf{A}$ the magnetic vector potential and $\nu$ the magnetofrictional coefficient (a parameter that modifies the velocity to ensure numerical convergence).

Stressed magnetic structures appear to form in the atmosphere self consistently due to collisional shearing acting at the cPIL (Figure~\ref{FIG_MF}a,b,c,d). The MF model captures the two bipole emergences initially as two separate conjugate flux systems (Figure~\ref{FIG_MF}a), which then interact as collision proceeds with time. The proper motions of the colliding polarities show P2 to overtake N1, slipping rapidly to the west. The connectivity between P2 and N1 in the low corona progressively delineates the PIL shape (Figure~\ref{FIG_MF}b,c,d). The location of the stressed magnetic structure coincides with the collisional PIL, the location where intermittent sub-flaring (flares up to C9.9 level) and major flare activity occurs in the corona (Figure~\ref{FIG_2} and Table~\ref{TABLE_1}). This localized activity lasts for as long as the collisional PIL exists. Note that the connectivity above self-PILs (i.e. conjugate bipoles N2 P2 and N1 P1) is nearly-potential (and consequently less stressed/energized) at all times (Figure~\ref{FIG_MF}a,b,c,d), in agreement with the observation that no flares occurred in the self-PILs of N1P1 and N2P2 (Table 1 and Figure~\ref{FIG_3}). However, the most striking finding from the MF model is the formation of a coronal magnetic flux rope structure at the collisional PIL only (Figure~\ref{FIG_MF2}) in agreement with our interpretation of the observations of collisional shearing. The MFR structure and the neighboring sheared arcade field lines were both traced from a region of strong electric field magnitude, $|\mathbf{E}|$, at the cPIL. That region of cPIL is both at a location where cancellation occurs (creating the MFR structure) and at a location where the MFR begins to interact with its immediate environment, i.e. with a simply sheared arcade (i.e. not a highly twisted structure) neighboring with the MFR. This finding from the model may also explain the confined transient activity observed above compact cPILs in complex emerging ARs \citep{Chintzoglou_etal_2015}.

\section{Discussion}\label{DISCUSSION}

Strong activity originates from complex emerging ARs throughout the solar cycle. Our observations of highly flare- and CME-productive (emerging) ARs are in contrast with the commonly investigated theoretical scenarios with (a) emergence-based data-inspired modeling, i.e. simulating eruptions during the emergence phase of simple bipolar AR, and (b) works based on the cancellation scenario in a simple emerging bipole or based on simple shear (no cancellation) in quadrupolar configurations. In the following paragraphs we discuss how our scenario and data-driven MF modeling is in contrast with commonly proposed scenarios to explain extreme flare activity and eruptions.  

First, virtually no eruption/significant activity is seen to originate from the ``self-PILs'' of the individual ``constituent'' bipoles. That is, no singular bipole during the emergence phase produced intense events in our observations. Emergence-based simulations of ARs that produced eruption(s) were obtained from the emergence of a single but highly twisted flux tube, forming a bipole with a compact self-PIL (e.g. \citealt{Fan_2001, Manchester_etal_2004, Gibson_Fan_2006, Archontis_Torok_2008, MacTaggart_Hood_2009, Archontis_Hood_2010, Roussev_etal_2012, Leake_etal_2013, Magara_2015, Syntelis_etal_2017}). Diffusion/cancellation-based simulations consider the conditions of shear and convergence in a single bipole (e.g. \citealt{vanBallegooijen_Martens_1989, Martens_Zwaan_2001, Amari_etal_2003b, Mackay_vanBallegooijen_2006a, Aulanier_etal_2010, Amari_etal_2011, Zuccarello_etal_2016}; note that we do not consider the case of inter-AR interactions, since at least one AR could be in its decaying phase). However, the cancellation-based models are, by construction, unable to capture the dynamics of the evolution of ARs as a result of the polarity proper motions (in such simuations the polarities are fixed and/or prescribed by the pre-emergence shape of the flux tubes). Instead, these models operate on an initial bipolar magnetic field distribution at the surface by varying surface flows (leading to diffusion, convergence and shearing).
While numerical setups require a certain degree of simplicity offered by the choice of a simple bipole, here we demonstrate that our attention needs to be shifted to the local physics of collisional-PILs (requiring at least two colliding nonconjugated polarities) and not in their respective self-PILs (i.e. originating in polarities belonging to the same bipole) in order to properly explain the associated solar activity. Shearing and convergence of opposite-signed flux required for the cancellation scenario is naturally provided by the proper motion of nonconjugated sunspots that drive the collision. The proper motions are prescribed by the self-separation and tilt of the bipoles.

\subsection{Are ``simple'' $\delta$-spot ARs truly quadrupolar or  bipolar?}\label{DELTA_SPOT}

There is evidence that $\delta$-spot ARs, which largely look bipolar, do not seem to host flares and spew eruptions simply due to their emergence. Such an example of a well-observed, seemingly ``bipolar'' delta spot (with polarities predominantly separated by a long and compact PIL) is the case of NOAA AR11429. \citet{Chintzoglou_etal_2015} showed that secondary and tertiary emergence of small bipoles are ``enriching'' the flaring and eruptive potential of this emerging ``bipolar'' AR. The emergence of a small bipole began at the north-east part of the PIL and continued to self-separate against the AR11429 PIL, in which collision and shear with the pre-existing nonconjugated polarity distributions naturally occurred due to the proper motions of the self-separating emerging bipole (\citealt{Chintzoglou_etal_2015}; see Figure 3 and discussion therein). Emergence did not seem to trigger an eruption immediately, but it took $\sim$ 1.5 days from the moment of this secondary emergence for the first CME to occur. However, in the meantime, the AR showed transient confined subflaring and flaring activity. An inherent limitation in such large compact PIL $\delta$-spots is that since the PIL is of such strong spatial gradients, the contrast of the spatial distribution in magnetic field values in individual elements along the PIL is rather low. AR11429 was a case observed continuously with modern, high resolution magnetograms (from \emph{SDO}/HMI) which allowed us to capture the course of the secondary bipole's self-separation along the compact-PIL. Note that this behavior is not captured in the simulation by~\citet{Takasao_etal_2015}. The latter work claims only a rough comparison of the surface evolution of a kink-unstable flux tube to the observed AR11429. We must stress here that the authors would not be able to consider the early phases of this AR's emergence, since it already appeared emerging and developing in the east limb. In addition, in \citet{Takasao_etal_2015}, Figure 16, where there is an attempted comparison between the compact PILs of the simulation with the observations, they do not consider the trailing negative sunspots in the east of AR11429. We argue here that the conclusion, suggesting that the overall evolution of this AR being due to the emergence of a single idealized kink unstable flux tube (such as in~\citealt{Linton_etal_1998, Linton_etal_1999}, must be taken with caution. The multipolarity seen in this AR, is certainly higher than that of a simple quadrupole. While at some stage the polarities could even be largely grouped as ``quadrupolar'' (and to some extent, even ``bipolar''), this AR was indeed the result of several \emph{individual} flux tubes, either emerging sequentially or simultaneously. This can be confirmed for as long as the individual polarities can be counted (up to the point they congregate with each other as it is seen later in the evolution of AR11429). We stress here the importance of using well-observed ARs (meaning from the very beginning of their emergence stage all the way to maturity) in order to draw safer conclusions about the drivers of intense solar activity. Since $\delta$-spot ARs are rare and our basic knowledge originates from conclusions that were drawn from previous sparse-in-time ground based observations (and low-resolution observations by SOHO/MDI, albeit continuous), it may not be safe to consider $\delta$-spots as simply ``bipolar'' or ``quadrupolar'' structures. Low-resolution SOHO/MDI observations and sporadic (sometimes only daily) ground-based magnetograms from solar cycles before the 24th, may be giving a wrong impression that ``bipolar'' $\delta$-spots or ``island''-$\delta$-spots are compatible with idealized eruptive bipole models (see following subsections and also \S~\ref{THEORY}). 
All the above suggest that numerical models need to account for the observed complexity (proper motions, multipolarity) that is often seen in flare-productive ARs. This now seems to become possible thanks to advances in computing power over the last decade.

\subsection{Implications for Previous Emergence-based Data-inspired Models}

As mentioned earlier, the characteristic way of how convergence and shearing occurs in two colliding bipoles is fundamentally different than the one often simulated by the emergence of a simple conjugate bipole. In such simulations of a simple conjugate bipole, the major activity occurs in its self-PIL (in this case there is no collisional PIL, since there is no collision happening in a single conjugate bipole). In ``emergence-based'' numerical experiments, all simulated major activity is typically emanating from the PIL in a solitary conjugate bipole. Here, both of our ARs under study did not exhibit conjugate bipoles with compact self PILs (except when the two bipoles collided and formed a \emph{collisional} PIL), and, unsurprisingly, no major activity (i.e. flare clusters, CMEs) occurred along the sPIL of any bipole of these two ARs.
 
A different approach has been taken by \citet{Toriumi_etal_2014}, in which the evolution of the well-observed AR11158 is modeled in two scenarios, first, ``Case 1'' where the AR evolves as a result of the emergence of a single flux tube with a dip in the middle of its length (causing two bipoles at the surface), and ``Case 2'' where the evolution is a result of two buoyant flux tubes that are relatively near to each other. These two cases in~\citet{Toriumi_etal_2014} were produced to infer the subphotospheric origin of AR11158 by comparing the proper motions and the non-potential shear at the PIL. Case 2 was rejected over Case 1 with the argument that the polarities N1 P2 did not collide and thus did not produce a compact PIL, as it happened in the actual AR11158. However, Case 1 did produce a compact PIL with non-potential shear and, due to that, it was presented as the best candidate for the subphotospheric origin and structure of AR11158. This simulation setup was also used in the work of~\citet{Fang_Fan_2015} and later on,~\citealt{Toriumi_Takasao_2017} who considered possible formation mechanisms for $\delta$-spot ARs. Our observations are in contrast with this scenario for two main reasons as we explain below. 

First, while the proper motions in Case 1 lead N1 P2 into a collision course consequently producing a compact PIL, at the moment of minimum centroid distance (i.e. maximum compression due to collision) they stop moving and remain interlocked for the remainder of the simulation. The multi-day observations for AR11158 (extending past the CME of February 15) reveal a different behavior for the proper motions between N1 P2 of Figure~\ref{FIG_2}. 
Our observations of AR11158 (also fully described in \citealt{Chintzoglou_Zhang_2013}) show that the polarities P2 and N1 do not stay interlocked after they collide but instead P2 slips rapidly by N1 and then migrates toward the west to meet polarity P1. In the \citet{Toriumi_etal_2014} the authors track the individual polarities with a flux-weighted centroid which it is a crude, yet adequate enough method to track the motion of the polarities. Our point is demonstrated by comparing Figure 5 (Case 1) and Figure 6 (Case 2) in \citet{Toriumi_etal_2014}. In their Figure 5, one can still see that the flux-weighted centroid tracking follows the leading portion of polarity P2 reasonably well. Consequently, one can see how the position of the leading edge of P2 changes with time, since it is also near the flux-weighted centroid. It is clear that in Case 1 the final distance (i.e. last panel at $t/\tau_0$=150) between the strong flux leading edge of P2 is as far as 40 $x/\mathrm{H_0}$ normalized distance units from the core of the P1 polarity. In comparison, the conjugate polarity separation of the individual bipoles is about 50 $x/\mathrm{H_0}$. In addition, P2 appears to interlock with N1 quickly, since P2 assumes its final position at $t/\tau_0$=100. For Case 2, the final distance between P2 and P1 is comparatively very small, $\sim$ 10 $x/\mathrm{H_0}$ with bipole separations of $\sim$ 70 $x/\mathrm{H_0}$, which is more consistent to the observed final separation of the P1 and P2 polarities (compare the strong flux leading edge of P2 polarity in the actual observations of AR11158 in our online movie1). This discrepancy between the observations and the simulated Case 1 can be well understood if we consider a physical limitation in the construction of Case 1. Because the colliding polarities P2 and N1 are sharing a ``dip'' at the bottom of the simulation's domain (i.e. since they were made out of a single progenitor flux
tube), this numerical setup strongly anchors these two polarities below the photosphere. Therefore, while this setup succeeds in bringing N1 and P2 together to collide, it also confines their photospheric proper motions not to move far from the location of the dip. Hence it is impossible to model the observed evolution of AR11158.

Second, in Figure 8a of \citet{Toriumi_etal_2014}, we can see that while the flux increases during the time of increasing self-separation distance, during collision, the flux in the colliding polarities is increasing even more, in excess to that of the non-colliding polarities. In our measurements presented in Figure~\ref{FIG_FLUX}a, in the beginning of February 14 during which polarity N1 and P2 are well into their collision phase, their conjugated non-colliding sunspots happen to reach the same amount of flux, i.e. $\Phi_\mathrm{P1}=|\Phi_\mathrm{N2}|=7\times10^{21}$\,Mx. At that time we can see that $\Phi_\mathrm{P1}>\Phi_\mathrm{P2}$, with their difference equivalent to the measured flux deficit which should follow equation~(\ref{DEFICIT_EQUIVALENCE_EQ}). However, according to Case 1, where before the onset of collision the bipoles are completely symmetric and balanced, after collision $\Phi_\mathrm{P2}>\Phi_\mathrm{P1}$ (\citealt{Toriumi_etal_2014}; Figure 8a), in contrast to our observations of AR11158. For Case 2, the disagreement with the observations seems to be somewhat milder, at least from the scope of proper motions. As the two parallel flux tubes emerge and the polarities self-separate, N1 moves to meet N2, and P2 separates from its conjugate N2 to meet the leading polarity P1. This is very consistent with the observations of AR11158 (Figure~\ref{FIG_2} and accompanying animation of Figure 2). While the authors rejected this case due to the absence of a compact PIL between N1 P2, we support that Case 2 describes the observations more closely (albeit without a compact cPIL), suggesting that AR11158 was a product of two individual emerging flux tubes.

In a recent simulation by \citet{Toriumi_Takasao_2017} the former two scenarios (Case 1 and 2) are reproduced in twice the spatial resolution (512$^3$ domain size) with the addition of two more scenarios. That is, a (linear) ``Quadrupole'' emergence (similar to Case 1), two flux tube emergence named as ``inter-AR'' emergence (but similar to Case 2), kink-unstable bipolar emergence referred to as ``spot-spot'' emergence, and parasitic bipole emergence called therein as ``spot-satellite'' emergence. In all cases, a current sheet forms above the PILs, with the strongest being the one of bipolar AR emergence (their ``spot-spot'' case, arising from a single highly twisted, kink-unstable flux tube). These simulated cases investigating $\delta$-spot formation produced high non-potential shear in the resulting compact PILs. Cancellation is not investigated in these cases and no eruption was produced. Nevertheless, the cases of simultaneous emergence (``inter-AR'') and parasitic emergence (``spot-satellite'') roughly capture the large-scale proper motions as shown in our observations of the two ARs studied here. We suggest that a reduction in the relative spacing between two colliding flux tubes in the simulations may be a promising approach for future emergence-based simulations.


\subsection{Implications for Previous Shear-based or Cancellation-based Data-Inspired Models}\label{DIFF}
Numerical models that operate by diffusion and shearing (i.e. cancellation-based and not emergence-based) seem more appropriate for the mature or decaying phase of ARs where turbulent diffusion and differential rotation drive the decay in ARs. These ARs, despite they are on their decaying phase, are observed to produce an eruption even although less energetic than those produced in emerging flux regions. In complex emerging ARs, it is not uncommon to have recurring eruptions (such as in the cases we present in this article). Our proposed scenario naturally explains the solar activity in the collisional PIL by means of cancellation but without the ad hoc assumptions of diffusion and shearing occurring in individual bipoles; the proper motions self-consistently drive cancellation for as long as the collision between opposite-signed polarities of nonconjugated bipoles persists. 

There is another class of eruptive models that energize the pre-eruptive structure solely by shearing, known as the ``Break out'' (introduced by \citealt{Antiochos_etal_1999}; also \citealt{Lynch_etal_2008, Wyper_etal_2017}). In the case that is relevant in forming CMEs (i.e. \citealt{Antiochos_etal_1999} and Figure 3 in~\citealt{Wyper_etal_2017}), four polarities are required to create a specific type of magnetic topology that could lead to an eruption. Similar to the other models in this category, this model completely disregards the occurrence of cancellation and considers just the shearing of an initial arcade. The arcade is driven at its footpoints to shear and as a result expands upwards~\citep{Mikic_Linker_1994}, forcing its reconnection with the overlying flux at a topological feature (i.e. a null point or a separator line). Then, as the sheared arcade is free to expand further, ``standard model'' reconnection is triggered underneath the expanded arcade and accelerates the eruption by adding more poloidal field around the escaping structure (a positive feedback resistive process forming an MFR). However, the simple shearing of the arcade is not able to explain the transient flaring and sub-flaring activity seen above the collisional PIL for as long as the collision takes place. Contrary to the emergence-based and cancellation-based models, the flux rope in the ``break out'' model forms \emph{after} the onset of the eruption and \emph{not} before the eruption. This is in contrast with our observations of collisional shearing (for example, see Figure~\ref{FIG_EUV1}, \ref{FIG_EUV2} and accompanying animations). While a quadrupolar configuration is the bare minimum for creating interesting magnetic topology in the corona (needed for the Break out model to operate), the amount of cancellation we measure due to the colliding flux tubes in ARs suggests that flux ropes naturally form prior to their eruption. The pre-existence of a magnetic flux rope forming at the collisional PIL is fully supported by our data-driven MF model of collisional shearing as discussed in \S~\ref{MODEL}.   

\subsection{Incompatibility with the CSHKP Standard Model and 2D Models in General}

The CSHKP Standard Eruptive Model (a collection of cartoons and 2D analytical modeling found in the works of \citealt{Carmichael_1964, Sturrock_1966, Hirayama_1974, Kopp_Pneuman_1976}) is essentially a simplistic 2D model that considers only one bipole to explain eruptive flares. We need to stress here that our proposed scenario for flare productive ARs shows the limitations of CSHKP, again for the same reasons as above (as in emergence-based single-bipole simulations), since more than one bipole was required to produce the eruptions, at least during the emergence phase. Also, due to its 2D nature, the CSHKP cannot account for the observed behavior in bipoles (i.e. shearing and/or cancellation). 

This eruptive model has been recently extended to 3D (\citealt{Aulanier_etal_2010}) in an attempt to include the mechanisms of cancellation and shearing, but again, only involving a single bipole. Note that the differences of this ``3D CSHKP'' model with diffusion-based models are minimal. We should acknowledge that the authors strive for realism, by composing an asymmetric magnetic distribution at the photosphere, something that distinguishes it from the previous diffusion-based models. As we already mentioned in \S\ref{DIFF}, the approach of a diffusion-based simulation is more appropriate for the conditions in decaying ARs. Not surprisingly, in \citet{Aulanier_etal_2010} a decaying AR was chosen, instead of an emerging AR, to demonstrate a crude similarity of the modeled coronal sigmoid with the observations. The authors do mention that the numerical setup for cancellation they used (i.e. diffusion) is not appropriate for emerging ARs but probably loosely appropriate for decaying ARs. With regards to shearing motions, a maximum shear velocity of 5\% of the Alfv\'{e}n speed at the photosphere was prescribed about the self PIL, peaking at 50\,m s$^{-1}$, which is a very slow speed as compared to the hundreds of m s$^{-1}$ of collisional shearing we observe in PILs. We thus consider this ``3D CSHKP'' model as an attempted simulation of an eruption inspired from a \emph{decaying} single bipole AR but \emph{not} consistent with the conditions found in a complex emerging, and still evolving AR. 

The above ``3D CSHKP'' model is in strong contrast with our observations of collisional shearing in emerging ARs. In the observations, the long period of inactivity before the onset of collision also suggests that in none of each bipole's self-PILs a flux rope was formed or was forming during their emergence. If this was not the case, we would have expected to either (a) see activity emanating from the self-PILs and/or from associated sigmoids/stressed magnetic structures about the self-PILs, or (b) expect the activity to scale in the entire PIL path, i.e. the self-PIL/collisional-PIL/self-PIL integral system (leading to a huge ribbon system during/after an eruptive flare). Moreover, if there were two flux ropes, each of them formed above the respective bipole's self-PIL, the ribbons from the eruptive flare should occur above the self-PILs. However, this is not the case in the observations. We do not see any ribbon brightenings or eruptive activity in the self-PILs, at least before the time the ARs enter their decay phase. In the decay phase, a collisional PIL does not exist but instead there is only one integral PIL, typical of those in diffuse decaying ARs. Therefore we conclude that the collisional-PIL is the location where physical conditions lead to eruption. This is a strong constraint for computational models since our observations disagree with all simulations that produce eruptions in the self-PIL of a single emerging bipole.

\citet{Chen_Shibata_2000} constructed a 2D eruptive model that in part involves the 2D standard model reconnection for the destabilization of a pre-existing ``MFR'' (we write MFR in quotes because an MFR is naturally a 3D structure and in a 2D model is not properly described). This model attempts to explain the observation that CMEs often occur in bipolar ARs that eventually become quadrupolar by the subsequent emergence of a new bipole inside or outside the initial self-PIL. The authors acknowledge the limitations of this 2D over-simplified approach, however they claim that this model reproduces key characteristics of solar eruptions. The assumption behind this model is that there is an ``MFR'' (i.e. O-loop system in 2D) pre-existing above the PIL, until secondary emergence (with orientation for the emerging poloidal field ``favoring'' reconnection in the 2D corona) causes a loss of equilibrium. In the case where emergence occurs inside the self-PIL, reconnection with the arcade below the ``MFR'' removes magnetic pressure and the flanks of the strapping flux move toward the PIL and form a current sheet underneath the ``MFR''. Then the ``MFR'' rises in the same fashion as in the CSHKP model. In the case where the emergence occurs outside of the self-PIL, reconnection of the emerging flux with the overlying flux (``strapping'' the ``MFR'' at a fixed height) weakens the latter and the MFR begins to lift. However, being a 2D model, the model by \citet{Chen_Shibata_2000} cannot possibly account for the shearing and the cancellation occurring in different configurations of magnetic distributions. Our observations of two ARs that are fundamentally different from each other, disagree with this model. The emergence of the bipoles (either simultaneous or sequential) drives the proper motions and forces collisional shearing, resulting in cancellation that creates the MFR. In the simultaneous case (AR11158), there was not an MFR above any segment of the PIL before the collision. This is simply because the conditions that could create a flux rope were not met in any part of the PIL before the collision. We conclude the same for the case of sequential emergence (AR12017), especially because the emergence of a parasitic bipole occurred in proximity to one of the two pre-existing polarities (i.e. N1) and a filament channel/MFR formed only during the collision in the photosphere. In both cases, multiple CMEs were produced during the period of collisional shearing (CME speeds of these are provided in Tables~\ref{TABLE_1} and~\ref{TABLE_2}), suggesting a different trigger mechanism for the CMEs, not inconsistent with that of ideal MHD instabilities (either the torus or the helical kink flux rope instabilities).


\subsection{Towards More Realistic Physics-Based Models}

In recent years, fully 3D radiative-convective MHD simulations of an emerging single flux tube have become possible using new sophisticated techniques and thanks to the increased computational power (e.g. \citealt{Cheung_etal_2010}). The realistic treatment of the radiative physics adds the missing complexity in the photospheric layers, such as reproducing the photospheric granulation as observed, and, account for the natural decay of polarities once they form and after they emerge. However, proper motions of polarities of emerging bipolar ARs in radiative-convective MHD simulations do not obey the self-separation seen in the observations: the polarities are seen to pop up in the photospheric surface at locations affixed by the footpoints of the emerging flux tube at the bottom of the computational domain (e.g. \citealt{Cheung_etal_2010}; see Figure 1 and space-time plot of Figure 7 therein). In addition, the characteristic cohesion of the conjugated polarities as they form a real bipolar AR is not reproduced in those models (i.e. polarities in those simulations look very dispersed or decayed-like even as they form during flux emergence). Thus, the next step for this kind of advanced models would be to simulate the collision of two individual emerging bipoles and reproduce the cancellation we detect with the $\Delta$, as well as confined and eruptive flares associated with the formation of a flux rope. This can be done in an idealized fashion, or by running data-driven simulations like the MF model studied here, but extended to fully-compressible 3D MHD with the relevant physics. 

\subsection{Implications for Improving the Prediction of Flares}

It is striking that for the time AR11158 was producing confined and eruptive flares, its magnetic configuration was classified by NOAA as a $\beta$ and it wasn't until February 14 when the AR was classified as $\beta\gamma$, and then February 16 when it was finally classified as $\beta\gamma\delta$. An AR with a $\beta\gamma\delta$ configuration has overall $\beta\gamma$ AR characteristics, but also has umbrae of opposite polarity within the same penumbra. As for AR12017, initially a $\beta$ on March 22 and 23, was a $\beta\gamma$ on March 24 without any flaring activity produced. After a long gap of activity, the AR was classified as $\beta$ even though a strong cluster of activity happened on March 28 (2 M-class flares and 2 CMEs). Main activity (including the X1.0 eruptive flare) happened on March 29 with the AR being in a $\beta\delta$ configuration. The AR was finally classified as $\beta\gamma\delta$ on March 30 when only one M-flare and one C-flare occurred, before decaying back to $\beta\gamma$ and $\beta$ for the two following days. It is evident from both of our flare- and CME-productive ARs that the Mt Wilson classification scheme captures the crude complexity of the AR but with a gross latency (such as in the first 5 days of AR11158; see evolution of flux and Mt Wilson class in Figure~\ref{FIG_FLUX}a and likewise in AR12017; Figure~\ref{FIG_FLUX}b). With this loosely defined classification scheme, it is possible to have ARs that are classified as $\beta$ or $\beta\gamma$ but being as flare productive as $\delta$ ARs. In the statistical sense, this introduces significant problems in blindly predicting solar activity based on the Mt. Wilson classification of magnetic configurations used by NOAA, since the underlying physics are not properly captured. Therefore in order to explain the activity of AR11158 and AR12017, it is necessary to look into the local physics around the PIL of the AR and identify the causality behind this activity.

An example of an approach where the local physics are considered for flare prediction was undertaken by \citet{Leka_Barnes_2007}. In that work they performed statistical tests on a large dataset of $\sim$ 500 ARs using daily vector magnetogram data from the ground-based Imaging Vector Magnetograph (IVM) of the University of Hawaii. They considered multiple photospheric variables (such as the total unsigned magnetic flux $\Phi_\mathrm{tot}$, vertical electric current $\mathrm{J_z}$, force-free parameter $\alpha$, etc) and combinations of them and they found that the quantity with the most predictive power is the total flux $\Phi_\mathrm{tot}$. The best success rate for $\Phi_\mathrm{tot}$ in that work was found to be 77\%, making it the best predictor for flare activity. 

Similarly, in the work by \citet{Welsch_etal_2009}, the authors explored several correlations between intensive and extensive parameters in flaring ARs from \emph{SOHO}/MDI magnetograms (low resolution and sensitivity as compared to \emph{SDO}/HMI). They noted that one extensive property associated with flows, $S_R$, a simplified estimate of the Poynting flux, was associated at a level compatible with the R parameter (which quantifies the flux near the PIL; \citealt{Schrijver_2007}). They also noted that $S_R$ was also more strongly associated with flares than $|\mathbf{B}|2$, implying that the square of the magnetic field strength is not solely responsible for the association. In conclusion, they mention that the flows (including shearing) are an essential factor in associating their quantity $S_R$ with flare activity. Our proposed interpretation, therefore, is still consistent with this aspect of \citet{Welsch_etal_2009} work. Furthermore, the authors considered additional parameters, such as the rate of the signed change of flux, $\dot{\Phi}/\Phi_\mathrm{tot}$ (i.e. denoting either emergence or cancellation in an AR). They found it to be more strongly correlated with flare power than the unsigned rate of change, $|\dot{\Phi}|/\Phi_\mathrm{tot}$, implying that flux emergence is more strongly correlated with flare energy release than flux cancellation, which is what the authors also suggest in their interpretation.

In our paper we show that for emerging ARs it is not possible to quantify cancellation by simply measuring $\dot{\Phi}$ (only meaningful for decaying ARs!), unless one applies our proposed flux deficit method (see  Figure~\ref{FIG_DEFICIT} and a discussion of it in \S~\ref{METHOD}). Our analysis for two observed cases shows a very good correlation for the onset of cancellation that precedes the onset of flare clusters. To our knowledge, such a direct correlation has not been reported by any previous statistical study of ARs or even in case studies of individual ARs.  We suggest that the set of previously used parameters in the correlations should be augmented with the flux deficit $\Delta$.

None of the aforementioned statistical works have demonstrated that $\Phi_\mathrm{tot}$ is a good predictor of X-flares, since given the same $\Phi_\mathrm{tot}$, there is still a wide distribution of flare classes. For example, if it were a simple $\beta$-type AR with a lot of $\Phi_\mathrm{tot}$, the region has low probability of flaring. Our observations, however, suggest that the best predictor $\Phi_\mathrm{tot}$ may be intimately related to the proper motions driving the shearing and cancellation between different bipoles nested within the same AR; since strong collision and rapid shearing develops as a result of the proper motions due to the emergence of each colliding bipole, the motions will last for as long as the emerging flux is increasing in each bipole. However, here, we show that activity occurs only when cancellation is ongoing due to the collision between nonconjugated polarities. $\Phi_\mathrm{tot}$ (calculated for the entire AR) does not discriminate between bipoles within the same AR. Thus $\Phi_\mathrm{tot}$ will continue to increase if additional flux emerges in the self-PIL of either of the colliding bipoles (compare $\Phi_\mathrm{tot}$ with the individual polarity fluxes in Figure~\ref{FIG_FLUX}). Note that it is not necessary that a reduction of flux due to cancellation would be detectable in the emergence phase of the $\Phi_\mathrm{tot}$ time profile, especially because the rate of emergence is much higher than the flux cancellation rate. Had it been different, the emergence of flux through the photosphere wouldn't occur on the Sun! Also, compare $\Phi_\mathrm{tot}$ in Figure~\ref{FIG_FLUX} with the corresponding $\Delta$ in Figures~\ref{FIG_4}c,~\ref{FIG_5}c. Therefore, we would expect that only a marginally strong correlation of flare activity with increasing $\Phi_\mathrm{tot}$ will be seen in the statistics, inhibiting the causality behind flaring activity. Our observations showing a correlation between cancellation and flares, suggest that the process of collisional shearing shall be taken into consideration for the improvement of solar activity prediction.

\subsection{The General Implications of Collisional Shearing}

The observation of continuous cancellation for the entire duration of the collision has significant implications. First, there is no evidence of significant (in terms of flux) discrete polarity patches cancelling at a distance from the collisional PIL, which means the cancellation must occur at the collision's contact layer, the collisional PIL. Quite conveniently, at that location, a necessary condition for cancellation is met: the distance between cancelling opposite polarities is less than $\approx$0.9 Mm~\citep{Parker_1979}. Second, the cancellation involves reconnection and topological changes of field lines between heights up to a few pressure scale heights above the photosphere~\citep{vanBallegooijen_Martens_1989}. This can happen when opposite polarity footpoints of line-tied magnetic fields are brought extremely close to each other (in our observations, collision enforces this condition) while at a certain orientation due to dragging/sliding (shearing motions, when collision progresses by shearing of the opposite polarity sunspots) along the PIL. The collisional shearing process meets these conditions in a self consistent way, as it is observed at the interface between the colliding polarities. According to the cancellation scenario, after the photospheric reconnection between colliding and sheared magnetic field line bundles, the topology gradually changes from sheared arcade field lines (green lines in Figure~\ref{FIG_6}) into a series of convex-up (red dashed line) and convex-down (black) field lines. The convex-down magnetic field is submerging into the SCZ due to a downward-pointing magnetic tension force exceeding the upward-pointing magnetic buoyancy force. The convex-up part stays in the atmosphere. This gradually forms a filament channel spanning in height from the the photosphere and chromosphere to the corona above the PIL where the cancellation occurs (Figure~\ref{FIG_6} c).

Another finding is that the activity in the corona appears to scale with cancellation rates - the faster the cancellation the more intense the clusters of activity. We make a quantitative comparison between the cancellation rates due to collisional shearing and due to that of the slow and intermittent cancellation at the external PIL between AR12017 and AR12018 (discussed in \S~\ref{OBS}). For the quantitative comparison we treat this number as an estimate for a nominal cancellation rate that occurs in PILs of decaying ARs. The rate of continuous cancellation of P1 flux at the external PIL was $d\Phi_\mathrm{P1}/dt$=-4.6$\times$10$^{15}$\,Mx s$^{-1}$ over a long period of $\sim$ 5 days. In the simultaneous emergence case (AR11158), the cancellation rate (in the first 24 hours during the early stage of the second episode) was 1.8$\times$ higher than the decaying AR cancellation rate. Similarly, for the sequential case (AR12017), during the emergence of N2P2 first episode, flux cancelled at a rate of 2.2$\times$ faster than AR11158 (for a duration over $\sim$12\,h) and additionally 2.9$\times$ faster after the onset of the second episode (duration $\sim$8\,h). If we compare those rates to the slow decaying AR cancellation, it is 4.0$\times$ faster for the first N2P2 episode and 5.2$\times$ faster for the second episode. Also, AR12017 produced twice as many events than AR11158 in a period of time of one day starting after the first flare event in each AR. However, the cancellation-related reconnection occurs at (or very close to) the photosphere and it is not expected to directly yield EUV or X-ray flare emission as in the case of typical reconnection in the corona. The key here is that cancellation largely progresses uniformly and not in discrete ``gusts'', since collisional shearing is ``feeding'' the cancellation continually with sheared flux along the collisional PIL. This means that the amount of submerged flux is reflecting the increase of poloidal flux wrapping around the filament channel~\citep{Green_etal_2011}. With more helical field added to the structure, the structure expands higher from the photosphere/chromosphere into the corona. Since the helical structure is not in vacuum, it would interact with the ambient field and/or magnetic topological features (such as magnetic null points, separators, or quasi-separatrix layers in the corona). This interaction by means of reconnection in the corona is effectively converting excess energy stored in the form of non-potential poloidal flux into EUV/X-ray radiation and plasma heating and as a result, weakening the overall MFR-like structure. This possibility has been suggested by~\citet{Chintzoglou_etal_2017} where an MFR structure loses poloidal flux (and thus free magnetic energy) due to reconnection with nearby flux systems producing transient brightenings as the MFR forms. The more the cancellation occurs, the more this fresh poloidal flux will expand and interact with the ambient field. When the resulting MFR has accumulated enough poloidal flux leading to a high twist, it may eventually become unstable and begin to lift. By lifting off, it may accumulate additional poloidal flux (e.g. \citealt{Patsourakos_etal_2013, Chintzoglou_etal_2015}) via confined flaring in X-rays and eventually launch CMEs. 

An interesting question is how much flaring activity and eruptions an AR hosts during its entire lifetime, from the emergence phase all the way to the end of the decay phase. Sporadic flare activity and eruptions indeed occur in decaying ARs. For the latter, it is understood to be a result of magnetic cancellation at the PIL that separates the decayed and diffuse opposite polarity distributions (sometimes giving the impresion of a ``bipolar'' decaying AR, while earlier in its life may have been multipolar). To distinguish this kind of ``slow'' cancellation in the decay phase from the rapid cancellation we observe in emerging ARs we call it ``self-cancellation'' in the case it happens in the internal PIL of a decaying AR. It is also possible to have cancellation in an external PIL forming between two different diffuse and decaying ARs (e.g. the ePIL between AR12017 and AR12018; Figure~\ref{FIG_1}e-h). However, this case (inter-AR cancellation) does not address the question of how much activity a \emph{singular} AR produces. Here, our observations suggest that the answer to this question is related to the total amount of flux cancelled in an isolated AR. To measure the total flux cancelled during the lifetime of an AR, one needs to sum the contributions from the cancellation during the emergence phase (the flux-deficit $|\Delta|$ measuring cancellation between two colliding BMRs \emph{within} the same AR) and from the cancellation at the decay phase which is measured by subtracting the end flux (i.e. the total AR flux measured at a time after the end of emergence, $t_\mathrm{end}$) from the peak flux at the end of the emergence phase, $max(\Phi_\mathrm{tot})$, as 

\begin{equation}
	\Phi^{total}_\mathrm{canceled}=\underbrace{\sum|\Delta|}_\mathrm{emergence\ phase}+\underbrace{\bigg[max\bigg(\Phi_\mathrm{tot}(t_\mathrm{start}<t<t_\mathrm{end})\bigg)-\Phi_\mathrm{tot}(t>t_\mathrm{end})\bigg]}_\mathrm{decay\ phase}.
\end{equation}

\noindent The emergence phase part of this formula would be composed of only one $|\Delta|$ in the case of collision between two nonconjugated polarities. At the event of additional emergence and collision from additional bipoles(s), additional $|\Delta|$ should be added to the total cancelled flux in the AR, and so on. This covers the case of multipolar ARs (AR11158 indeed hosted additional sequential emergence on February 13 and on February 16 that we did not consider in this work). The decay phase part of this formula also includes the contribution of cancellation that occurs along external PILs during the decay phase of an AR, when the AR flux spreads and can interact with neighboring flux systems external to the AR (inter-AR cancellation). Also, emergence can happen in the decaying phase of a pre-existing bipole (such as in AR12017). Thus proper care must be taken to ensure the correct characterization of cancelled flux in the emergence and decay phase of a bipole. Note that during emergence, small cancellation events occasionally occur in the self-PIL of individual bipolar ARs. These typically occur when the self-PIL contains intermixed polarities (salt-and-pepper appearance of small magnetic elements in magnetograms) which motions lead them to head-on collision (shearing is not necessary in their PILs as they cancel). These events are not measured by the flux-deficit $\Delta$, since cancellation occurs in the self-PIL of a bipole that is balanced (no collision occurs between two different bipoles and thus no deficit is produced). In essence, it would have an impact to the total flux $\Phi_\mathrm{tot}$, however the rate of flux emergence is overpowering the flux decrease in the evolution of the total flux. Such small-scale cancellation events are often associated with Ellerman bombs (e.g.~\citealt{Georgoulis_etal_2002}) and UV bombs (e.g.~\citealt{Peter_etal_2014}), whose contribution to the overall energy release budget from the emerging AR is confined to the heating of the low atmosphere (chromosphere and transition region).   

We summarize the phenomenology of collisional shearing with a cartoon model in Figure~\ref{FIG_7} demonstrating the essence behind the two possibilities of (a) simultaneous and (b) sequential cases of emerging bipoles, nested within the same AR. The confined flaring following the onset of collision and the duration of the flare clusters for periods of ongoing photospheric cancellation points to the formation of MFR structures in the corona along the cPILs (pre-existing the eruption; e.g. \citealt{Patsourakos_etal_2013, Chintzoglou_etal_2015}). Reconnection may also occur between adjacent sheared arcade field higher in the corona (as in ``tether-cutting'' reconnection;~\citealt{Moore_etal_2001}) and/or other weakly twisted MFR structures co-hosted along the length of the same cPIL, which with additional reconnection may conglomerate into larger and more monolithic structures that eventually launch as CMEs~\citep{Chintzoglou_etal_2015}. The interaction of the newly forming MFR structures with their immediate environment, i.e. near the collisional PIL as well as potential interaction with the magnetic topology (e.g. interaction of a confined MFR with the topology of quadrupolar configuration; ~\citealt{Chintzoglou_etal_2017}), may also play a role in enriching the activity of the flare clustering. 



\section{Conclusion}\label{CONCLUSION}

In this article we presented the analysis of observations of highly flare- and CME-productive ARs in support of a new scenario for the origin of major solar activity. We examine the amount of flux cancelled in collisional PILs and the timing of flux cancellation with respect to flare activity. Using two well observed flare- and CME-productive ARs composed of two conjugate bipoles that emerged (a) simultaneously (AR11158) and (b) sequentially (AR12017), we show that clusters of intense flare activity correlate with the onset and duration of magnetic cancellation in collisional PILs. The cancellation begins at the onset of collision between the opposite-signed nonconjugated polarities and not at the self-PILs of the individual emerging bipoles/flux tubes.

Both the emergence-based and cancellation-based simulations aim to create the conditions that lead to eruptions from first principles. However, despite the soundness of their simplistic approach (the latter largely imposed by computational power restrictions in the previous decades), the results from our observations are in contrast with these scenarios, at least during the phases of simple bipolar AR emergence (i.e. no activity in the self-PILs of individual bipoles) and as ARs mature (before they enter the decay phase and cancellation of diffused polarities takes place which may result to flaring and eruptive activity). It is also important to note that even the standard eruptive flare model, the CSHKP model, invokes a simple bipole, also not consistent with our findings.

Our proposed scenario considers cancellation occurring between colliding flux tubes that emerge simultaneously of sequentially, forming a single flare-productive AR (Figure~\ref{FIG_7}, $t_0$ panels). During the emergence stage, each flux tube manifests itself at the photospheric surface in the form of two conjugate polarities of opposite sign. As the emergence stage
progresses, these conjugate polarities undergo a mutual selfseparation,
following the canonical diverging polarity motions in each bipole as seen on the surface. As a result, collision (convergence) and shearing may occur between oppositesigned nonconjugated polarities, forming a collisional PIL and
lasting for as long they are in close contact and while the bipoles self-separate (Figure~\ref{FIG_7}, $t>t_\textrm{\small collision onset}$). Activity follows shortly after the onset of collision ($\approx$12\,h) and persists for the duration of the collision. This process develops self-consistently due to the proper motions of colliding polarities. To emphasize the essence of these proper motions leading to collision and to differentiate from the cancellation scenario that involves only a single conjugated bipole, we introduce the term ``\emph{collisional shearing}'' to refer to the observed process. As a result of collisional shearing, cancellation occurs in the collisional PIL between the two individual conjugate bipoles (i.e. collision between the opposite-signed nonconjugated polarities; Figures~\ref{FIG_2},~\ref{FIG_6}). Doppler measurements show a dominance of average redshifts over average blueshifts at the collisional PIL (Figure~\ref{FIG_DOPPLER}). This is consistent with photospheric cancellation, by submergence of flux at the collisional PIL. Cancellation is followed by activity clusters of flaring and confined activity (Figures~\ref{FIG_2},~\ref{FIG_3}, \ref{FIG_4}, \ref{FIG_5}). We emphasize that virtually no flaring or eruptive activity occurs at the self-PILs separating the conjugated bipoles (Figures~\ref{FIG_6},~\ref{FIG_7}). The intense flaring and eruptive activity produced in the corona is rooted at the collisional PIL (Tables 1 and 2, Figures~\ref{FIG_2},~\ref{FIG_3}), effects consistent with the formation of stressed and twisted magnetic segments (i.e. magnetic flux ropes) that progressively reconnect and coalesce into more monolithic flux systems (e.g.~\citealt{Chintzoglou_etal_2015}). Observational evidence (Figures~\ref{FIG_EUV1}, \ref{FIG_EUV2}) as well as data-driven modeling (Figures~\ref{FIG_MF}, \ref{FIG_MF2}) provide additional support to the creation of stressed and twisted magnetic structures at the collisional PIL, \emph{before} they drive eruptions.

\begin{figure*}
      \epsscale{1.0}
       \plotone{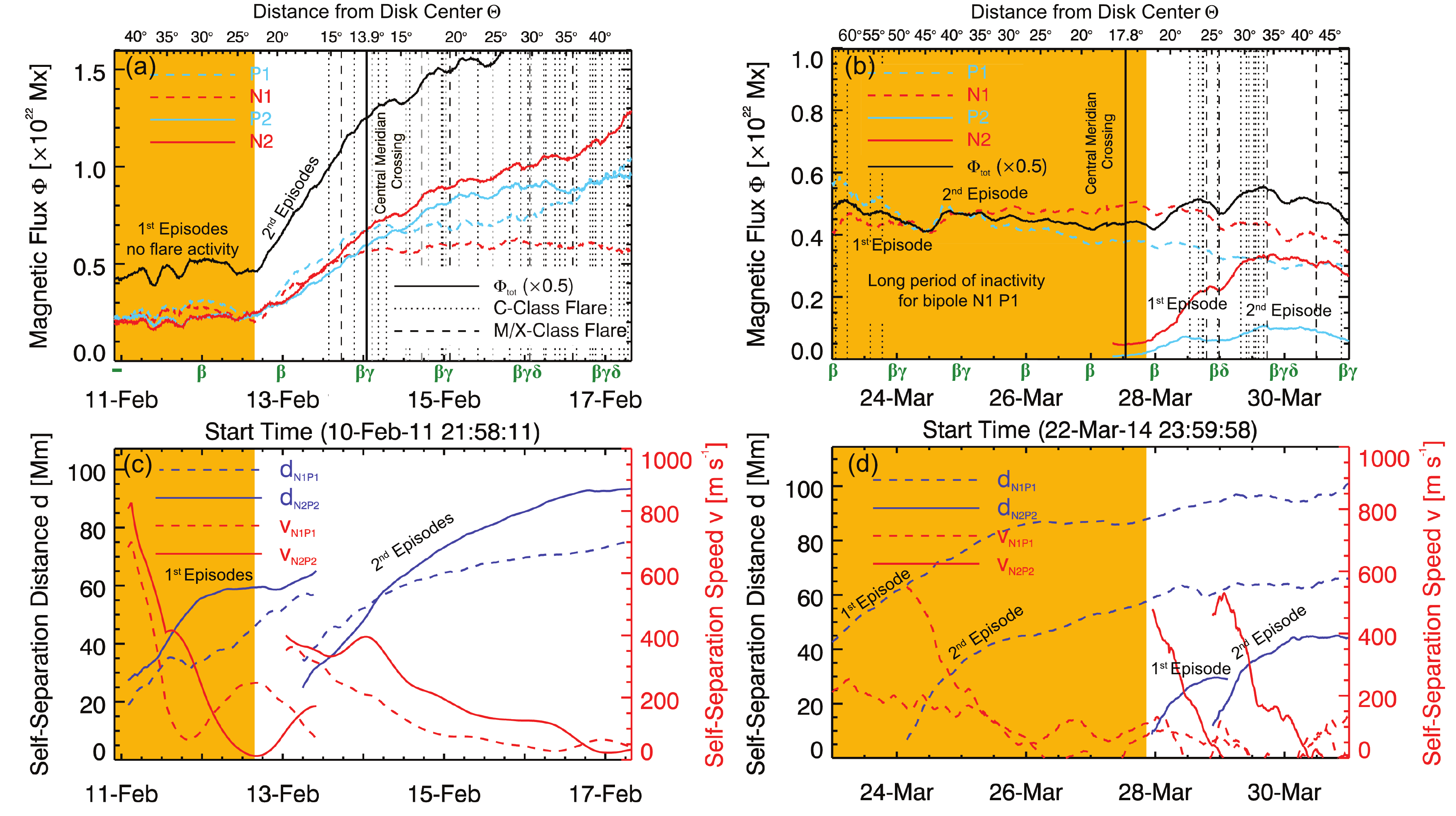}
       \caption{ 
	Top panels: The time evolution of the magnetic flux for each of the four polarities (dashed lines for bipole 1, solid for bipole 2) with the flare events (vertical lines) produced during disk transit for AR11158 (a) and AR12017 (b). The dark curve shows the total unsigned magnetic flux (multiplied by 0.5). The orange background denotes the period over which virtually no flare activity occurred in the ARs (corresponding to the end of the first episode in AR11158 and the emergence of the first bipole of AR12017 which only produced four minor C-class flares). A striking similarity between each AR is that each individual bipole undergoes a two-stage evolution (two episodes). The evolution of the daily Mt Wilson sunspot classification for the ARs is also shown with green characters in the time axis. Bottom panels: The time evolution of the self-separation distance and the self-separation speed for each bipole (dashed lines for bipole 1, solid for bipole 2) of AR11158 (c) and AR12017 (d). Note: for presentation clarity, in panel (d) we don't display the self-separation distance and speed for the north-south oriented bipole and we only show the first and second episodes of the east-west bipole that composes N2 P2. Typical to the behavior of simple emerging bipoles, each individual bipole self-separates rapidly during the early stages of its emergence. Individual episodes introduce additional magnetic flux and also further extend the time period of rapid polarity motions within the same AR.
}\label{FIG_FLUX}
\end{figure*}

\begin{figure*}
	\epsscale{0.8}
	\plotone{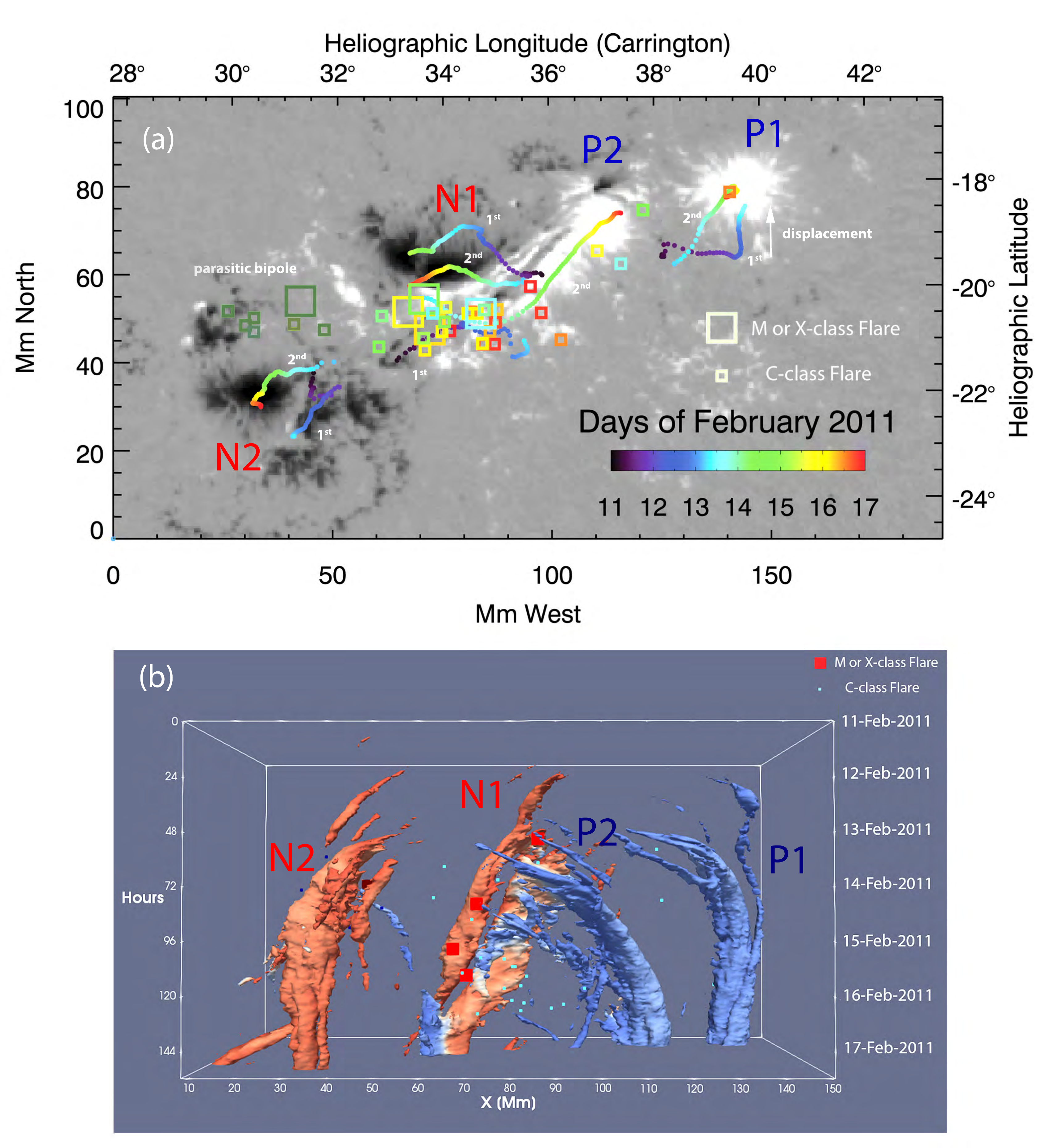}
	\caption{3D collision of emerging flux-tubes at the photosphere as a result of a simultaneous emergence process (AR11158). Panel a: Multiple day tracking of magnetic polarities (color-coded traces on the Heliographic plane) superimposed on a CEA magnetogram map of an intermediate time instance (at 15-Feb-2011 21:58 UT). The emergence occurred in two episodes per flux tube (marked as 1$^\mathrm{st}$ and 2$^\mathrm{nd}$ episodes) with the 2$^\mathrm{nd}$ episode of emergence being stronger as compared to the 1$^\mathrm{st}$. Proper motions of polarities are more dramatic during the self-separation associated with 2$^\mathrm{nd}$ episode per each bipole. As a result, a collision occurs between polarities N1 and P2. Overlaid boxes mark the centroids of flaring activity (large box: M and X class flares, small box: C-class flares). Note the containment of the activity along the collisional PIL. The dark green boxes denote events correlated with the emergence of the parasitic bipole north of N2 (see Figure~\ref{FIG_1}) and are not included in our analysis (minor events in a different PIL; also Table 1 marked with asterisk). Panel b: 3D space-time representation of the collision using isosurfaces of $|\textbf{B}|$=1,200 G as viewed from the South. The squares show the location and time of activity, which is predominantly clustered at the collision site. (An animation of this figure as a 3D flyby of panel b is available in the online version of the journal.)}
\label{FIG_2}
\end{figure*}

\begin{figure*}
	\epsscale{0.8}
	\plotone{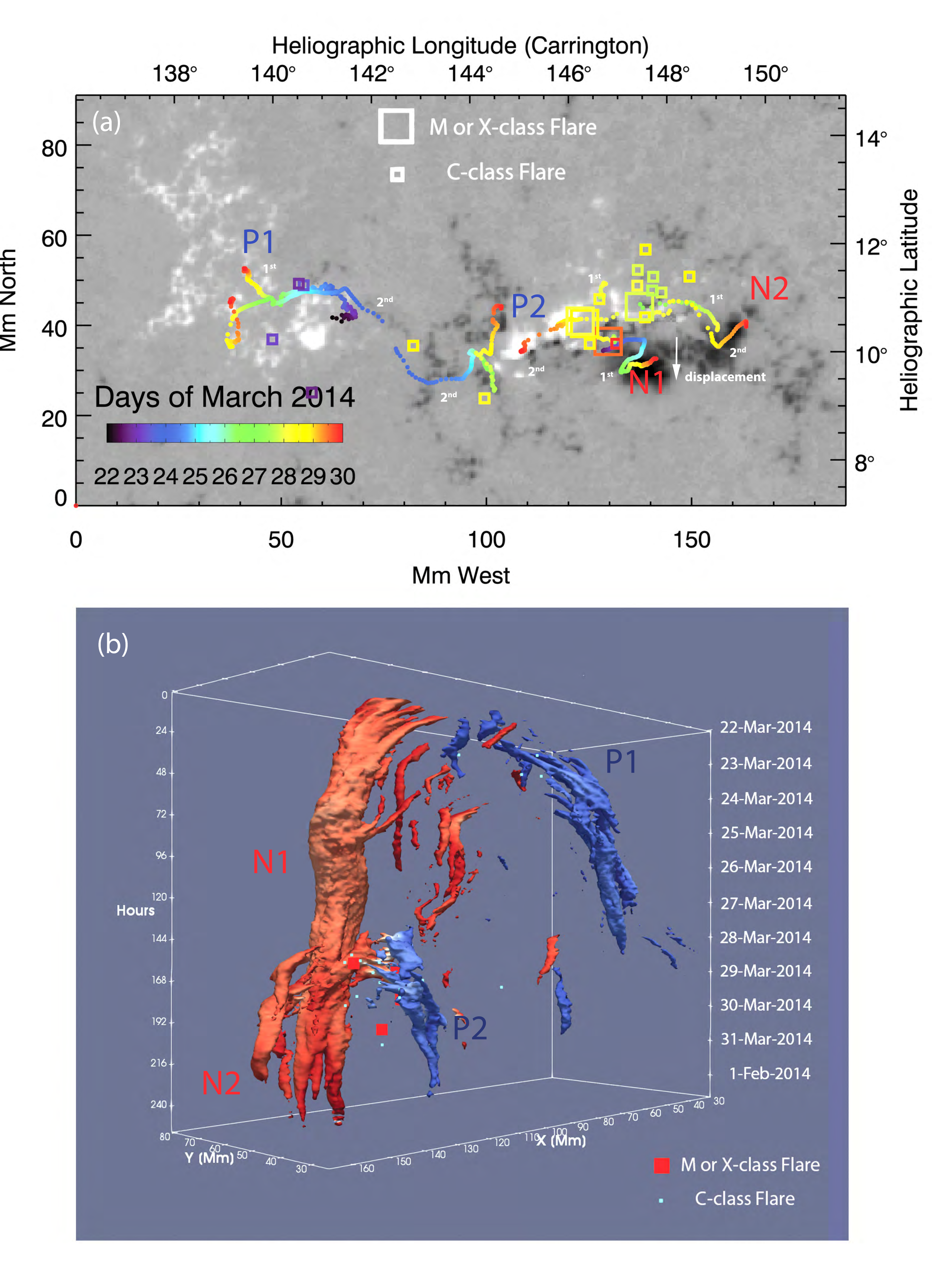}
        \caption{3D collision of emerging flux-tubes at the photosphere as a result of a sequential emergence process (AR12017). Panel a: Multiple day polarity tracking (background magnetogram at 30-Mar-2014 02:59:46 UT). Flaring activity occurs at the location of the newly bipole N2P2 emerging right by the pre-existing polarity N1. Each big box was also a CME. Panel b: 3D view of the collision for AR12017. The view is from the northwest looking towards the southeast.
(An animation of this figure as a 3D fly-by of panel b is available in the online version of the journal.) 
}\label{FIG_3}
\end{figure*}

\begin{figure*}
        \epsscale{0.8}
        \plotone{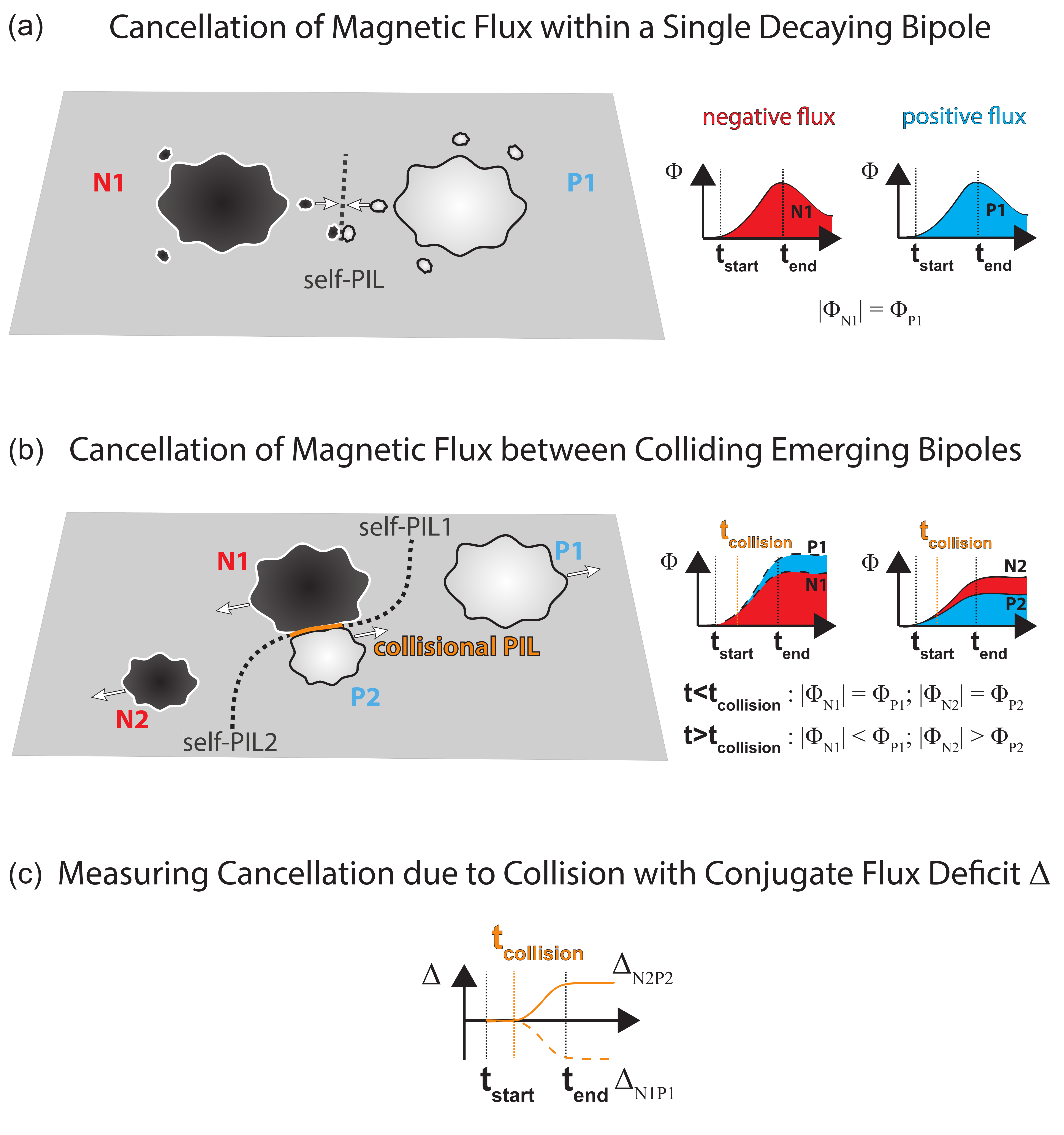}
        \caption{ 
	Cartoon illustrating the observed conjugate flux deficit, $\Delta$, as a result of flux imbalance between two colliding conjugate bipoles. (a) The case of a single bipole emerging in isolation from other flux concentrations. As a result, the negative flux is in balance with the positive, yielding zero conjugate flux deficit. After the end of the emergence phase, the decay phase becomes apparent in the unsigned fluxes due to cancellation of opposite polarity fragments at the self-PIL. (b) The case of two emerging conjugate bipoles (N1P1 and N2P2 with white arrows showing the self-separation) comprising a single AR, where collision occurs between the opposite-signed nonconjugated polarities (N1 and P2). After the collision onset ($\mathrm{t_{collision}}$) an imbalance develops between the negative and the positive flux in each conjugate bipole (N1 flux less than P1 flux and inversely, P2 flux less than N2). (c) The imbalance is the non-zero conjugate flux deficit $\Delta$ for each conjugate bipole (orange curves), a result of flux cancellation at the collisional PIL. Note that the flux is always balanced for the quadrupolar AR as a whole (summing the deficits should yield zero).
}\label{FIG_DEFICIT}
\end{figure*}

\begin{figure*}
        \includegraphics[width=\linewidth]{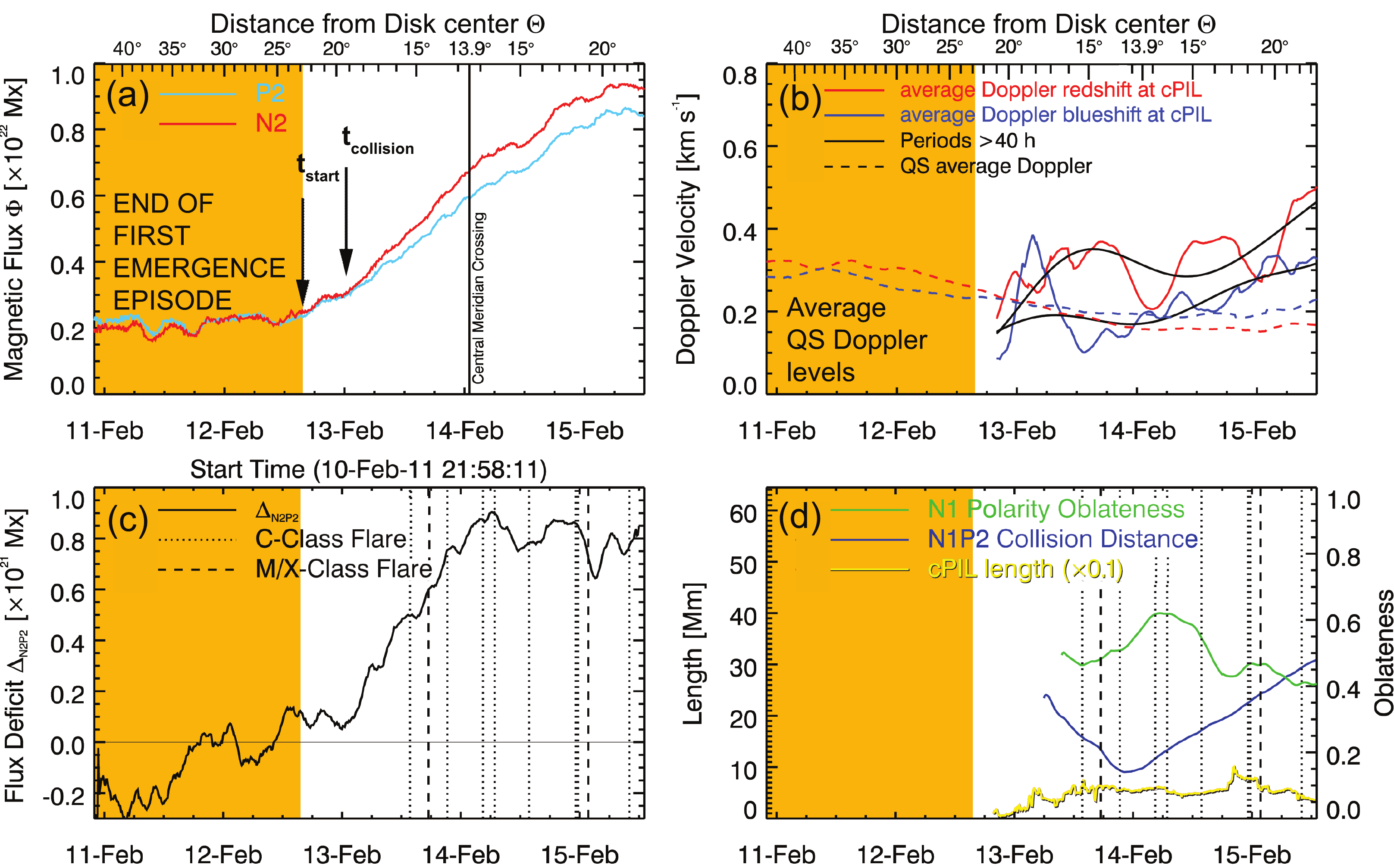}
	\caption{
		Panel a: Time evolution of the magnetic flux for bipole 2 of AR11158 (conjugated polarities N2 and P2). The orange background marks the time range up until the onset of the second emergence episode. Panel b: Average photospheric Doppler velocities for the nearby QS (dashed) in addition to average Doppler velocities in the collisional PIL (solid curves). Note the relative dominance of average redshifts over blueshifts, and also as compared to the QS average velocities. Panel c: Magnetic flux deficit with C-class flares (purple), M-class and X-class (red) overplotted. Note that a cluster of flare activity begins after the ascending phase of the deficit in both bipoles suggesting correlation of the flares with photospheric cancellation. The deficit drops due to smaller emergence events contaminating the surface used in the calculation of the magnetic flux. Panel d: Plot of the oblateness of sunspot N1, centroid distance between the colliding N1 and P2 and the resulting collisional PIL length with time. Note the very high oblateness ($>$60\%) reached around the time of the minimum collision distance related to the higher compression of N1 and P2 sunspots. See text for more discussion.
    }\label{FIG_4}
\end{figure*}

\begin{figure*}
        \includegraphics[width=\linewidth]{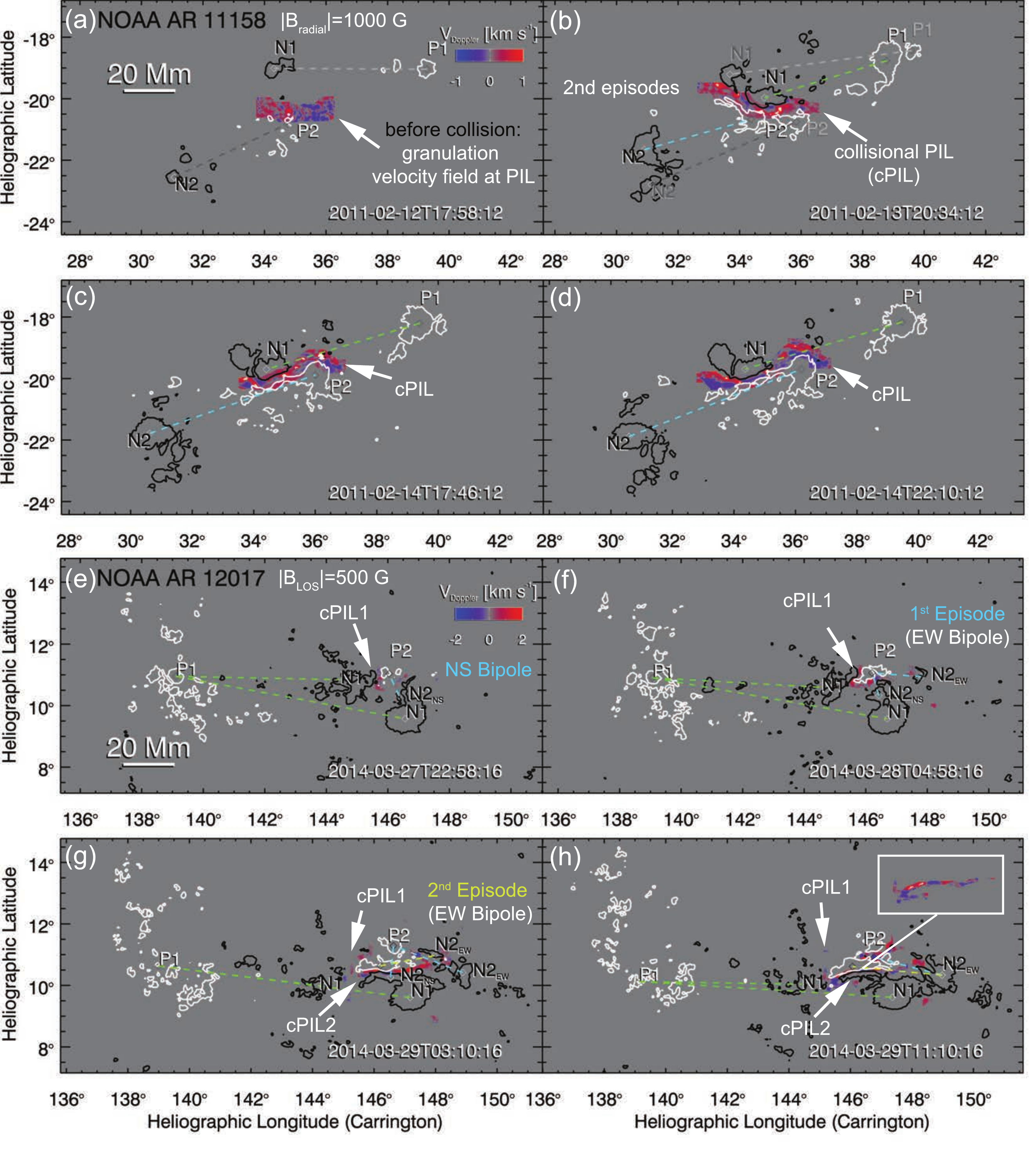}
        \caption{Doppler velocities at the location of the collisional PIL for AR11158 (panels a-d) and AR12017 (panels e-h). Different polarities (positive/white, negative/black magnetic field contours) are annotated, including these produced by different flux emergence episodes. To improve visual clarity and minimize masking from the magnetic field contours, the collisional PIL masks are either augmented in thickness to show larger areas of the velocity patterns within them, or shown in an inset panel (panel h). }
\label{FIG_DOPPLER}
\end{figure*}

\begin{figure*}
        \includegraphics[width=\linewidth]{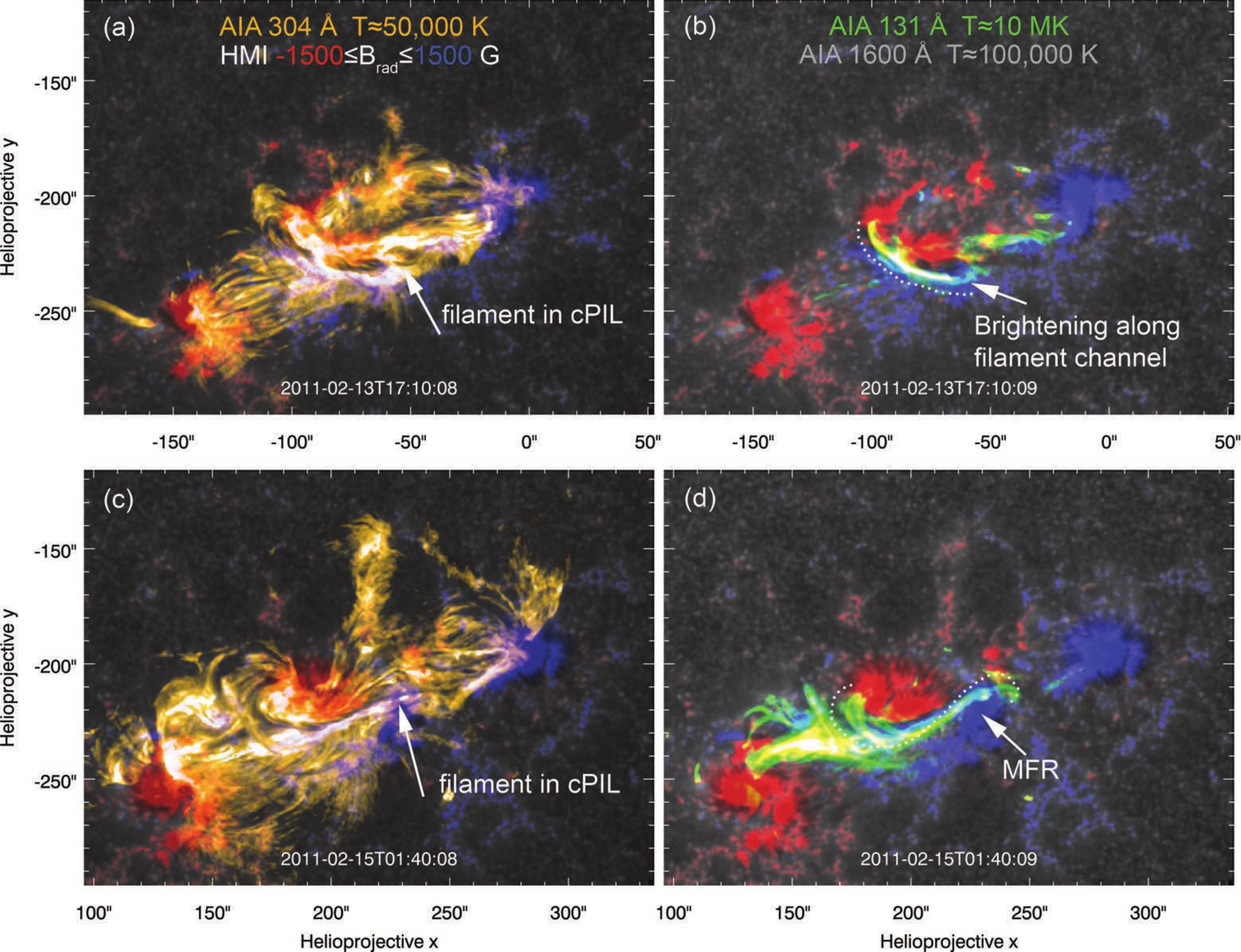}
	\caption{Selected composite images from \emph{SDO}/AIA and HMI observations for AR11158. Left: composite of 1600\,\AA\ (grey), the corresponding HMI magnetogram (red/negative, blue/positive) and 304\,\AA\ (orange) showing the cool filament material (dark lane) above the cPIL. Right: composite of 1600\,\AA\, HMI magnetogram, and 131\,\AA\ (green) showing plasma emission in 10\,MK. In panel d a clear sigmoidal structure is seen above the cPIL (traced with a dotted line) signifying a pre-existing MFR. (A movie is available in the online version of the journal.)}
\label{FIG_EUV1}
\end{figure*}

\begin{figure*}
	\includegraphics[width=\linewidth]{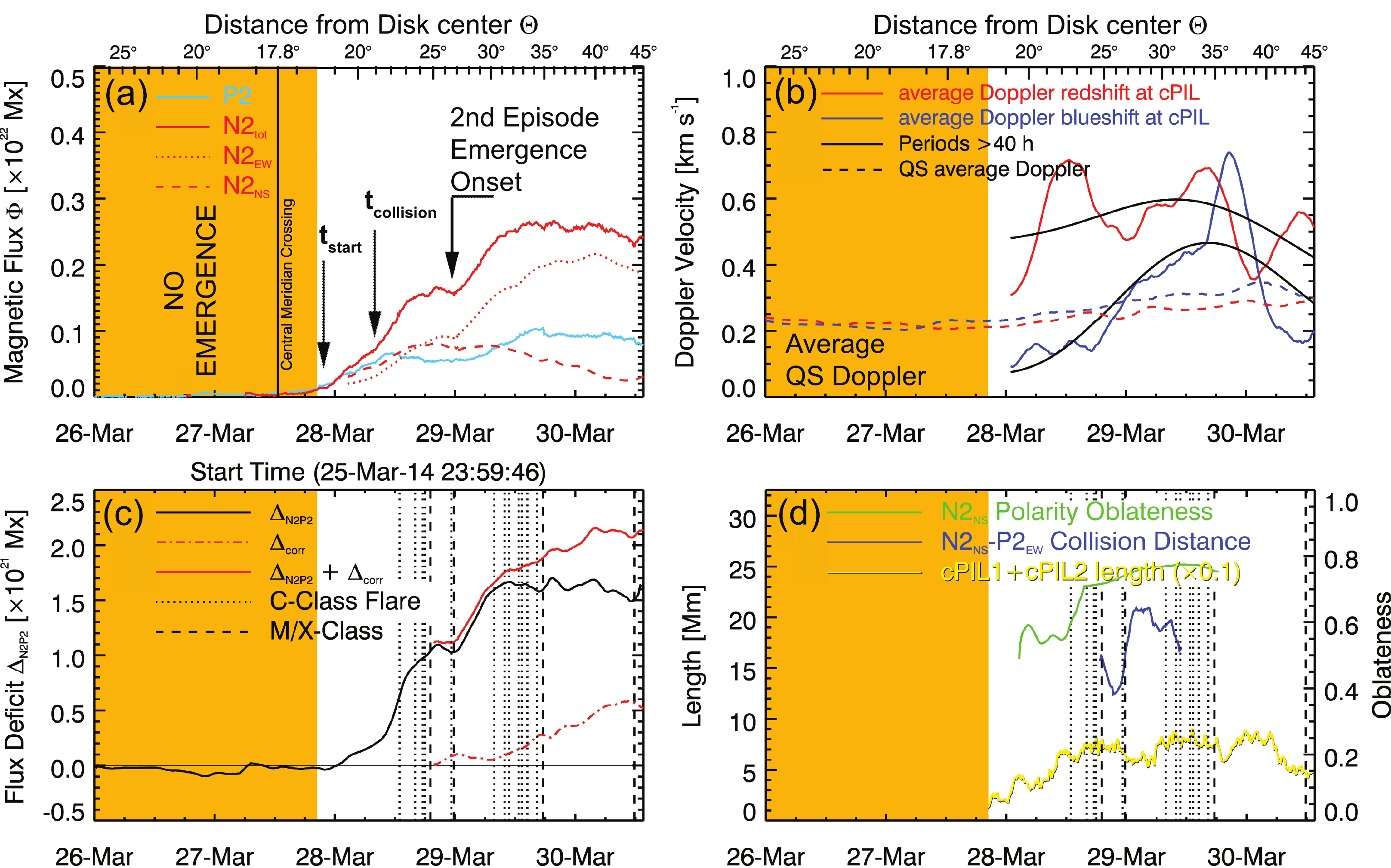}
        \caption{Panel a: Time evolution for the magnetic flux of the secondary emerging bipole of AR12017 (conjugated polarities N2 and P2). The flux measurements are base-differenced to remove the bias of the pre-existing negative flux in the area where the parasitic emergence occurs (also see Figure~\ref{FIG_FLUX} for the bias). The pre-existing bipole N1 P1 was fully emerged for four days before N2 P2 began to emerge and it is in its decay phase throughout the time range shown. The orange background marks the time range where emergence of the second bipole hasn't yet started. Bipole 2 carries half the magnetic flux of bipole 1. Panel b: Average photospheric Doppler velocities for the nearby QS (dashed) in addition to average Doppler velocities in the collisional PIL (solid curves). The average redshifts in the collisional PIL dominate both the blueshifts and QS average velocities for the entire period of emergence. Panel c: The evolution of the flux deficit for bipole 2 including the deficit correction $\Delta_\textrm{corr}$ (solid red line; see text for details). Note that a cluster of flare activity begins only when the deficit is increasing. Panel d: Plot of the oblateness of polarity N2$^\mathrm{NS}$ (green), collision distance between N2$^\mathrm{NS}$ and the second episode's P2 (blue) the PIL length (yellow) with time. Note the rapid increase of the oblateness (from 50\% to 80\%) for N2$^\mathrm{NS}$ after the onset of collision with the second episode within bipole 2. See text for more discussion.
        }\label{FIG_5}
\end{figure*}

\begin{figure*}
        \includegraphics[width=\linewidth]{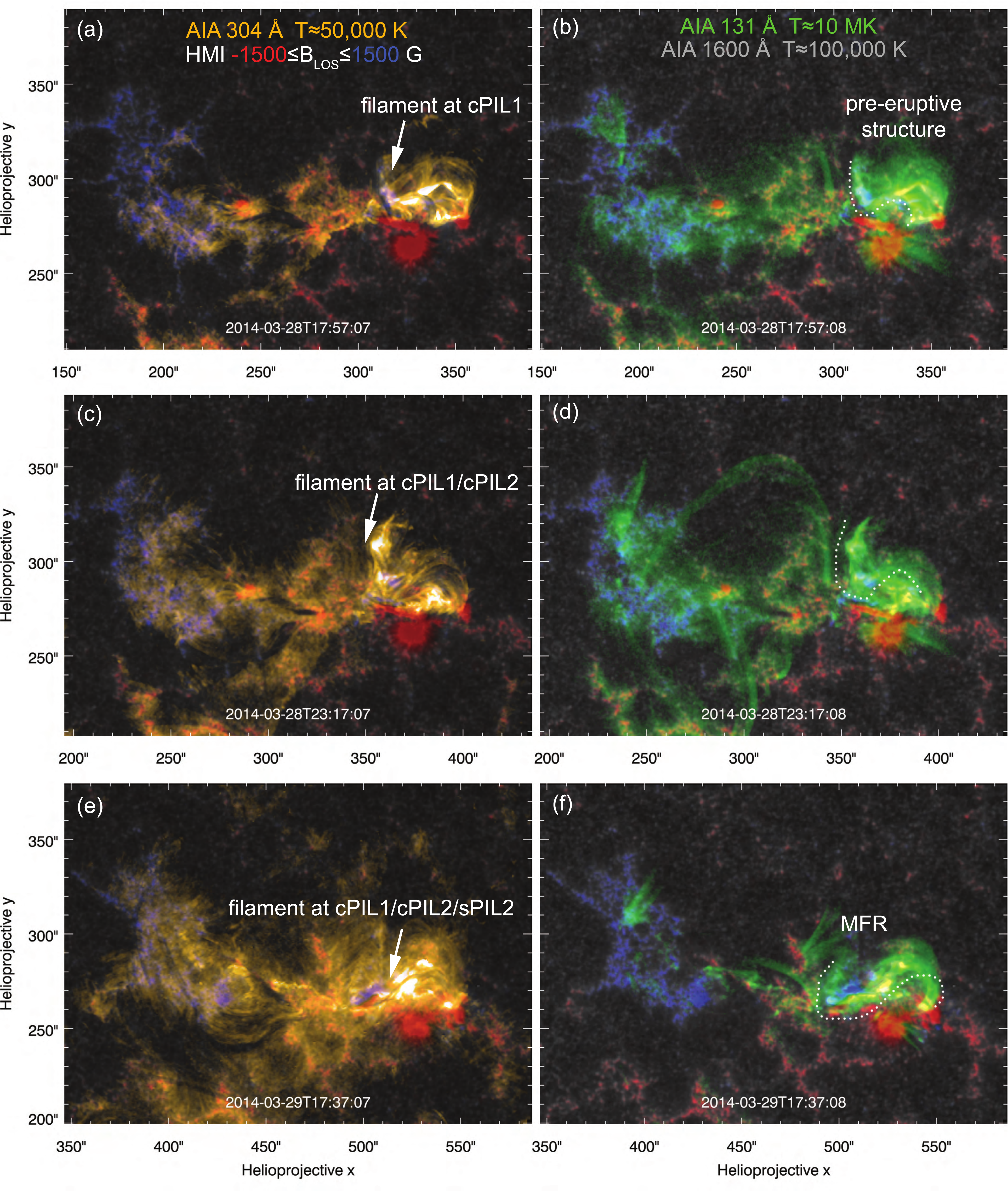}
        \caption{Selected composite images from \emph{SDO}/AIA and HMI observations for AR12017. Note that the formation of the filament occurs in the cPIL1, progressively extending to the conjoined cPIL1/cPIL2/sPIL2 system for the second and third eruption. (A movie is available in the online version of the journal.) 
}
\label{FIG_EUV2}
\end{figure*}

\begin{figure*}
        \epsscale{0.8}
        \plotone{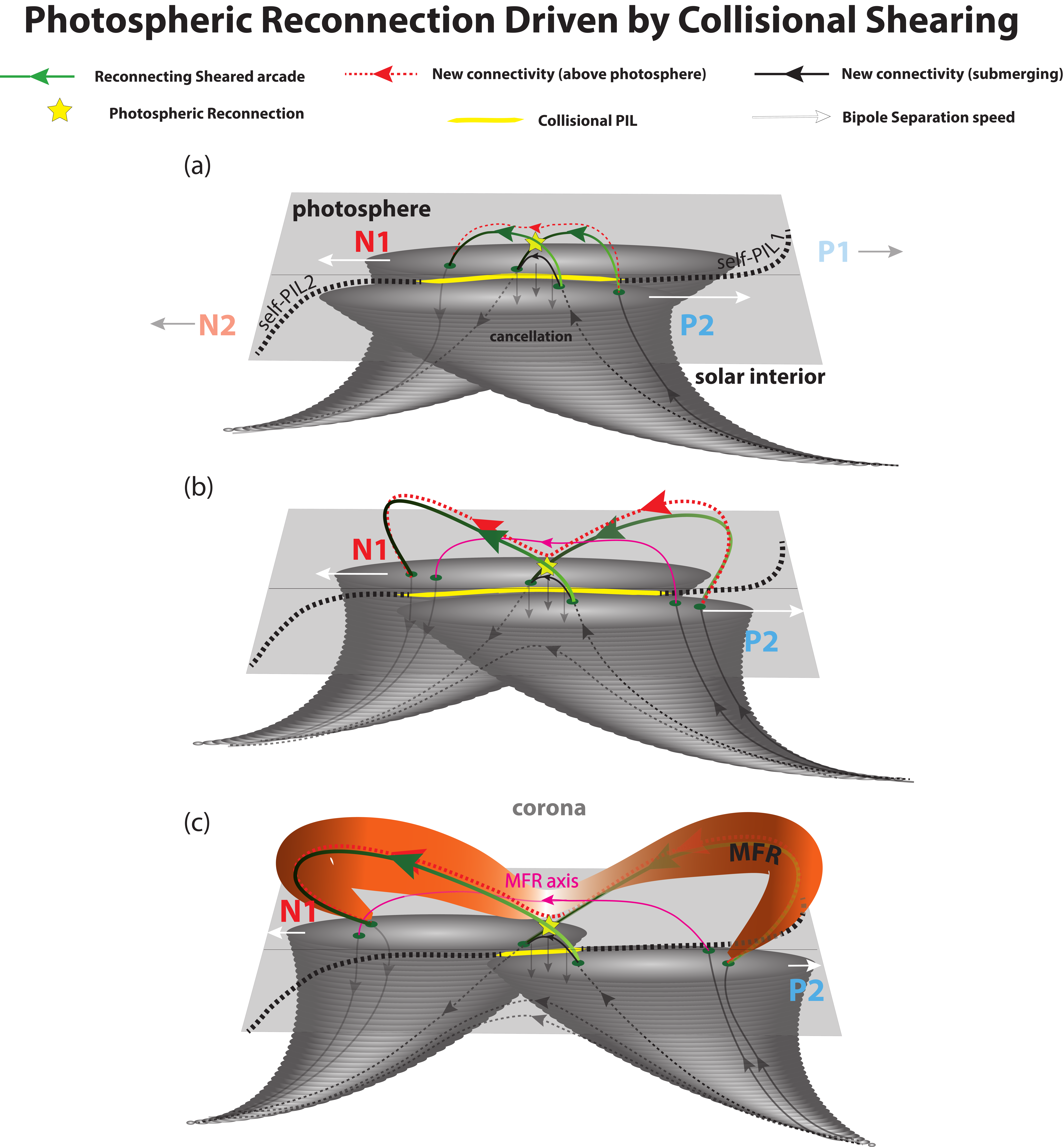}
        \caption{
		Cartoon demonstrating the physical processes at the collisional PIL between two nonconjugated polarities (their conjugate polarities are not shown to increase clarity). Proper motions (shown with white arrows) are due to the self-separation between the conjugate polarities, bringing them to a collision course and shearing as they overtake each other. As a result, the arcade connectivity near the collisional PIL shears and once it reconnects at photospheric heights flux submerges (panel a; small black loop). This manifests as a deficit in the balance of magnetic flux if colliding flux (e.g. P2) is compared against its conjugate non-colliding polarity flux (e.g. N2). In the mean time, PIL-aligned connectivity appears low in the corona (red dashed line). With ongoing cancellation (panel b), more flux reconnects below the PIL-aligned connectivity with equal amounts submerged below the photosphere and added as poloidal in the structure above the cPIL eventually forming a magnetic flux rope (MFR; panel c). This process is also accompanied by confined sub-flaring and flaring activity in the corona above the collisional PIL, suggesting additional increase in the poloidal flux due to reconnection below the flux rope \citep{Patsourakos_etal_2013, Chintzoglou_etal_2015}.
}\label{FIG_6}
\end{figure*}

\begin{figure*}
        \epsscale{1}
        \plotone{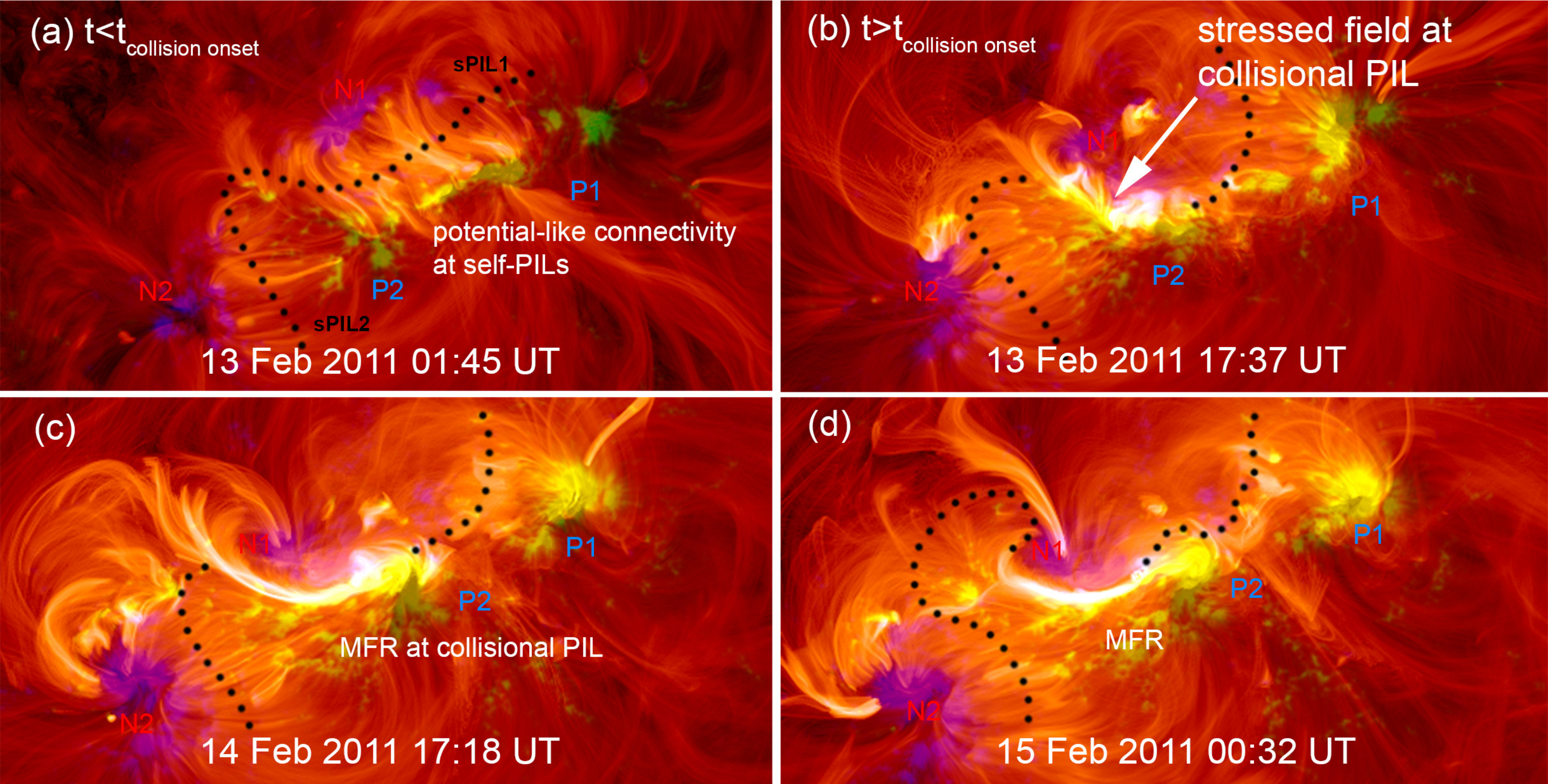}
        \caption{
		Results from the data-driven 3D MF simulation for AR11158 showing the coronal magnetic field at four different times (panels a, b, c and d) along the evolution of the AR. The view is top-down. The red-orange volumetric rendering shows the square of the electric current, $|\mathbf{J}|^2$, which represents spatial deformities in the coronal field (obtained by tracing millions of field lines at each snapshot). The maps are overlayed to the HMI magnetogram showing the magnetic polarities (purple: negative, green: positive) and the PIL is shown with a dark dotted line for reference. Before the collision onset (panel a) the connectivity is not sheared and resembles closely a potential configuration. After the collision onset (panels b,c,d), shear builds up at the collisional PIL due to the proper motions of P2 (moves rapidly towards P1) and N1 (mostly static) and the line-tied coronal field becomes stressed and aligned with respect to the PIL. The field connectivity between the conjugate polarities (N1P1 and N2P2) is potential-like at all times. This energization of the coronal field lasts for as long as the collisional PIL exists. 
		(An animation of this figure is available in the online version of the journal.)
}\label{FIG_MF}
\end{figure*}

\begin{figure*}
        \epsscale{1}
        \plotone{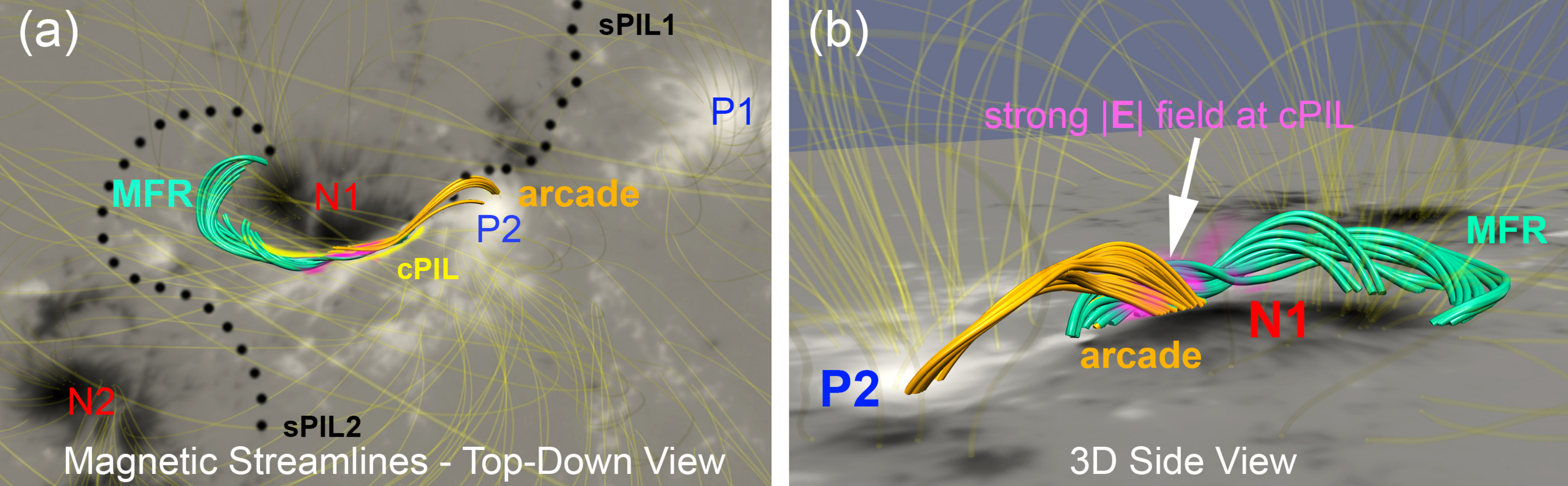}
        \caption{
		Panel a: streamlines of the 3D magnetic field corresponding to the time in Figure~\ref{FIG_MF} d. Translucent yellow lines show the ambient coronal field. The green (Magnetic Flux Rope) and orange (simply-sheared arcade) field lines were traced from the very same region of strong electric field magnitude right at the collisional PIL (magenta volumetric rendering). Panel b: perspective view of the MFR above the collisional PIL.
                (An animation of this figure is available in the online version of the journal.)
}\label{FIG_MF2}
\end{figure*}

\begin{figure*}
        \includegraphics[width=\linewidth]{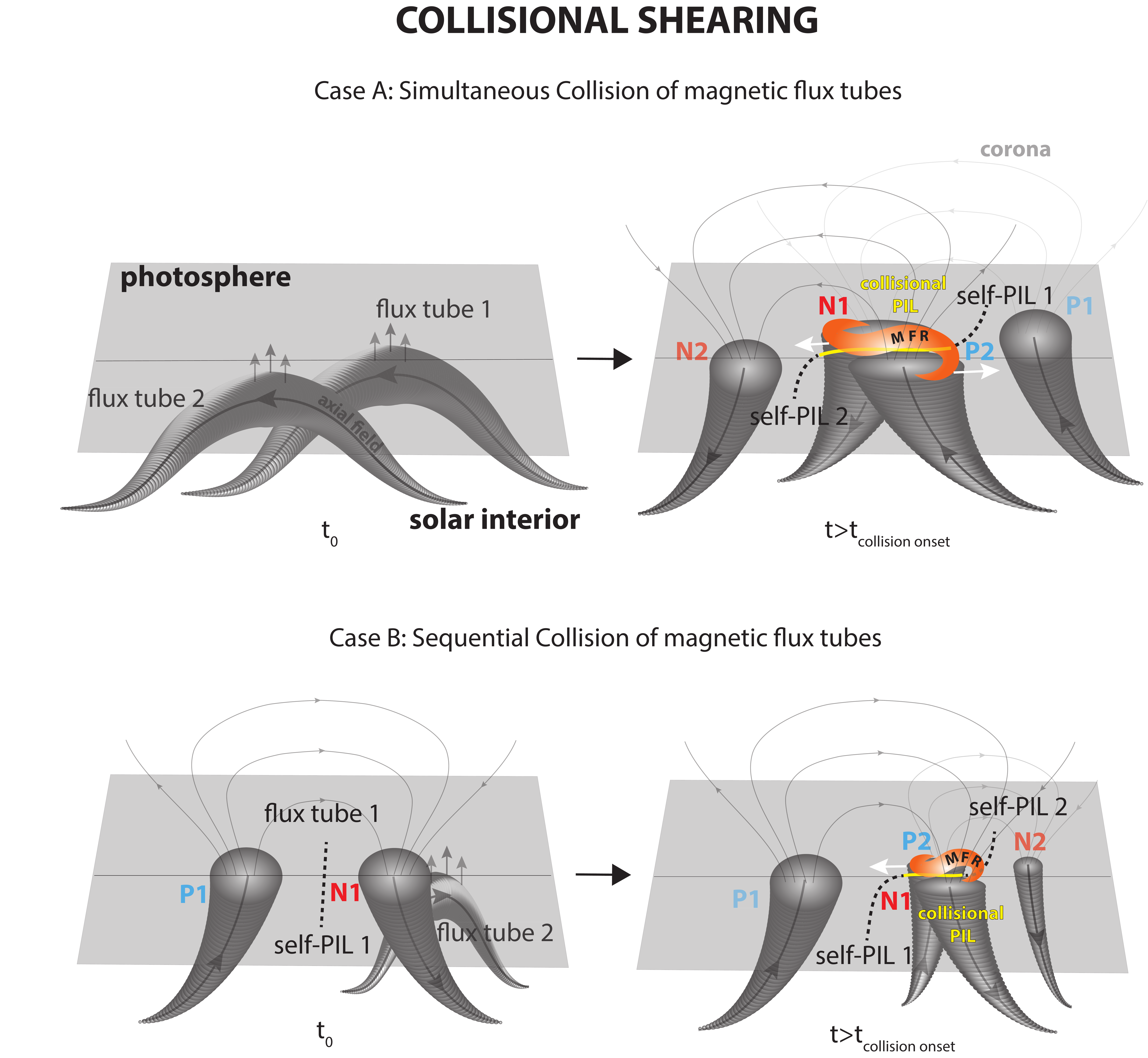}
        \caption{
		Cartoon representation of the simultaneous and sequential collisional shearing processes. Time $t_0$ shows the AR just before it becomes quadrupolar. For each flux tube that emerges a self-PIL is separating the opposite conjugated polarities (i.e. N1 P1 for flux tube 1 and N2 P2 for flux tube 2). However, if the proximity of the two flux tubes emerging within the same AR is close, this may result into a collision between opposite nonconjugated polarities (at $\mathrm{t=t_{collision\ onset}}$). The collision forms a neutral line, the collisional PIL, linking self-PIL1 with self-PIL2. The orientation and shape of the long neutral line system, self-PIL1/collisional-PIL/self-PIL2, changes with time due to the proper motions of the polarities relative to each other. Magnetic cancellation occurs at the collisional PIL accompanied by rich flaring activity in the corona, in addition to the formation of MFRs above the collisional PILs. The MFRs may eventually become unstable and escape as CMEs. Note that sequential collisions may occur at any time during the lifetime of an AR (e.g. it occurs during the simultaneous collision in AR11158), enriching its flaring and eruptive potential.    
}\label{FIG_7}
\end{figure*}

Traditional measurements of the unsigned magnetic flux
taken over the entire surface of a complex multipolar AR are
naturally unable to reveal the action and the role of the
cancellation process during emergence (that is, measuring the
net reduction of flux even though an AR is emerging; see $\Phi_\mathrm{tot}$
in Figure~\ref{FIG_FLUX}).To overcome this limitation, in this paper we
introduced a novel measurement technique, the \emph{conjugate flux
deficit method} (illustrated in Figure~\ref{FIG_DEFICIT}), which allows the
measurement of the total amount of canceled magnetic flux and
also the rate of magnetic cancellation at the collisional PIL
(Figures~\ref{FIG_4},~\ref{FIG_5})). The ``flux deficit'' method reveals when and for
how long magnetic cancellation occurs by means of recording
the flux imbalance (in fact, a deficit) that develops gradually
between nonconjugated polarities upon collision. The fact that
cancellation is measured between nonconjugated polarities in
emerging ARs has significant implications in understanding the
flare clustering and the origin of eruptive flares. Apart from the
usual shearing of magnetic fields in emerging ARs, our
identification of magnetic cancellation in such ARs once again
points to the formation of magnetic flux ropes on the Sun
before they destabilize and erupt. Therefore, revealing the
physical processes that take place in collisional PILs is of
outmost importance and it is suggestive that our findings shall
be used to constrain numerical models of eruptive ARs as we
move toward more accurate flare and CME prediction schemes.

With the above we conclude that ``bipole-bipole interaction'' in the photosphere by means of collisional shearing is a culprit behind intense solar activity. The activity is hosted at the collision interface, the collisional PIL. On the other hand, the self-PILs, (i.e. the PIL segments separating the polarities of conjugate bipoles \emph{within the same ARs}) do not seem important in hosting the activity clusters. This view, given it requires at least two bipoles, is inconsistent with the CSHKP standard model, and the MHD emergence-based and cancellation-based, ``data-inspired'' 3D models of a single bipole. Sequential emergence in ARs classified as $\delta$ ARs has been resolved in fine structure inside their compact PILs (as discussed, for instance, in the present paper and \citealt{Chintzoglou_etal_2015}). This could not have been possible in systematic observations of lower spatial resolution magnetograms, which was the case before the SDO/HMI era. Finally, we stress that in order to understand the formation of compact, flare-productive PILs (and essentially $\delta$ ARs), it is imperative to study and use as examples only well-observed emerging ARs, meaning they are observed from the very beginning of their birth all the way to their maturity. 

This work \emph{underlines the necessity} for future vector magnetograph instruments positioned in different vantage points in the solar system, capable of providing the true radial magnetic field at the solar surface. One possibility is the L5 Lagrangian point at Earth's orbit, which could allow us to maximize the observational coverage of the emergence phase of ARs (which often rotate into Earth view already developed) and with that, our predictive capability for the near-Earth space weather.   


\acknowledgements

We thank an anonymous referee for comments that impoved the manuscript. G.C. would like to thank C. Schrijver, B. Welsch, and P. Demoulin for fruitful discussions. G.C. acknowledges support from NASA Earth \& Space Science Fellowship (NESSF) Grant NNX12AL73H that supported him in the preparation of this work as part of his PhD thesis. G.C. also acknowledges support by NASA contract NNG04EA00C (SDO/AIA). J.Z. is supported by NSF grants AGS-1249270 and AGS-1156120. M.C.M.C. acknowledges support by NASA's Heliophysics Grand Challenges Research grant \emph{Physics and Diagnostics of the Drivers of Solar Eruptions} (80NSSC18K0025, formerly NNX14AI14G to LMSAL) and grant 80NSSC18K0024 (formerly NNX13AJ96G) to LMSAL. M.K. is supported by the Coronal Global Evolutionary Model (CGEM) award NSF AGS 1321474. HMI and AIA are instruments on board SDO, a mission for NASA's Living with a Star program. We thank Dr. P. H. Scherrer of Stanford University for useful discussion and for providing help in the proper calibration of the Doppler measurements from the HMI instrument.


\clearpage

\clearpage

\clearpage

\clearpage

\clearpage

\clearpage

\clearpage

\clearpage

\clearpage

\clearpage

\clearpage

\clearpage

\clearpage

\clearpage

\end{document}